\documentclass[
  aps,
  prx,
  twocolumn,
  superscriptaddress,
  amsmath,
  amssymb,
  nofootinbib,
  10pt
]{revtex4-2}
\usepackage{array}[=2016-10-06]
\usepackage{dcolumn}
\usepackage[caption=false]{subfig}
\usepackage{graphicx}
\usepackage{epstopdf}
\usepackage{amsmath}
\numberwithin{equation}{section}
\renewcommand{\theequation}{\arabic{section}.\arabic{equation}}
\usepackage{physics}
\usepackage{verbatim}
\usepackage{ragged2e}
\usepackage{amsfonts,amssymb,amscd,mathtools}
\usepackage{mathalfa}
\usepackage{textcomp}
\usepackage{quantikz}
\usepackage{amsthm}
\usepackage{tabularx}
\usepackage{enumerate}
\usepackage{braket}
\usepackage{microtype}

\usepackage[dvipsnames]{xcolor}

\usepackage{tikz}
\usepackage[lmargin=.7in,rmargin=.7in,tmargin=.7in,bmargin=1in]{geometry}
\usepackage{newtxtext,newtxmath}
\usepackage{booktabs}
\theoremstyle{plain}

\usepackage{aliascnt}
\newtheorem{thm}{Theorem}
\newaliascnt{cor}{thm}
\newtheorem{cor}[cor]{Corollary}
\aliascntresetthe{cor}
\newtheorem{lem}{Lemma}
\newaliascnt{pro}{thm}
\newtheorem{pro}[pro]{Proposition}
\aliascntresetthe{pro}
\newaliascnt{claim}{thm}

\aliascntresetthe{claim}

\newtheorem{definition}{Definition}
\newaliascnt{remark}{thm}

\aliascntresetthe{remark}
\newaliascnt{ex}{thm}

\aliascntresetthe{ex}
\newaliascnt{question}{thm}

\aliascntresetthe{question}

\newcommand{\restatementtitle}{Restatement}
\newtheorem*{restatementinner}{\restatementtitle}

\DeclareMathOperator{\supp}{supp}

\newcommand{\inc}{\mathbb{D}}
\newcommand{\dep}{\operatorname{depth}}
\newcommand{\codedep}{\operatorname{depth}}

\newcommand{\F}{\mathbb{Z}_2}
\newcommand{\Z}[1]{\mathbb{Z}_{#1}}
\newcommand{\Zm}{\mathbb{Z}_{2^m}}
\newcommand{\Ph}{\mathcal{P}}
\newcommand{\CH}{\mathcal{C}}
\newcommand{\Hcode}{\mathcal{H}_{\mathrm{c}}}
\newcommand{\Hphys}{\mathcal{H}_{\mathrm{p}}}
\newcommand{\Lg}{\mathcal{L}}
\newcommand{\Trv}{\mathcal{I}}
\DeclareMathOperator{\im}{im}

\newcommand{\pair}[2]{\langle #1,#2\rangle}
\newcommand{\T}{\mathsf{T}}
\newcommand{\ii}{i}
\newcommand{\lpol}[1]{\overline{#1}}

\newcommand{\ba}{\begin{eqnarray}}
\newcommand{\ea}{\end{eqnarray}}
\newcommand{\bann}{\begin{eqnarray*}}
\newcommand{\eann}{\end{eqnarray*}}
\newcommand{\bal}{\begin{equation}\begin{aligned}}
\newcommand{\eal}{\end{aligned}\end{equation}}

\newcolumntype{L}[1]{>{\raggedright}p{#1}}
\newcolumntype{C}[1]{>{\centering}p{#1}}
\newcolumntype{R}[1]{>{\raggedleft}p{#1}}
\newcolumntype{Y}{>{\centering\arraybackslash}X}

\definecolor{Gblue}{HTML}{4285F4}
\definecolor{Gred}{HTML}{EA4335}
\definecolor{Gyellow}{HTML}{FBBC05}
\definecolor{Ggreen}{HTML}{34A853}

\usepackage{bm}
\newcommand{\CZ}{\mathrm{CZ}}
\newcommand{\CS}{\mathrm{CS}}
\newcommand{\CCZ}{\mathrm{CCZ}}
\newcommand{\CPh}[1]{\mathsf C^{|#1|-1}Z_{#1}^{1/2^{m-|#1|}}}
\newcommand{\CPhpow}[2]{\bigl(\CPh{#1}\bigr)^{#2}}
\newcommand{\Ug}[2][m]{U^{(#1)}_{#2}}
\newcommand{\Ugd}[2][m]{U^{(#1)\dagger}_{#2}}

\newcommand{\calS}{\mathcal{S}}

\newcommand{\codepar}[1]{\ensuremath{[\![#1]\!]}}

\newcommand{\eq}[1]{(\hyperref[eq:#1]{\ref*{eq:#1}})}

\renewcommand{\sec}[1]{\hyperref[sec:#1]{Section~\ref*{sec:#1}}}
\newcommand{\thrm}[1]{\hyperref[thrm:#1]{Theorem~\ref*{thrm:#1}}}
\newcommand{\lemm}[1]{\hyperref[lemm:#1]{Lemma~\ref*{lemm:#1}}}
\newcommand{\prop}[1]{\hyperref[prop:#1]{Proposition~\ref*{prop:#1}}}
\newcommand{\corr}[1]{\hyperref[corr:#1]{Corollary~\ref*{corr:#1}}}
\newcommand{\fig}[1]{\hyperref[fig:#1]{~\ref*{fig:#1}}}
\newcommand{\deff}[1]{\hyperref[deff:#1]{~\ref*{deff:#1}}}
\renewcommand{\appendix}{\par
  \setcounter{section}{0}
  \setcounter{subsection}{0}
  \gdef\thesection{\Alph{section}}
  \gdef\theequation{\Alph{section}.\arabic{equation}}
}

\usepackage[hypertexnames=false]{hyperref}
\hypersetup{
    pdfstartview={FitH},
    colorlinks=true, 
    linkcolor=Violet, 
    citecolor=OliveGreen, 
    urlcolor=Magenta 
}
\usepackage{cleveref}

\crefname{section}{Section}{Sections}
\Crefname{section}{Section}{Sections}
\crefname{subsection}{Section}{Sections}
\Crefname{subsection}{Section}{Sections}
\crefname{appsec}{Supplemental Materials}{Supplemental Materials}
\Crefname{appsec}{Supplemental Materials}{Supplemental Materials}
\newcommand{\ofsupp}{ of the Supplemental Materials~\cite{SupplementalMaterial}}
\crefformat{appendix}{Section~#2#1#3\ofsupp}
\crefmultiformat{appendix}{Sections~#2#1#3}{ and~#2#1#3\ofsupp}{, #2#1#3}{, and~#2#1#3\ofsupp}
\Crefformat{appendix}{Section~#2#1#3\ofsupp}
\Crefmultiformat{appendix}{Sections~#2#1#3}{ and~#2#1#3\ofsupp}{, #2#1#3}{, and~#2#1#3\ofsupp}
\crefformat{subappendix}{Section~#2#1#3\ofsupp}
\crefmultiformat{subappendix}{Sections~#2#1#3}{ and~#2#1#3\ofsupp}{, #2#1#3}{, and~#2#1#3\ofsupp}
\Crefformat{subappendix}{Section~#2#1#3\ofsupp}
\Crefmultiformat{subappendix}{Sections~#2#1#3}{ and~#2#1#3\ofsupp}{, #2#1#3}{, and~#2#1#3\ofsupp}
\crefformat{subsubappendix}{Section~#2#1#3\ofsupp}
\crefmultiformat{subsubappendix}{Sections~#2#1#3}{ and~#2#1#3\ofsupp}{, #2#1#3}{, and~#2#1#3\ofsupp}
\Crefformat{subsubappendix}{Section~#2#1#3\ofsupp}
\Crefmultiformat{subsubappendix}{Sections~#2#1#3}{ and~#2#1#3\ofsupp}{, #2#1#3}{, and~#2#1#3\ofsupp}
\crefname{thm}{Theorem}{Theorems}
\crefname{lem}{Lemma}{Lemmas}
\crefname{cor}{Corollary}{Corollaries}
\crefname{pro}{Proposition}{Propositions}
\crefname{claim}{Claim}{Claims}
\crefname{remark}{Remark}{Remarks}
\crefname{ex}{Example}{Examples}
\crefname{question}{Question}{Questions}
\crefname{definition}{Definition}{Definitions}
\crefname{condition}{Condition}{Conditions}
\crefname{figure}{FIG.}{FIGs.}
\Crefname{figure}{FIG.}{FIGs.}

\crefname{equation}{Eq.}{Eqs.}
\Crefname{equation}{Eq.}{Eqs.}
\makeatletter
\newcommand{\backsection}[1]{%
  \par\vspace*{0.8cm}%
  \@startsection{section}{1}{\z@}{1sp}{0.5cm}{\normalfont\small\bfseries\centering}*{#1}%
}
\makeatother

\begin{document}

\title{Homological classification of logical diagonal gates in the Clifford hierarchy}
\author{Koki Ehara}
\email{kokieha0331@g.ecc.u-tokyo.ac.jp}
\affiliation{Department of Basic Science, The University of Tokyo, Tokyo, Japan}

\author{Junichi Haruna}
\email{j.haruna1111@gmail.com}
\affiliation{Graduate School of Informatics, Kyoto University, Sakyo-ku, Kyoto, 606-8501, Japan}
\affiliation{Center for Quantum Information and Quantum Biology, University of Osaka, Japan}

\begin{abstract}
Logical diagonal gates play a central role in fault-tolerant quantum computation.
However, no general framework has been established that covers them across various codes and gate types.
In this work, we give a unified classification of the logical diagonal gates in the Clifford hierarchy that applies to general qubit Calderbank--Shor--Steane codes.
We prove that, at each level, the group of these gates is isomorphic to that of phase functions on the first homology group of the code.
We also determine a complete set of generators of the logical identity operators, which gives all the diagonal circuits without ancillas for each logical gate.
As an application, we determine the minimum depth of the intra-block logical controlled-Z gate on the toric code.
\end{abstract}

\maketitle
\addtocontents{toc}{\protect\iffalse}

\section{Introduction}\label{sec:intro}

Quantum error-correcting codes~\cite{Shor1995,Steane1996PRL} protect quantum information by encoding it into a larger Hilbert space.
They are an essential ingredient of fault-tolerant quantum computation (FTQC)~\cite{Shor1996FT,AharonovBenOr1997,Kitaev1997,KnillLaflammeZurek1996,Preskill1998}, in which large-scale quantum algorithms are executed reliably with noisy physical operations.
In such a computation, the encoded information is processed by logical gates, each implemented by a physical circuit that preserves the code space.
Universal quantum computation further requires a set of logical gates whose products approximate every logical unitary to arbitrary accuracy.

Although every logical gate can be implemented by some physical circuit, fault tolerance requires a circuit that limits the propagation of errors, such as a transversal or constant-depth circuit~\cite{NielsenChuang2010,ZengCrossChuang2011,BravyiKonig2013}.
However, the Eastin--Knill theorem rules out a universal set of transversal gates on any code that detects local errors~\cite{EastinKnill2009,ZengCrossChuang2011}, and the level of the Clifford hierarchy~\cite{GottesmanChuang1999} reachable by transversal or constant-depth circuits is bounded by properties of the code~\cite{BravyiKonig2013,Pastawski_2015,JochymOConnor2018}.
A basic problem is therefore to determine all the circuits for each logical gate of a given code and to search for a fault-tolerant one among them.

This problem has become more relevant with recent progress in quantum hardware~\cite{Bluvstein2024} and the development of new code families such as quantum low-density parity-check (qLDPC) codes~\cite{BreuckmannEberhardt2021}, which reduce the overhead of FTQC~\cite{BravyiBB2024,XuAtomArrays2024}.
Qubit Calderbank--Shor--Steane (CSS) codes~\cite{CalderbankShor1996,Steane1996PRL} form a broad class that includes many qLDPC codes, such as the bivariate bicycle codes~\cite{BravyiBB2024}, as well as topological codes such as the surface code~\cite{Kitaev2003,BravyiKitaev1998,Dennis2002,BombinMartinDelgadoOptimal2007} and the color code~\cite{BombinMartinDelgado2006}.
Even for this well-studied class, no unified theory of the circuits that implement the logical gates without ancillas has been established.

Among logical gates, diagonal gates have attracted particular attention for two reasons.
First, they include non-Clifford gates such as $\mathrm{T}$ and $\CCZ$, which together with the Clifford gates form a universal gate set~\cite{NielsenChuang2010}, whether applied directly or through magic state injection~\cite{BravyiKitaev2005,BravyiHaah2012}.
Second, they multiply each computational basis state by a phase, and therefore their logical action is specified by the phases on the logical basis states.

The diagonal gates in the Clifford hierarchy~\cite{GottesmanChuang1999} have been classified at each level~\cite{CGK2017,Rengaswamy2019}, and they include $\mathrm{S}$ and $\CZ$ at level two and $\mathrm{T}$, $\CS$, and $\CCZ$ at level three.
Previous studies have extensively studied which codes admit transversal circuits for logical diagonal gates at a given level, through divisibility and triorthogonality conditions on the check matrices~\cite{Ward1981,BravyiHaah2012,HaahTowers2018,Rengaswamy2020,HuLiangCalderbank2022,ShiTriorthogonal2024,JainAlbert2024}, parity constraints across the hierarchy~\cite{CampbellAnwarBrowne2012,AndersonDuclosCianciPoulin2014,AndersonJochymOConnor2016,CampsMoreno2026}, and the XP stabilizer formalism and self-dual codes~\cite{WebsterXP2022,WebsterQuintavalleBartlett2023,Tansuwannont2025,Tansuwannont2026}.
For certain families of qLDPC codes, transversal or constant-depth logical gates have been constructed from products of algebraic codes~\cite{GolowichLin2025}, cup products and cohomology operations~\cite{BreuckmannCupsGates2024,BreuckmannCupsGatesII2026,Kobayashi_2026,HsinKobayashiZhu2024,ZhuSikanderPortnoyCrossBrown2023}, and $ZX$ dualities~\cite{BreuckmannBurton2024}.
In addition, Ref.~\cite{Bauer2026} proposed an algorithm that finds all the code-preserving diagonal circuits of a given CSS code within a fixed set of physical gates.

However, each of these conditions and constructions is specific to a class of codes or a type of gate, and the algorithm treats one code at a time.
It remains open which property of a CSS code determines its logical diagonal gates, and a unified theoretical framework for these gates is still lacking.

Homology theory~\cite{Kitaev2003} is a promising candidate for such a framework.
Indeed, homology describes the logical Pauli operators of every CSS code in a unified way~\cite{BombinHomological2007}.
The first homology group $H_1$ and the first cohomology group $H^1$ label the logical $X$ and $Z$ operators, respectively.
It is therefore natural to expect homology to give a unified treatment of the logical diagonal gates at the higher levels of the Clifford hierarchy as well.

Consistent with this expectation, several recent studies have described logical gates in terms of homology.
Gauge fields on the chain complex of the code can be used to construct logical gates explicitly~\cite{HarunaGauge2025}.
The transversal implementability of logical phase gates has been characterized by a homological obstruction~\cite{Haruna_2026}.
A chain-map hierarchy has also been developed whose homology classes give multi-controlled-$\mathrm{Z}$ logical gates~\cite{SahayLongKhemani2026}.

Although these results apply to broad classes of CSS codes, each of them covers a particular family of gates or circuits.
The remaining problem is to determine all the logical gates implemented by diagonal circuits and all the diagonal circuits that implement each of them.
From the set of all the circuits for a logical gate, we can decide whether the gate has a transversal or constant-depth circuit on a given code and, conversely, derive the condition on a code under which the gate has such a circuit.

In this work, we classify the logical diagonal gates of any qubit CSS code at every level of the Clifford hierarchy.
We describe a diagonal circuit by its phase function, which assigns a phase to each computational basis state.
In terms of the phase function, we formulate the conditions for a circuit to preserve the code space and to act trivially on it.
By solving these conditions, we determine all the logical gates and all the circuits that act trivially, which we call logical identity operators, as described in the two theorems below.

Our first theorem relates the logical gates to the homology of the code.
We prove that the code-preserving diagonal circuits of a given level, taken modulo the logical identity operators, form a group isomorphic to that of the phase functions of the same level on the first homology group $H_1$ (\Cref{thm:cliffclass,thm:hier-classification}).
For a given circuit, we can test whether it is code-preserving using a finite set of equations on the $X$ checks (\Cref{thm:cliffcriterion,thm:hier-finite}), and obtain its logical gate from its phase function on the homology classes.
In particular, on every CSS code, we can implement every logical diagonal gate of a given level by a physical circuit of the same level.

Our second theorem determines all the circuits that implement a given logical gate.
Since two such circuits differ by a logical identity operator, it suffices to determine the group of these operators at each level (\Cref{thm:cliffkernel,thm:generalkernel}).
We show that this group is generated by the circuits that apply a phase according to the parity of a $Z$ check under the control of fewer physical qubits than the level.
These circuits act trivially on the code space because this parity vanishes there.
At level two, for example, a generator is either the circuit of $\mathrm S$ and $\CZ$ gates on the support of a $Z$ check or a $Z$ check controlled by one qubit.

We also give an explicit physical circuit for every logical gate and every logical identity operator in \Cref{sec:physical-implementation}.
Since every circuit that implements a logical gate differs from the one we give by logical identity operators, we can search for a transversal or constant-depth circuit by choosing these operators.
As applications, we reproduce the transversal $\mathrm S$ and $\CZ$ gates on the Steane code and determine the minimum depth of the intra-block logical $\CZ$ gate on the toric code (\Cref{sec:examples}).

The remainder of this paper is organized as follows.
In \Cref{sec:prelim}, we review CSS codes, introduce phase functions, and characterize the code-preserving diagonal circuits.
In \Cref{sec:clifford}, we give the classification at level two and determine the logical identity operators at that level.
In \Cref{sec:hierarchy}, we extend the classification and the generators of the logical identity operators to every level of the Clifford hierarchy.
In \Cref{sec:discussion}, we examine the scope and extensions of the classification, describe the physical circuits of the logical gates and the logical identity operators, discuss transversality and circuit depth, and give a chain-map interpretation.
In \Cref{sec:examples}, we apply these results to the Steane and toric codes and discuss the use of the classification in magic state distillation.
In \Cref{sec:conclusion}, we summarize the results and discuss open questions.
The proofs and the details of the examples are given in the Supplemental Materials~\cite{SupplementalMaterial}.

\section{Preliminaries}
\label{sec:prelim}
In this section, we summarize the basic notions used in the rest of this paper.
In \Cref{sec:prelim-css}, we review CSS codes and fix the encoded basis and the logical coordinates that label it.
In \Cref{sec:prelim-logical}, we define diagonal gates and state when such a gate preserves the code space and when it acts there as the identity.
In \Cref{sec:prelim-normal}, we introduce the phase functions of level $m$, their normal form, and their polarizations.

\subsection{CSS codes and logical coordinates}
\label{sec:prelim-css}
To identify the encoded basis states on which diagonal gates act, we first define CSS codes and derive their logical coordinates.
We denote the binary field $\mathbb Z/2\mathbb Z$ by $\Z{2}$.
\begin{definition}[CSS code]\label{def:css}
A CSS code is specified by binary parity-check matrices $H_X\in\Z{2}^{r_X\times n}$ and $H_Z\in\Z{2}^{r_Z\times n}$ satisfying $H_XH_Z^{\T}=0$.
The physical Hilbert space is $\Hphys=(\mathbb C^2)^{\otimes n}$. The code space $\Hcode\subseteq\Hphys$ is the common $+1$ eigenspace of the Pauli checks that apply $X$, respectively $Z$, at the nonzero entries of each row of $H_X$, respectively $H_Z$~\cite{CalderbankShor1996,Steane1996PRL}.
\end{definition}
We write $\xi_\beta$ for the $\beta$-th row of $H_X$ and $\zeta_\alpha$ for the $\alpha$-th row of $H_Z$. We regard these rows as column vectors when applying linear maps.

We label computational basis states by bit strings $x\in\Z{2}^n$ and denote binary addition by $\oplus$.
The $Z$ checks restrict the strings appearing in a code state to the kernel of $H_Z$, which we denote by $Z_1=\ker H_Z$. The $X$ checks require equal amplitudes on strings differing by an element of the image of $H_X^{\T}$, which we denote by $B_1=\im H_X^{\T}$.
Since the condition $H_XH_Z^{\T}=0$ implies $B_1\subseteq Z_1$, an encoded basis is therefore
\begin{equation}
 \ket{[x]}=|B_1|^{-1/2}\sum_{b\in B_1}\ket{x\oplus b},
 \qquad x\in Z_1,
 \label{eq:codebasis}
\end{equation}
where $[x]=x\oplus B_1$ denotes the coset $\{x\oplus b:b\in B_1\}$ and labels a class in the homology group $H_1=Z_1/B_1$, and $|\cdot|$ denotes the number of elements of the set $\cdot$. Thus the code encodes $k=\dim H_1=n-\rank H_X-\rank H_Z$ qubits. Choosing a basis of the homology group $H_1$ identifies $[x]$ with the bit string of these $k$ logical qubits and fixes an isomorphism $\Hcode\cong(\mathbb C^2)^{\otimes k}$.

We can describe these spaces by the chain complex~\cite{BombinHomological2007}
\begin{equation}
 C_2=\Z{2}^{r_X}\xrightarrow{\ \partial_2=H_X^{\T}\ }
 C_1=\Z{2}^n\xrightarrow{\ \partial_1=H_Z\ }C_0=\Z{2}^{r_Z},
\end{equation}
 with $\partial_1\partial_2=0$.
Here we use the cycle group $Z_1$, the boundary group $B_1$, and the homology group $H_1$. In the encoded basis \eqref{eq:codebasis}, a cycle is a bit string that appears in a code state, and two cycles that differ by a boundary appear in the same encoded basis state.

We say that a unitary operator $U$ on the physical space $\Hphys$ preserves the code space $\Hcode$, or is \emph{code-preserving}, when it maps every code state to a code state.
Its restriction $U|_{\Hcode}:\Hcode\to\Hphys$ then acts within $\Hcode$.
A \emph{logical identity operator} is a unitary operator on $\Hphys$ that acts as the identity on every code state. We use $I$ for the identity on the relevant space.

For binary vectors $u,z\in C_1$, the physical Pauli strings $X(u),Z(z)$ act on the computational basis state $\ket{x}$ with $x\in \Z{2}^n$ as
\begin{equation}
 X(u)\ket{x}=\ket{x\oplus u},\qquad
 Z(z)\ket{x}=(-1)^{\pair{z}{x}}\ket{x},
\end{equation}
where $\pair{z}{x} \coloneqq \sum_i z_ix_i \bmod{2} \in\Z{2}$ is the inner product over the binary field $\Z{2}$.

The logical $X$ operators are labeled by the homology group $H_1$. Similarly, with
\begin{equation}
 Z^1=\ker H_X,\qquad B^1=\im H_Z^{\T},\qquad H^1=Z^1/B^1,
\end{equation}
the cocycles in $Z^1$ specify logical $Z$ operators, and adding a coboundary in $B^1$ multiplies them by a product of $Z$ checks.
The distinct logical $Z$ operators are therefore labeled by the first cohomology group $H^1$.

We choose logical $X$ representatives $\gamma_a\in Z_1$ whose classes form a basis of the homology group $H_1$, and logical $Z$ representatives $z_a\in Z^1$ with $\pair{z_a}{\gamma_b}=\delta_{ab}$, which we call the dual representatives.
When a logical $Z$ operator $Z(z_a)$ acts on an encoded basis state $\ket{[x]}$, the eigenvalue is determined by the inner product of the representative $z_a$ with the cycle $x$ as
\begin{equation}
 Z(z_a)\ket{[x]}=(-1)^{\pair{z_a}{x}}\ket{[x]},\qquad x\in Z_1.
 \label{eq:logical-Z-eigenvalue}
\end{equation}
The exponent is unchanged when a boundary is added to $x$, and therefore it depends only on the class $[x]$.
These exponents are the bits of the $k$ logical qubits, and we define the \emph{logical coordinates} on the homology group $H_1$ by
\begin{equation}
 \lambda_a([x]) = \pair{z_a}{x},\qquad x\in Z_1,
 \label{eq:logical-coordinates}
\end{equation}
 for $a=1,\ldots,k$.
The code-space construction, the logical $Z$ operators, and the logical coordinates are detailed in \Cref{app:css-foundations}.

\subsection{Diagonal gates on the code space}
\label{sec:prelim-logical}
With the encoded basis of \Cref{sec:prelim-css} fixed, we now ask when a diagonal physical gate preserves the code space and when it acts there as the identity.
A diagonal gate changes the phases of computational basis states. Throughout this paper, we fix its global phase so that it leaves the state $\ket{0}^{\otimes n}$ unchanged.

These gates are classified into the Clifford hierarchy~\cite{Gottesman1997}, which is defined for general unitary gates as follows.
\begin{definition}[Clifford hierarchy]\label{def:hierarchy}
The first level $\CH^{(1)}$ is the group of Pauli strings $e^{\ii\theta}X(u)Z(z)$ with $\theta\in\mathbb R$ and $u,z\in C_1$.
For $m\geq2$, the level $\CH^{(m)}$ consists of the unitaries $U$ on the physical space $\Hphys$ satisfying
\begin{equation}
 U\,\CH^{(1)}\,U^\dagger\subseteq\CH^{(m-1)}.
\end{equation}
\end{definition}
The second level consists of the Clifford gates, and higher levels include non-Clifford gates such as $\mathrm T$ and $\CCZ$. The diagonal gates at each level form an abelian group~\cite{CGK2017}.

In this paper, we focus on the diagonal gates, which are defined as follows.
\begin{definition}[Diagonal gates]\label{def:gates}
For $m\geq1$ and $f:C_1\to\Zm$ with $f(0)=0$, we define the unitary on the physical space $\Hphys$ by
\begin{equation}
 \Ug{f}=\sum_{x\in C_1}
 \exp\!\left(\frac{2\pi\ii f(x)}{2^m}\right)\ket{x}\bra{x},
 \label{eq:Uf}
\end{equation}
where $\Zm=\mathbb Z/2^m\mathbb Z$.
\end{definition}
We call $f$ a \emph{phase function}, and its value $f(x)$ records the phase in units of $2\pi/2^m$. Multiplying diagonal gates adds their phase functions. The level superscript fixes this unit, and for example $\Ug[m+1]{2f}=\Ug{f}$.

The elementary gates at level $m$ are controlled powers of the Pauli $Z$. On a set $\calS\subseteq[n]=\{1,\ldots,n\}$ of qubit indices with $1\leq|\calS|\leq m$, they take the form
\begin{equation}
 \CPh{\calS}=\Ug{2^{|\calS|-1}\prod_{i\in \calS}x_i}.
 \label{eq:elementary-gate}
\end{equation}
Here bits on the right are evaluated as $0$ or $1$, and the gate applies a phase $\exp(\pi\ii/2^{m-|\calS|})$ when all qubits in the subset $\calS$ are in the state $\ket1$. The symbol $\mathsf C^{|\calS|-1}$ specifies the number of controls, and $\calS$ specifies the participating qubits. We omit the symbol $\mathsf C^0$ when $|\calS|=1$ and write $Z_{\calS}^1$ as $Z_{\calS}$ when $|\calS|=m$.
At level two these gates are $\mathrm S_i=Z_i^{1/2}$ and $\CZ_{ij}$, and at level three they are $\mathrm T_i=Z_i^{1/4}$, $\CS_{ij}$, and $\CCZ_{ijk}$.

To preserve the code space, a diagonal gate must give the same phase to all strings in each encoded basis state \eqref{eq:codebasis}.
The following lemma gives the necessary and sufficient conditions for a diagonal gate to preserve the code space and to be a logical identity operator.
\begin{lem}[{\cite{Hu2022,WebsterQuintavalleBartlett2023}}]\label{thm:logical-clifford}
We consider a function $f:C_1\to\Zm$ satisfying $f(0)=0$.
\begin{enumerate}
\item[(a)] The circuit $\Ug{f}$ preserves $\Hcode$ if and only if the phase function $f$ is constant on every coset $x\oplus B_1$, $x\in Z_1$. Equivalently, $f$ must satisfy
\begin{equation}
\label{eq:coset-condition}
    f(x\oplus\xi_\beta)=f(x),
\end{equation}
for every $x\in Z_1$ and every $X$ check $X(\xi_\beta)$.
\item[(b)] $\Ug{f}|_{\Hcode}=I$ if and only if 
\begin{equation}
\eval{f}_{Z_1}=0.
\end{equation}
\end{enumerate}
\end{lem}
The proof is given in \Cref{app:phase-properties}.

\subsection{Phase functions and their normal form}
\label{sec:prelim-normal}
The conditions in \Cref{sec:prelim-logical} are stated in terms of physical phase functions. To apply them at a fixed level of the Clifford hierarchy, we now introduce the functions that describe diagonal gates at level $m$.
They are defined on bit strings of a general length, that is, on a binary vector space $V$, and take values in $\Zm$ to record phases in units of $2\pi/2^m$.
We write $\Ph_m(V)$ for the group of these functions.
For the physical qubits, we take $V$ to be the chain group $C_1=\Z{2}^n$, whose elements are the bit strings of the $n$ physical qubits.

To characterize these functions, we examine how the phase changes when a bit string is shifted.
For bit strings $x,u\in V$, shifting $x$ by $u$ gives the string $x\oplus u$.
A function $f:V\to\Zm$ assigns the phases $f(x)$ and $f(x\oplus u)$ to the two strings, and we define the difference of $f$ in the direction $u$ as
\begin{equation}
 (\Delta_u f)(x)=f(x\oplus u)-f(x).
\end{equation}
We define the phase functions at level $m$ in terms of iterated differences as follows~\cite[Definition~1.2]{TaoZiegler2012}.
\begin{definition}[Phase functions at level $m$]\label{def:phase}
The group $\Ph_m(V)$, with pointwise addition, consists of the functions $f:V\to\Zm$ satisfying $f(0)=0$ and
\begin{equation}
 \Delta_{u_1}\cdots\Delta_{u_{m+1}}f=0\pmod{2^m}
 \label{eq:phase-difference-definition}
\end{equation}
for all $u_1,\ldots,u_{m+1}\in V$.
\end{definition}

These functions have the following normal form~\cite{CGK2017,CampbellHoward2017}.
\begin{lem}[{\cite{Zhegalkin1927}, \cite[Secs.~2.2.1 and~2.2.4]{Carlet2021}, \cite{CGK2017}}]\label{lem:phase}
Every $f\in\Ph_m(V)$ has a unique expression
\begin{equation}
 f(x)=\sum_{\substack{\calS\subseteq[\dim V]\\1\leq|\calS|\leq m}}
 2^{|\calS|-1}c_\calS x^\calS,\qquad c_\calS\in\Z{2^{m-|\calS|+1}},
 \label{eq:normalform}
\end{equation}
where $x^\calS=\prod_{i\in \calS}x_i$ uses the $0/1$ values of the bits, with $x^\varnothing=1$. Conversely, every such expression belongs to $\Ph_m(V)$.
\end{lem}
Throughout this paper, we omit the set braces in subscripts and write, for example, $c_i$ and $c_{ij}$ for $c_{\{i\}}$ and $c_{\{i,j\}}$.

We now apply the normal form to the physical qubits, where $V=C_1$.
Each term $2^{|\calS|-1}x^\calS$ in \eqref{eq:normalform} is the phase function of the elementary gate $\CPh{\calS}$ in \eqref{eq:elementary-gate}.
Since multiplying diagonal gates adds their phase functions, the diagonal gate $\Ug{f}$ of a phase function $f\in\Ph_m(C_1)$ is the product of elementary gates
\begin{equation}
 \Ug{f}=\prod_{\substack{\calS\subseteq[n]\\1\leq|\calS|\leq m}}\CPhpow{\calS}{c_\calS}.
 \label{eq:gate-normalform}
\end{equation}
Thus the coefficient $c_\calS$ is the power to which the elementary gate $\CPh{\calS}$ on the qubits in $\calS$ appears in the circuit.
These are exactly the diagonal gates in the $m$-th level $\CH^{(m)}$ up to a global phase~\cite{CGK2017}. The normal form and its uniqueness are reviewed in detail in \Cref{app:phase-properties}.

We next introduce the parity functions, which we use to build the physical phase functions in the later sections.
For a binary vector $z\in\Z{2}^n$, we define the \emph{parity function} $\pi_z$ of $z$ by letting $\pi_z(x)\in\{0,1\}$ be the parity $\pair{z}{x}$ regarded as an integer.
It is the sum modulo two of the bits $x_i$ with $i\in\supp z=\{i:z_i=1\}$, and it labels the eigenvalue of the Pauli string $Z(z)$ through $Z(z)\ket{x}=(-1)^{\pi_z(x)}\ket{x}$.

As an integer polynomial in the bits, the parity reads
\begin{equation}
 \pi_z(x)=\sum_{\varnothing\ne \calS\subseteq\supp z}(-2)^{|\calS|-1}x^\calS.
 \label{eq:parityexpansion}
\end{equation}
The expansion has the form \eqref{eq:normalform}, and hence the parity function $\pi_z$ belongs to the group $\Ph_m(\Z{2}^n)$ with values reduced modulo $2^m$.

The weighted products $2^{\ell-1}\prod_{j=1}^{\ell}\pi_{z_j}$ for every $\ell\geq1$ also belong to this group.
This follows from the general rule that composing a phase function with a linear map gives a phase function again.
For binary vector spaces $V$ and $W$ and a linear map $\varphi:V\to W$ over $\Z{2}$, this rule reads
\begin{equation}
 g\in\Ph_m(W)\quad\Longrightarrow\quad g\circ\varphi\in\Ph_m(V).
 \label{eq:pullback}
\end{equation}
The weighted products above are obtained in this way with $V=\Z{2}^n$ and $W=\Z{2}^\ell$, from the monomial $g(y)=2^{\ell-1}y_1\cdots y_\ell$ in the bits $y_1,\ldots,y_\ell$ of $W$ and the map $\varphi(x)=(\pair{z_j}{x})_{j=1}^{\ell}$.
We refer to \eqref{eq:pullback} as the substitution rule.
These facts are detailed in \Cref{app:phase-properties}.
In circuit synthesis, sums of parities $\sum_z a_z\pi_z$ are called \emph{phase polynomials}~\cite{Amy,AmyMosca2019,CampbellHoward2017}.

To read off the coefficients $c_\calS$ of the normal form, we take iterated differences and evaluate them at the zero vector. We use the following notation.
\begin{definition}[Polarization]\label{def:finite}
For $f:V\to\Zm$ with $f(0)=0$ and $d\geq1$, we define its $d$-th polarization~\cite{Vojtechovsky2004} by
\begin{equation}
 \delta_d f(u_1,\ldots,u_d)
 \coloneqq \eval{\Delta_{u_1}\cdots \Delta_{u_d}f(x)}_{x=0}.
 \label{eq:hier-pol}
\end{equation}
\end{definition}
Expanding the differences expresses it through the values of $f$ on sums of the directions.
\begin{lem}\label{lem:pol-sum}
For $f:V\to\Zm$ with $f(0)=0$ and $d\geq1$,
\begin{equation}
 \delta_d f(u_1,\ldots,u_d) = \sum_{J\subseteq[d]}(-1)^{d-|J|} f\left(\bigoplus_{j\in J}u_j\right),
 \label{eq:hier-pol-sum}
\end{equation}
where the sum over the empty set is the zero vector.
\end{lem}

Let $e_i\in V$ denote the $i$-th standard basis vector, whose $i$-th entry is one and whose other entries are zero.
The coefficient $c_\calS$ of the normal form \eqref{eq:normalform} then satisfies
\begin{equation}
 2^{|\calS|-1}c_\calS=\delta_{|\calS|}f((e_i)_{i\in \calS}).
\end{equation}
The proof and further difference identities are given in \Cref{app:phase-properties,app:polarization-details}.

So far, we have described the diagonal gates of level $m$ on the physical qubits through their phase functions.
In the following sections, we classify their actions on the logical qubits.
For level-$m$ phase functions on physical bit strings, we define
\begin{align}
 \Lg_m(\Hcode)&=\{f\in\Ph_m(C_1):\Ug{f}\Hcode\subseteq\Hcode\},\\
 \Trv_m(\Hcode)&=\{f\in\Ph_m(C_1):\Ug{f}|_{\Hcode} = I\}.
 \label{eq:vanishingkernel}
\end{align}
These groups distinguish the implemented logical gate from the freedom to change its physical implementation.

\section{Classification at level two}
\label{sec:clifford}
In this section, we investigate the logical diagonal gates and logical identity operators at level two, which are implemented by physical diagonal Clifford circuits.
Such a circuit is a product of single-qubit $\mathrm S$ gates and two-qubit $\CZ$ gates.
We first ask when it preserves the code space, then determine its logical gate and describe the circuits that can be added without changing its logical action.

Let us write such a circuit as
\begin{equation}
\Ug[2]{f} \coloneqq \prod_i\mathrm S_i^{c_i}\prod_{i<j}\CZ_{ij}^{M_{ij}},
\label{eq:clifford-gate-product}
\end{equation}
where $c_i\in\mathbb Z_4$ is the exponent of $\mathrm S_i$, and $M_{ij}\in\Z{2}$ with $i<j$ is the exponent of $\CZ_{ij}$.
For later convenience, we extend these exponents to a symmetric matrix $M=M^{\T}\in\Z{2}^{n\times n}$ by setting $M_{ji}=M_{ij}$ and $M_{ii}=c_i\bmod2$.
This extension does not change the circuit \eqref{eq:clifford-gate-product}, which contains only the entries $M_{ij}$ with $i<j$.
The phase function $f$ on the left-hand side is defined below in \eqref{eq:quadraticnormal}.

Since the gate $\mathrm S_i$ multiplies the state $\ket{x}$ by $\ii^{x_i}$ and $\CZ_{ij}$ multiplies it by $\ii^{2x_ix_j}$, this circuit is the diagonal gate of the phase function $f\in\Ph_2(C_1)$ given by
\begin{equation}
f(x)=\sum_i c_i x_i+2\sum_{i<j}M_{ij}x_ix_j.
\label{eq:quadraticnormal}
\end{equation}

An $X$ check $X(\xi_\beta)$ transforms under conjugation by this gate as
\begin{equation}
\Ug[2]{f}X(\xi_\beta)\Ugd[2]{f}
=\ii^{f(\xi_\beta)}X(\xi_\beta)Z(M\xi_\beta).
\label{eq:clifford-check-conjugation}
\end{equation}
With the diagonal entries $M_{ii}=c_i\bmod2$, the $Z$ operator produced by the odd powers of $\mathrm S$ is included in $Z(M\xi_\beta)$.

For a code-preserving gate, the extra $Z$ operator must be a product of $Z$ checks, and the phase must be $+1$.
The following proposition expresses these requirements in terms of the matrix $M$ and the phase function $f$.

\begin{pro}[Logical condition at level two~{\cite{DehaeneDeMoor2003,Rengaswamy2019}}]\label{thm:cliffcriterion}
For $f\in\Ph_2(C_1)$, the circuit $\Ug[2]{f}$ preserves $\Hcode$ if and only if
\begin{subequations}
\label{eq:cliffcriterion}
\begin{align}
\text{(L1)}\quad&f(\xi_\beta)=0\bmod{4},
\label{eq:L1}\\
\text{(L2)}\quad&M B_1\subseteq B^1,
\label{eq:L2}
\end{align}
\end{subequations}
where (L1) is required for every $X$ check $X(\xi_\beta)$.
\end{pro}
\begin{proof}
The (L1) condition is obtained by requiring the phase in \eqref{eq:clifford-check-conjugation} to vanish modulo four.
As for the (L2) condition, it corresponds to the requirement that the $Z$ operator $Z(M\xi_\beta)$ in \eqref{eq:clifford-check-conjugation} should be a product of the $Z$ checks.
\end{proof}

From the viewpoint of phase functions, the two conditions express the coset condition \eqref{eq:coset-condition} of \Cref{thm:logical-clifford}(a) through the polarizations of $f$.
Substitution into \eqref{eq:quadraticnormal} gives
\begin{equation}
\delta_1 f(u)=f(u),\qquad
\delta_2 f(u,v)=2\pair{u}{Mv}.
\label{eq:matrixpolar}
\end{equation}
By the definition of the polarization $\delta_2 f$ ($\delta_2 f(u,v)=f(u\oplus v)-f(u)-f(v)$), the coset condition $f(x\oplus\xi_\beta)=f(x)$ reads
\begin{equation}
\delta_1 f(\xi_\beta)+\delta_2 f(x,\xi_\beta)=0
\qquad(x\in Z_1)
\label{eq:clifford-coset-condition}
\end{equation}
for every $X$ check $X(\xi_\beta)$.
Taking $x=0$ gives $\delta_1 f(\xi_\beta)=0$, which is (L1).
The remaining condition $\delta_2 f(x,\xi_\beta)=2\pair{x}{M\xi_\beta}=0$ for every $x\in Z_1$ holds exactly when $M\xi_\beta\in Z_1^\perp=B^1$, which is (L2).
Here $W^\perp$ denotes the binary vectors orthogonal to every element of $W$.
The two conditions therefore take the common form
\begin{equation}
\delta_\ell f(u_1,\ldots,u_{\ell-1},\xi_\beta)=0
\qquad(\ell=1,2)
\label{eq:clifford-polar-condition}
\end{equation}
for all cycles $u_1,\ldots,u_{\ell-1}\in Z_1$ and every $X$ check $X(\xi_\beta)$.
As we will see in \Cref{sec:hierarchy}, the level-$m$ condition has the same form with $\ell=1,\ldots,m$.

We now turn to the logical gate implemented by a code-preserving circuit.
When (L1) and (L2) hold, the coset condition \eqref{eq:coset-condition} shows that $f(x)$ for $x\in Z_1$ depends only on the homology class $[x]$.
The circuit therefore multiplies the encoded basis state $\ket{[x]}$ by $\ii^{\lpol f([x])}$, where the logical phase function $\lpol f$ is the function on the homology group $H_1$ defined by
\begin{equation}
\lpol f([x])\coloneqq \eval{f(x)}_{x\in Z_1}.
\label{eq:clifford-logical-phase}
\end{equation}
Its polarizations are obtained by restricting those of the physical phase function $f$ to cycles.
For cycles $x,y,z\in Z_1$, \eqref{eq:matrixpolar} gives
\begin{align}
\delta_1 \lpol f([x]) & = f(x), \\
\label{eq:logicalpolar}
\delta_2 \lpol f([x],[y]) & = 2\pair{x}{My}, \\
\delta_3 \lpol f([x],[y],[z]) & = 0.
\end{align}
The last line holds because $f$ is a level-two phase function, and it shows that $\lpol f$ is a level-two phase function on $H_1$, that is, $\lpol f\in\Ph_2(H_1)$.

Recall that phase functions on any $\Z{2}$ vector space have a normal form, as we reviewed in \Cref{sec:prelim-normal}.
This fact applies to those on the homology group $H_1$ since it is isomorphic to $\Z{2}^k$.
With the logical coordinates $\lambda_a$ defined in \eqref{eq:logical-coordinates}, the logical phase function $\lpol f$ can be written in the normal form
\begin{equation}
\lpol f([x])=\sum_a \bar{c}_a\lambda_a([x])+2\sum_{a<b}\bar{c}_{ab}\lambda_a([x]) \lambda_b ([x])
\label{eq:clifford-logical-coefficients}
\end{equation}
with coefficients $\bar{c}_a\in\Z{4}$ and $\bar{c}_{ab}\in\Z{2}$.
Since $\lambda_a([\gamma_b])=\delta_{ab}$, evaluating the polarizations on the classes $[\gamma_a]$ gives
\begin{align}
\bar{c}_a &= \delta_1 \lpol f([\gamma_a]) = f(\gamma_a), \\
2\bar{c}_{ab} &= \delta_2 \lpol f([\gamma_a],[\gamma_b]) = 2\pair{\gamma_a}{M\gamma_b}.
\end{align}

These coefficients specify the logical gate.
The normal form has two kinds of terms, $\lambda_a$ and $2\lambda_a\lambda_b$ with $a<b$, and we write $\Ug[2]{\lambda_a}$ and $\Ug[2]{2\lambda_a\lambda_b}$ for the gates on the code space with these phase functions.
We identify them by their action on the encoded basis.
The state $\ket{[\gamma_b]}$ has the $b$-th logical qubit in $\ket{1}$ and the others in $\ket{0}$, and the state $\ket{[\gamma_c\oplus\gamma_d]}$ with $c\neq d$ has the $c$-th and $d$-th logical qubits in $\ket{1}$.
Since $\lambda_a([\gamma_b])=\delta_{ab}$, we find
\begin{align}
\Ug[2]{\lambda_a}\ket{[\gamma_b]}&=\ii^{\delta_{ab}}\ket{[\gamma_b]},
\label{eq:clifford-logical-S}\\
\Ug[2]{2\lambda_a\lambda_b}\ket{[\gamma_c\oplus\gamma_d]}&=(-1)^{\delta_{ac}\delta_{bd}+\delta_{ad}\delta_{bc}}\ket{[\gamma_c\oplus\gamma_d]}.
\label{eq:clifford-logical-CZ}
\end{align}
They are therefore the logical $\mathrm S$ gate $\overline{\mathrm S}_a$ on the $a$-th logical qubit and the logical $\CZ$ gate $\overline{\CZ}_{ab}$ on the pair $(a,b)$.

Since multiplying diagonal gates adds their phase functions, the normal form \eqref{eq:clifford-logical-coefficients} decomposes the action of the circuit on the code space as
\begin{equation}
\Ug[2]{f}\big|_{\Hcode}
=\prod_a\overline{\mathrm S}_a^{\,\bar{c}_a}\prod_{a<b}\overline{\CZ}_{ab}^{\,\bar{c}_{ab}}.
\label{eq:clifford-logical-gate-product}
\end{equation}
Thus $\bar{c}_a$ is the exponent of the logical $\mathrm S$ gate on the $a$-th logical qubit and $\bar{c}_{ab}$ is the exponent of the logical $\CZ$ gate on the pair $(a,b)$.
Hence every logical gate implemented by a code-preserving circuit is a product of logical $\mathrm S$ and $\CZ$ gates.

Conversely, we can ask whether every phase function in $\Ph_2(H_1)$ is the logical phase function $\lpol f$ of some code-preserving circuit.
The following theorem answers this affirmatively and establishes the homological classification of logical diagonal Clifford gates.
\begin{thm}
\label{thm:cliffclass}
The logical-action map $f\mapsto\lpol f$, where $\lpol f([x])=f(x)$, is a surjective homomorphism from $\Lg_2(\Hcode)$ to $\Ph_2(H_1)$ with kernel $\Trv_2(\Hcode)$.
It induces the isomorphism of groups
\begin{equation}
 \Lg_2(\Hcode)/\Trv_2(\Hcode)\cong\Ph_2(H_1),
 \label{eq:cliffclassification}
\end{equation}
and the group $\Ph_2(H_1)$ of logical phase functions has the structure
\begin{equation}
 \Ph_2(H_1)\cong\Z{4}^{k}\oplus\Z{2}^{\binom{k}{2}}.
 \label{eq:cliffclassification-structure}
\end{equation}
\end{thm}
\begin{proof}
As shown above, a code-preserving phase function $f$ gives $\lpol f\in\Ph_2(H_1)$.
The map $f\mapsto\lpol f$ is additive, and its kernel is $\Trv_2(\Hcode)$ by \Cref{thm:logical-clifford}(b).

Conversely, we take any phase function $\lpol f\in\Ph_2(H_1)$.
Using the dual representatives $z_a$ in \eqref{eq:logical-coordinates}, we set
\begin{multline}
 f_0=\sum_a \delta_1\lpol f([\gamma_a])\,\pi_{z_a}\\
    +\sum_{a<b} \delta_2\lpol f([\gamma_a],[\gamma_b])\,\pi_{z_a}\pi_{z_b}.
 \label{eq:clifflift}
\end{multline}
The substitution rule \eqref{eq:pullback} puts the phase function $f_0$ in $\Ph_2(C_1)$, and the corresponding physical circuit of $\mathrm S$ and $\CZ$ gates is given in \Cref{sec:physical-implementation}.
Since $\pi_{z_a}(x)=\lambda_a([x])$ for $x\in Z_1$, the normal form \eqref{eq:clifford-logical-coefficients} gives $f_0(x)=\lpol f([x])$ for every $x\in Z_1$.
Hence $f_0$ is constant on every coset $x\oplus B_1$, and therefore it preserves the code space by \Cref{thm:logical-clifford}(a) and induces the given logical phase function, proving surjectivity.

Finally, the uniqueness of the normal form in \Cref{lem:phase} identifies $\Ph_2(H_1)$ with the group of coefficients $(\bar{c}_a,\bar{c}_{ab})$ in the direct sum $\Z{4}^{k}\oplus\Z{2}^{\binom{k}{2}}$, which completes the proof.
\end{proof}

An expanded proof is given in \Cref{app:clifford-proof}.

We can also use the theorem to describe all the physical implementations of a given logical phase function.
Every implementation has the form $f=f_0+g$ with $g\in\Trv_2(\Hcode)$, that is, $\Ug[2]{f}=\Ug[2]{f_0}\Ug[2]{g}$, where the additional circuit $\Ug[2]{g}$ acts as the identity on the code space.
For fixed representatives $\gamma_a$ and $z_a$, the assignment $\lpol f\mapsto f_0$ is additive and therefore also gives the direct-sum decomposition
$\Lg_2(\Hcode)\cong\Trv_2(\Hcode)\oplus\Ph_2(H_1)$.

The remaining task is to determine the group $\Trv_2(\Hcode)$ of these additional circuits.
By \Cref{thm:logical-clifford}(b), its elements are the phase functions that vanish on every cycle.
The next theorem shows that these phase functions are generated by two families.
\begin{thm}
\label{thm:cliffkernel}
The group $\Trv_2(\Hcode)$ of the logical identity operators implemented by the diagonal Clifford circuits is generated by the phase functions
\begin{equation}
\pi_{\zeta_\alpha},\qquad 2\pi_{e_i}\pi_{\zeta_\alpha}
\label{eq:cliffkernel}
\end{equation}
for all $i\in[n]$ and $\alpha\in[r_Z]$.
Its group structure is the direct sum
\begin{equation}
\Trv_2(\Hcode)\cong\mathbb Z_4^{\,\rank H_Z}\oplus
\mathbb Z_2^{\binom n2-\binom{\dim Z_1}{2}}.
\label{eq:cliffkernelstructure}
\end{equation}
\end{thm}
\begin{proof}
Each displayed phase belongs to $\Ph_2(C_1)$ by the substitution rule \eqref{eq:pullback} and vanishes on the cycle group $Z_1$, and therefore it lies in $\Trv_2(\Hcode)$ by \Cref{thm:logical-clifford}(b).

Conversely, we take $f\in\Trv_2(\Hcode)$ and solve the condition $f|_{Z_1}=0$ in the normal form.
We write $f$ in the normal form
\begin{multline}
f=\sum_{j\leq r}a_j\pi_{\zeta_j}+\sum_{i\in T}b_i\pi_{e_i}
+2\sum_{j<j'\leq r}a_{jj'}\pi_{\zeta_j}\pi_{\zeta_{j'}}\\
+2\sum_{j\leq r}\sum_{i\in T}a_{ji}\,\pi_{e_i}\pi_{\zeta_j}
+2\sum_{\substack{i<i'\\i,i'\in T}}b_{ii'}\pi_{e_i}\pi_{e_{i'}}.
\label{eq:cliffkernel-normalform}
\end{multline}
Here $\zeta_1,\ldots,\zeta_r$ are independent rows of $H_Z$ with $r=\rank H_Z$, and $T\subseteq[n]$ is a set of $n-r$ physical qubits such that the parities $\pi_{\zeta_j}$, $j\leq r$, together with the parities $\pi_{e_i}$, $i\in T$, are binary coordinates on the chain group $C_1$, that is, their values determine $x\in C_1$ uniquely.
Such a set $T$ exists because independent vectors can be completed to a basis of $\Z{2}^n$ by standard basis vectors.
By the substitution rule \eqref{eq:pullback} and \Cref{lem:phase}, the normal form in these coordinates exists and is unique, with coefficients $a_j,b_i\in\Z{4}$ and $a_{jj'},a_{ji},b_{ii'}\in\Z{2}$.

The cycle group $Z_1$ is given by $\pi_{\zeta_1}=\cdots=\pi_{\zeta_r}=0$, and the parities $\pi_{e_i}$, $i\in T$, are binary coordinates on it.
On $Z_1$, every term that contains a check parity vanishes, and the restriction is
\begin{equation}
f|_{Z_1}=\sum_{i\in T}b_i\pi_{e_i}+2\sum_{\substack{i<i'\\i,i'\in T}}b_{ii'}\pi_{e_i}\pi_{e_{i'}},
\end{equation}
which is itself a normal form on $Z_1$.
By the uniqueness of the normal form, $f|_{Z_1}=0$ holds exactly when $b_i=0$ and $b_{ii'}=0$, and hence
\begin{multline}
f=\sum_{j\leq r}a_j\pi_{\zeta_j}
+2\sum_{j<j'\leq r}a_{jj'}\pi_{\zeta_j}\pi_{\zeta_{j'}}\\
+2\sum_{j\leq r}\sum_{i\in T}a_{ji}\,\pi_{e_i}\pi_{\zeta_j}
\label{eq:cliffkernel-solution}
\end{multline}
with arbitrary coefficients $a_j\in\Z{4}$ and $a_{jj'},a_{ji}\in\Z{2}$.
The first and third sums consist of the generators in \eqref{eq:cliffkernel}.
For the second sum, the expansion \eqref{eq:parityexpansion} gives
\begin{equation}
2\pi_{\zeta_j}\pi_{\zeta_{j'}}
=\sum_{i\in\supp\zeta_{j'}}2\pi_{e_i}\pi_{\zeta_j}\pmod4,
\end{equation}
and therefore the two families in \eqref{eq:cliffkernel} generate $\Trv_2(\Hcode)$.
Finally, the coefficients in \eqref{eq:cliffkernel-solution} are independent.
There are $r$ coefficients $a_j\in\Z{4}$ and $\binom r2+r(n-r)=\binom n2-\binom{n-r}{2}$ coefficients $a_{jj'},a_{ji}\in\Z{2}$, and $n-r=\dim Z_1$, which gives \eqref{eq:cliffkernelstructure}.
\end{proof}

The two families of generators correspond to the following physical circuits.
The phase function $2\pi_{e_i}\pi_{\zeta_\alpha}$ gives the $Z$ check $Z(\zeta_\alpha)$ controlled by the $i$-th physical qubit.
The phase function $\pi_{\zeta_\alpha}$ gives the circuit that consists of $\mathrm S$ gates on the qubits in the support of the $Z$ check and $\CZ$ gates on all pairs of them.
Since the controlled checks satisfy relations such as $\sum_{i\in\supp\zeta_{\alpha'}}2\pi_{e_i}\pi_{\zeta_\alpha}=\sum_{i\in\supp\zeta_\alpha}2\pi_{e_i}\pi_{\zeta_{\alpha'}}$, the exponent of $\mathbb Z_2$ in \eqref{eq:cliffkernelstructure} is smaller than the number of these generators.
We discuss these circuits in detail in \Cref{sec:physical-implementation}.

To summarize this section, a physical circuit of $\mathrm S$ and $\CZ$ gates is described by its phase function $f$, and the classification at level two consists of three steps.
First, the circuit preserves the code space exactly when (L1) and (L2) hold (\Cref{thm:cliffcriterion}).
Second, its logical gate is the product \eqref{eq:clifford-logical-gate-product} of logical $\mathrm S$ and $\CZ$ gates, whose exponents are determined by $f$ on the representatives $\gamma_a$, and every such product is realized by some physical circuit (\Cref{thm:cliffclass}).
Third, two circuits implement the same logical gate exactly when they differ by an element of the group $\Trv_2(\Hcode)$, which is generated by the parities of the $Z$ checks and their products with single bits (\Cref{thm:cliffkernel}).

We can extend all of these steps to the higher levels of the Clifford hierarchy.
The logical condition takes the form \eqref{eq:clifford-polar-condition} in terms of the polarizations of $f$, the logical gate is read off from the normal form of the logical phase function, and the logical identity operators are obtained by solving $f|_{Z_1}=0$ in the normal form.
In the next section, we use these formulations to classify the circuits of $\mathrm T$, $\CS$, and $\CCZ$ gates and the circuits at every higher level.

\section{Generalization to higher Clifford levels}
\label{sec:hierarchy}
In this section, we extend the classification of logical diagonal gates and logical identity operators from level two to an arbitrary level $m$.
We follow the same three steps as in \Cref{sec:clifford}.
We ask when a circuit preserves the code space (\Cref{sec:hier-logical}), then determine its logical gate (\Cref{sec:hier-class}) and describe the circuits that can be added without changing its logical action (\Cref{sec:hier-trivial}).

\begin{table}[t]
\caption{Physical and logical descriptions of a code-preserving diagonal circuit at level $m$. The logical description is obtained from the physical one by replacing the $n$ physical bits by the $k$ logical coordinates.}
\label{tab:physical-logical}
\centering
\setlength{\tabcolsep}{3.5pt}
\begin{tabular}{@{}lll@{}}
\hline\hline
 & Physical & Logical\\
\hline
Bit strings & $x\in C_1=\Z{2}^n$ & $[x]\in H_1\cong\Z{2}^k$\\
Coordinates & $x_i$, $i\in[n]$ & $\lambda_a([x])$, $a\in[k]$\\
Phase function & $f\in\Ph_m(C_1)$ & $\lpol f\in\Ph_m(H_1)$\\
Normal form & $\sum_{\calS}2^{|\calS|-1}c_\calS\prod_{i\in\calS}x_i$ & $\sum_{J}2^{|J|-1}\bar{c}_J\prod_{a\in J}\lambda_a$\\
Coefficient & $\delta_{|\calS|}f((e_i)_{i\in\calS})$ & $\delta_{|J|}f((\gamma_a)_{a\in J})$\\
Gate & $\prod_{\calS}\bigl(\CPh{\calS}\bigr)^{c_\calS}$ & $\prod_{J}\bigl(\overline{\CPh{J}}\bigr)^{\bar{c}_J}$\\
\hline\hline
\end{tabular}
\end{table}

\subsection{Logical condition at level \texorpdfstring{$m$}{m}}
\label{sec:hier-logical}\label{sec:hier-pol}
A physical circuit at level $m$ is the product \eqref{eq:gate-normalform} of elementary gates on up to $m$ qubits, and its phase function $f\in\Ph_m(C_1)$ is the normal form \eqref{eq:normalform}.
At level three, these gates are the $\mathrm T$, $\CS$, and $\CCZ$ gates.

We first consider the logical condition.
By \Cref{thm:logical-clifford}(a), the circuit $\Ug{f}$ preserves the code space if and only if the phase function $f$ satisfies the coset condition \eqref{eq:coset-condition}, that is, the phase difference $f(x\oplus\xi_\beta)-f(x)$ vanishes for every cycle $x\in Z_1$ and every $X$ check $X(\xi_\beta)$.

In terms of circuits, an $X$ check $X(\xi_\beta)$ and the level-$m$ diagonal circuit $\Ug{f}$ satisfy
\begin{equation}
  \Ugd{f} X(\xi_\beta) \Ug{f} X(\xi_\beta) = \Ug{\Delta_{\xi_\beta} f}.
\end{equation}
For $\Ug{f}$ to preserve the code space, the diagonal gate on the right-hand side must act as the identity on the code space.
That is, $\Delta_{\xi_\beta}f$ must vanish on the cycle group $Z_1$, which is the coset condition again.

Using the polarizations $\delta_\ell f$ of the phase function, we reduce the coset condition on all the cycles to conditions on a basis of the cycle group $Z_1$.
The polarization of order $m$ is $2^{m-1}$ times a symmetric $m$-linear form over $\Z{2}$.
For $m=2$, this form is given by the matrix $M$ of \Cref{sec:clifford}, $\delta_2 f(u,v)=2\pair{u}{Mv}$.
The polarizations describe the conditions and the exponents of the logical gates in the same form at every order and every level.

We write a cycle as a sum $x=u_1\oplus\cdots\oplus u_t$ of distinct elements of a basis of the cycle group $Z_1$.
The phase difference then expands in the polarizations of the phase function $f$ as (see \Cref{app:polarization-details})
\begin{align}
(\Delta_{\xi_\beta}f)(x) =\sum_{J\subseteq[t]}\delta_{|J|+1}f\bigl((u_j)_{j\in J},\xi_\beta\bigr).
\label{eq:hier-expansion}
\end{align}
The term with $J=\emptyset$ is $\delta_1 f(\xi_\beta)$, and a term with $|J|=\ell-1$ is the polarization of order $\ell$ evaluated on $\ell-1$ of the $u_j$ and the $X$ check.
The polarizations of order greater than $m$ vanish for a level-$m$ phase function, and therefore the sum contains the orders $\ell=1,\ldots,m$.
The coset condition holds for every cycle exactly when each of these terms vanishes, as shown in the proof below.
We thus obtain the following proposition, with one condition (L$\ell$) for each order $\ell$ up to $m$.

\begin{pro}[Logical condition at level $m$]\label{thm:hier-finite}
For $f\in\Ph_m(C_1)$, the circuit $\Ug{f}$ preserves $\Hcode$ if and only if
\begin{equation}
\text{(L$\ell$)}\quad
\delta_\ell f(u_1,\ldots,u_{\ell-1},\xi_\beta)=0
\qquad(\ell=1,\ldots,m)
\label{eq:hier-finite}
\end{equation}
for every $X$ check $X(\xi_\beta)$ and all distinct $u_1,\ldots,u_{\ell-1}$ in a basis of the cycle group $Z_1$.
\end{pro}
\begin{proof}
By the expansion \eqref{eq:hier-expansion}, the conditions (L$\ell$) imply the coset condition.
Conversely, inverting the expansion as in \Cref{lem:pol-sum} gives
\begin{multline}
\delta_\ell f(u_1,\ldots,u_{\ell-1},\xi_\beta)
\\ =\sum_{J\subseteq[\ell-1]}(-1)^{\ell-1-|J|}\Delta_{\xi_\beta}f\Bigl(\bigoplus_{j\in J}u_j\Bigr).
\end{multline}
The coset condition states that $\Delta_{\xi_\beta}f$ vanishes on $Z_1$, and therefore every term on the right vanishes.
\end{proof}
For $m=2$, $\delta_1 f(\xi_\beta)=f(\xi_\beta)$ and $\delta_2 f(u,\xi_\beta)=2\pair{u}{M\xi_\beta}$, and therefore these conditions are (L1) and (L2) of \Cref{thm:cliffcriterion}.
At level three, a circuit of $\mathrm T$, $\CS$, and $\CCZ$ gates satisfies in addition the condition (L3), $\delta_3 f(u,u',\xi_\beta)=0$.

\subsection{Logical gates and their classification}
\label{sec:hier-class}
We now turn to the logical gate implemented by a code-preserving circuit.
When the conditions (L$\ell$) hold, the phase $f(x)$ of a cycle $x\in Z_1$ depends only on its homology class $[x]$.
We can therefore define the logical phase function $\lpol f$ on the homology group $H_1$ by \eqref{eq:clifford-logical-phase} as at level two, and the circuit multiplies the encoded basis state $\ket{[x]}$ by $\exp(2\pi\ii\,\lpol f([x])/2^m)$.
The iterated differences of $\lpol f$, and hence its polarizations, are those of the physical phase function $f$ evaluated on cycles.
The logical phase function $\lpol f$ is therefore a level-$m$ phase function on $H_1$, that is, an element of the group $\Ph_m(H_1)$.

By \Cref{lem:phase}, the logical phase function has the normal form
\begin{equation}
\lpol f([x])=\sum_{\substack{J\subseteq[k]\\1\leq|J|\leq m}}
2^{|J|-1}\bar{c}_J\prod_{a\in J}\lambda_a([x])
\label{eq:hier-logical-coefficients}
\end{equation}
with coefficients $\bar{c}_J\in\Z{2^{m-|J|+1}}$.
It is the normal form \eqref{eq:normalform} with the $n$ physical bits $x_i$ replaced by the $k$ logical coordinates $\lambda_a$.
Evaluating the polarizations of the logical phase function $\lpol f$ on the classes $[\gamma_a]$ gives the coefficients as
\begin{equation}
2^{|J|-1}\bar{c}_J=\delta_{|J|}f\bigl((\gamma_a)_{a\in J}\bigr).
\label{eq:hier-logical-exponents}
\end{equation}

These coefficients specify the logical gate.
The gate on the code space with phase function $2^{|J|-1}\prod_{a\in J}\lambda_a$ multiplies $\ket{[x]}$ by $\exp(\pi\ii/2^{m-|J|})$ when all logical qubits in the subset $J$ are in the state $\ket{1}$, and acts as the identity otherwise.
It is therefore the elementary gate \eqref{eq:elementary-gate} acting on the logical qubits in $J$, which we denote by $\overline{\CPh{J}}$.
Since multiplying diagonal gates adds their phase functions, the circuit acts on the code space as
\begin{equation}
\Ug{f}\big|_{\Hcode}
=\prod_{\substack{J\subseteq[k]\\1\leq|J|\leq m}}
\Bigl(\overline{\CPh{J}}\Bigr)^{\bar{c}_J},
\label{eq:hier-logical-gate-product}
\end{equation}
which is the product \eqref{eq:gate-normalform} with the physical qubits replaced by the logical qubits.
Thus $\bar{c}_J$ is the exponent of the logical elementary gate on the logical qubits in $J$.
\Cref{tab:physical-logical} collects the correspondence between the physical and logical descriptions.

For example, at level three a circuit of $\mathrm T_i$, $\CS_{ij}$, and $\CCZ_{ijk}$ gates acts on the code space as
\begin{equation}
\Ug[3]{f}\big|_{\Hcode}
=\prod_a\overline{\mathrm T}_a^{\,\bar{c}_a}
\prod_{a<b}\overline{\CS}_{ab}^{\,\bar{c}_{ab}}
\prod_{a<b<c}\overline{\CCZ}_{abc}^{\,\bar{c}_{abc}},
\end{equation}
with $\bar{c}_a=f(\gamma_a)$, $2\bar{c}_{ab}=\delta_2 f(\gamma_a,\gamma_b)$, and $4\bar{c}_{abc}=\delta_3 f(\gamma_a,\gamma_b,\gamma_c)$.
The first two formulas have the same form as at level two, and the third one is new.

The following theorem shows that every phase function in $\Ph_m(H_1)$ is the logical phase function $\lpol f$ of some code-preserving circuit and gives the classification at every level, in the same form as \Cref{thm:cliffclass}.
\begin{thm}\label{thm:hier-classification}
For every $m\geq1$, the logical-action map $f\mapsto\lpol f$, where $\lpol f([x])=f(x)$, is a surjective homomorphism from $\Lg_m(\Hcode)$ to $\Ph_m(H_1)$ with kernel $\Trv_m(\Hcode)$.
It induces the isomorphism of groups
\begin{equation}
\Lg_m(\Hcode)/\Trv_m(\Hcode)\cong\Ph_m(H_1),
\label{eq:hier-classification}
\end{equation}
and the group $\Ph_m(H_1)$ of logical phase functions has the structure
\begin{equation}
\Ph_m(H_1)\cong\bigoplus_{\ell=1}^{\min(m,k)}
\left(\Z{2^{m-\ell+1}}\right)^{\binom{k}{\ell}}.
\label{eq:hier-classification-structure}
\end{equation}
\end{thm}
\begin{proof}
The proof follows that of \Cref{thm:cliffclass}, with \eqref{eq:hier-logical-coefficients} in place of the normal form \eqref{eq:clifford-logical-coefficients}.
The kernel of the logical-action map consists of the phase functions with $f|_{Z_1}=0$, which are the logical identity operators $\Trv_m(\Hcode)$ by \Cref{thm:logical-clifford}(b).
For surjectivity, we construct a physical circuit for each logical elementary gate in \Cref{sec:physical-implementation}.
\end{proof}

We illustrate the construction at level three.
The logical gates are generated by $\overline{\mathrm T}_a$, $\overline{\CS}_{ab}$, and $\overline{\CCZ}_{abc}$, whose logical phase functions are $\lambda_a$, $2\lambda_a\lambda_b$, and $4\lambda_a\lambda_b\lambda_c$.
On a cycle $x\in Z_1$, the logical coordinate is the parity of the dual representative, $\lambda_a([x])=\pi_{z_a}(x)$.
The three gates are therefore implemented by the physical phase functions
\begin{equation}
\pi_{z_a},\qquad
2\pi_{z_a}\pi_{z_b},\qquad
4\pi_{z_a}\pi_{z_b}\pi_{z_c}.
\label{eq:level-three-lift}
\end{equation}

We obtain the corresponding circuits by expanding each parity as a polynomial in the physical bits.
For the logical $\CCZ$ gate, the prefactor four reduces each parity to the sum of its bits, and
\begin{equation}
4\pi_{z_a}\pi_{z_b}\pi_{z_c}
=4\sum_{i\in\supp z_a}\sum_{j\in\supp z_b}\sum_{k\in\supp z_c}x_ix_jx_k\pmod8.
\label{eq:level-three-ccz}
\end{equation}
When the three supports are disjoint, as for logical qubits in three code blocks, the circuit applies a $\CCZ$ gate to every triple with one qubit in each support.
This is the round-robin circuit for the logical $\CCZ$ gate~\cite{YoderTakagiChuang2016}.
For the logical $\mathrm T$ gate, the parity $\pi_{z_a}$ expands modulo eight into the circuit \eqref{eq:t-gate-implementation} of $\mathrm T$, $\CS^{-1}$, and $\CCZ$ gates on the qubits, pairs, and triples in the support $\supp z_a$.
This construction of the logical $\mathrm T$ gate was given in the gauge field formalism~\cite{HarunaGauge2025}.

Any logical diagonal gate at level three is a product of powers of these gates, with exponents defined modulo eight, four, and two, which gives the direct sum $\Z{8}^{k}\oplus\Z{4}^{\binom k2}\oplus\Z{2}^{\binom k3}$ of groups in \eqref{eq:hier-classification-structure}.
The same construction applies at every level of the Clifford hierarchy, and an expanded proof is given in \Cref{app:general-proof}.

\subsection{Logical identity operators and their generators}\label{sec:hier-trivial}
Finally, we determine the logical identity operators at level $m$.
By \Cref{thm:logical-clifford}(b), this amounts to solving $f|_{Z_1}=0$ for the level-$m$ phase functions $f$.

At level two, we solved this condition in the normal form whose coordinates are the parities of independent $Z$ checks together with single bits that complete them to coordinates on the chain group $C_1$.
A term that contains the parity of a $Z$ check vanishes on every cycle, and the condition removes all the other terms.
The solutions were therefore generated by the parity $\pi_{\zeta_\alpha}$ of a $Z$ check and its product $2\pi_{e_i}\pi_{\zeta_\alpha}$ with a single bit (\Cref{thm:cliffkernel}).

The same argument applies at level $m$, where the normal form has products of up to $m$ parities, and therefore the parity of a $Z$ check is multiplied by up to $m-1$ single bits.
The following theorem states that these products generate every logical identity operator and gives the group structure of $\Trv_m(\Hcode)$.

\begin{thm}\label{thm:generalkernel}
For every $m\geq1$, the group $\Trv_m(\Hcode)$ is generated by
\begin{equation}
2^{|\calS|}\pi_{\zeta_\alpha}\prod_{i\in\calS}\pi_{e_i}
\label{eq:generalkernel}
\end{equation}
for all subsets $\calS\subseteq[n]$ with $|\calS|\leq m-1$ and all $\alpha\in[r_Z]$.
Its group structure is the direct sum
\begin{equation}
\Trv_m(\Hcode)\cong\bigoplus_{\ell=1}^{\min(m,n)}
\left(\Z{2^{m-\ell+1}}\right)^{\binom n\ell-\binom{n-\rank H_Z}\ell},
\label{eq:kernelgroup}
\end{equation}
where $\binom{n-\rank H_Z}\ell=0$ for $\ell>n-\rank H_Z$.
\end{thm}
\begin{proof}
In the coordinates $\pi_{\zeta_j}$, $j\leq r$, and $\pi_{e_i}$, $i\in T$, used in the proof of \Cref{thm:cliffkernel}, the normal form of a phase function $f\in\Ph_m(C_1)$ is
\begin{equation}
f=\sum_{A,\calS}2^{|A|+|\calS|-1}c_{A,\calS}\prod_{j\in A}\pi_{\zeta_j}\prod_{i\in\calS}\pi_{e_i},
\end{equation}
where the sum runs over the subsets $A\subseteq[r]$ and $\calS\subseteq T$ satisfying $1\leq|A|+|\calS|\leq m$, and the coefficients are $c_{A,\calS}\in\Z{2^{m-|A|-|\calS|+1}}$.
The cycle group $Z_1$ is given by $\pi_{\zeta_1}=\cdots=\pi_{\zeta_r}=0$. Therefore, the terms with $A\neq\varnothing$ vanish on $Z_1$, and the condition $f|_{Z_1}=0$ holds exactly when $c_{\varnothing,\calS}=0$ for every $\calS$.
Expanding all but one check parity of each term into single bits by \eqref{eq:parityexpansion} gives a combination of the generators \eqref{eq:generalkernel} modulo $2^m$.
Counting the terms with $A\neq\varnothing$ together with the orders of their coefficients gives \eqref{eq:kernelgroup}.
The complete proof is given in \Cref{app:parity-products}.
\end{proof}

As at level two, we now explain the physical circuit of each generator.
For $\calS=\varnothing$, the phase function $\pi_{\zeta_\alpha}$ gives a circuit on the support of the $Z$ check, with a gate on every subset of at most $m$ qubits.
For a general subset $\calS$, the factor $2^{|\calS|}$ lowers the level by $|\calS|$, and the product $\prod_{i\in\calS}\pi_{e_i}$ adds the qubits in $\calS$ as controls.
The generator therefore gives this circuit at level $m-|\calS|$, controlled by the qubits in $\calS$.
For $|\calS|=m-1$, it is the $Z$ check $Z(\zeta_\alpha)$ controlled by the qubits in $\calS$.

For example, at level three, the parity $\pi_{\zeta_\alpha}$ gives the circuit of $\mathrm T$, $\CS^{-1}$, and $\CCZ$ gates on the qubits, pairs, and triples in the support of the check.
The generator $2\pi_{e_i}\pi_{\zeta_\alpha}$ gives the level-two circuit of $\mathrm S$ and $\CZ$ gates controlled by the $i$-th qubit, which consists of the gates $\CS_{ij}$ and $\CCZ_{ijk}$ for the qubits $j$ and the pairs $j<k$ in the support.
When the $i$-th qubit belongs to the support, a repeated index is counted once, so that $\CS_{ii}=\mathrm S_i$ and $\CCZ_{iik}=\CZ_{ik}$.
The generator $4\pi_{e_i}\pi_{e_j}\pi_{\zeta_\alpha}$ gives the $Z$ check controlled by the $i$-th and $j$-th qubits, which is a product of $\CCZ$ gates.
The circuit of each generator is constructed in the same way at every level of the Clifford hierarchy, as given in \Cref{sec:physical-implementation}.

To summarize this section, the three steps of \Cref{sec:clifford} extend to every level $m$ of the Clifford hierarchy.
A physical circuit preserves the code space exactly when its phase function satisfies the conditions (L1), \ldots, (L$m$) (\Cref{thm:hier-finite}), its logical gate is specified by the logical phase function in $\Ph_m(H_1)$ (\Cref{thm:hier-classification}), and the circuits for the same logical gate differ by logical identity operators, which are generated by the parities of the $Z$ checks multiplied by at most $m-1$ single bits (\Cref{thm:generalkernel}).

\section{Discussion}
\label{sec:discussion}

In this section, we examine the scope of the classification and its consequences for physical circuit implementations.
In \Cref{sec:applicability}, we discuss the gates and codes to which the method applies, and in \Cref{sec:physical-implementation}, we expand logical gates and logical identity operators into elementary gates.
We then use the freedom to choose physical implementations to study circuit depth in \Cref{sec:hier-depth} and relate the classification to chain maps in \Cref{sec:chain-map}.

\subsection{Scope and extensions of the classification}
\label{sec:applicability}

We first examine the scope of the classification of \Cref{sec:hierarchy}.
It covers the products of the elementary gates at a fixed level $m$, which are the diagonal gates in the $m$-th level $\CH^{(m)}$ of the Clifford hierarchy up to a global phase.
Conversely, every logical diagonal gate in $\CH^{(m)}$ has a diagonal physical implementation (\Cref{thm:hier-classification}).

Logical gates diagonal in the $X$ basis, such as $\sqrt{X}$ and the $X$-controlled-$X$ gate $\mathrm{XCX}_{ab}=\mathrm{H}_a\mathrm{H}_b\,\CZ_{ab}\,\mathrm{H}_a\mathrm{H}_b$, are classified in the same way.
Applying Hadamard gates on all physical qubits turns such gates into $Z$-diagonal gates and exchanges the $X$ checks with the $Z$ checks, and therefore the results of \Cref{sec:hierarchy} apply with the check matrices $H_X$ and $H_Z$ exchanged.
The logical $X$ and $Z$ operators are exchanged as well.
Conjugating each logical elementary gate $\overline{\CPh{J}}$ by Hadamard gates on the logical qubits in $J$ gives a generating set for the group of level-$m$ logical $X$-diagonal gates.
This group is therefore isomorphic to the group in \eqref{eq:hier-classification-structure}.
To obtain generators for the physical $X$-diagonal circuits at level $m$ that act as the identity on the code space, we apply \Cref{thm:generalkernel} to the code with $H_X$ and $H_Z$ exchanged and conjugate the resulting generator circuits by $\mathrm H^{\otimes n}$.
The $Z$ checks used in those circuits thus become the $X$ checks $X(\xi_\beta)$ of the original code.

At level two, we obtain every logical Clifford gate by combining the logical gates diagonal in the $Z$ basis with those diagonal in the $X$ basis.
The logical $\overline{\mathrm{S}}_a$ and $\overline{\CZ}_{ab}$ have $Z$-diagonal physical implementations by \Cref{thm:cliffclass}, and their $X$-basis counterparts, $\overline{\sqrt{X}}_a$ and $\overline{\mathrm{XCX}}_{ab}$, have $X$-diagonal physical implementations by the previous paragraph.
Since $\mathrm{H}=e^{-\ii\pi/4}\,\mathrm{S}\sqrt{X}\mathrm{S}$ and $\mathrm{CNOT}_{ab}=\mathrm{H}_b\CZ_{ab}\mathrm{H}_b$, every logical Clifford gate on the $k$ logical qubits is a product of $Z$-diagonal and $X$-diagonal logical gates.
The corresponding statement fails for $m\geq3$ because the level $\CH^{(m)}$ is not closed under products.

The results hold for every CSS code.
The code enters only through its check matrices $H_X$ and $H_Z$, without sparsity, geometry, or self-duality assumptions.
Intra-block and inter-block gates are treated alike, since several code blocks form one CSS code whose check matrices are block diagonal, as in the single-block and two-block Steane examples of \Cref{sec:steane}.

We expect the classification to extend in the following three directions.
\begin{itemize}
\item CSS codes on prime-dimensional qudits have a chain complex over $\mathbb F_p$, and their diagonal gates in the Clifford hierarchy have an analogous normal form~\cite{CGK2017}. Therefore, the group of code-preserving phase functions modulo the logical identity operators should classify their logical diagonal gates.
\item A permutation of the physical qubits that preserves the code space maps cycles to cycles and boundaries to boundaries, and therefore it acts on the encoded basis $\ket{[x]}$ by an invertible linear map on the homology group $H_1$, that is, as a circuit of logical $\mathrm{CNOT}$ gates.
A code-preserving diagonal circuit followed by such a permutation then implements a logical diagonal gate followed by this circuit, which extends the classification to automorphism gates~\cite{GrasslRoetteler2013}.
\item Every non-CSS stabilizer code is mapped to a CSS code on the same qubits by an $X$-type Pauli operator and a diagonal Clifford operator~\cite[App.~C]{WebsterQuintavalleBartlett2023}.
The diagonal Clifford operator commutes with diagonal circuits, and conjugation by the Pauli operator preserves each level of the diagonal gates. Therefore, the classification applied to the corresponding CSS code would classify the logical diagonal gates of the original non-CSS code.
\end{itemize}
We are investigating the classification of the logical Clifford gates of qudit stabilizer codes, and the details will be reported in a forthcoming paper~\cite{HarunaMatsudaInPrep}.

\subsection{Physical circuit implementation of logical gates and logical identity operators}

\label{sec:physical-implementation}
We now write down the logical gates and the logical identity operators as circuits of elementary gates explicitly.
Both are obtained by substituting parity functions for the bits in the phase function of an elementary gate.
By the definition \eqref{eq:elementary-gate} of the elementary gates, the gate $\CPh{\calS}$ on the qubits in $\calS$ is a controlled power of $Z$ with the phase function $2^{|\calS|-1}\prod_{i\in\calS}x_i$, with the bits evaluated as integers.

The logical elementary gate $\overline{\CPh{J}}$ on the logical qubits in $J$ has the logical phase function $2^{|J|-1}\prod_{a\in J}\lambda_a$, which is the same monomial with the physical bits $x_i$ replaced by the logical coordinates $\lambda_a$.
Since $\lambda_a([x])=\pi_{z_a}(x)$ for every cycle $x$, substituting the parity function $\pi_{z_a}$ for each logical coordinate $\lambda_a$ gives a physical phase function that implements the logical gate.
If we instead substitute the parity $\pi_{\zeta_\alpha}$ of a $Z$ check for one bit, the phase function vanishes on every cycle. Therefore, its gate is a logical identity operator, and the qubits corresponding to the remaining bits act as controls.
In both cases, expanding the parities into monomials of the bits by \eqref{eq:parityexpansion} gives the circuit of elementary gates.

\begin{figure*}[tb]
\centering
\begin{tikzpicture}[x=1cm,y=1cm,line width=0.5pt,>=stealth,
  dot/.style={circle,fill,inner sep=1.2pt},
  gate/.style={draw,fill=white,inner sep=1.8pt,font=\footnotesize},
  lab/.style={font=\small}]
\begin{scope}
\node[lab,anchor=west] at (-0.1,1.85) {(a)\ \ $\overline{\mathrm S}_a$};
\foreach \y in {0.85,0.3,-0.25}{\draw (0,\y)--(2.1,\y); \node[gate] at (0.4,\y) {$\mathrm S$};}
\draw (1.0,0.85)--(1.0,0.3); \node[dot] at (1.0,0.85) {}; \node[dot] at (1.0,0.3) {};
\draw (1.4,0.85)--(1.4,-0.25); \node[dot] at (1.4,0.85) {}; \node[dot] at (1.4,-0.25) {};
\draw (1.8,0.3)--(1.8,-0.25); \node[dot] at (1.8,0.3) {}; \node[dot] at (1.8,-0.25) {};
\draw (2.18,0.95)--(2.26,0.95)--(2.26,-0.35)--(2.18,-0.35);
\node[lab,anchor=west] at (2.28,0.3) {$\supp z_a$};
\end{scope}
\begin{scope}[shift={(4.2,0)}]
\node[lab,anchor=west] at (-0.1,1.85) {(b)\ \ $\overline{\CZ}_{ab}$};
\foreach \y in {1.2,0.8,0.4,-0.2,-0.6}{\draw (0,\y)--(2.3,\y);}
\foreach \x/\ya/\yb in {0.25/1.2/-0.2,0.61/1.2/-0.6,0.97/0.8/-0.2,1.33/0.8/-0.6,1.69/0.4/-0.2,2.05/0.4/-0.6}{
  \draw (\x,\ya)--(\x,\yb); \node[dot] at (\x,\ya) {}; \node[dot] at (\x,\yb) {};}
\draw (2.38,1.3)--(2.46,1.3)--(2.46,0.3)--(2.38,0.3);
\node[lab,anchor=west] at (2.48,0.8) {$\supp z_a$};
\draw (2.38,-0.1)--(2.46,-0.1)--(2.46,-0.7)--(2.38,-0.7);
\node[lab,anchor=west] at (2.48,-0.4) {$\supp z_b$};
\end{scope}
\begin{scope}[shift={(8.6,0)}]
\node[lab,anchor=west] at (-0.1,1.85) {(c)\ \ $\Ug[2]{\pi_{\zeta_\alpha}}$};
\foreach \y in {1.05,0.5,-0.05,-0.6}{\draw (0,\y)--(2.85,\y); \node[gate] at (0.4,\y) {$\mathrm S$};}
\foreach \x/\ya/\yb in {0.95/1.05/0.5,1.28/1.05/-0.05,1.61/1.05/-0.6,1.94/0.5/-0.05,2.27/0.5/-0.6,2.6/-0.05/-0.6}{
  \draw (\x,\ya)--(\x,\yb); \node[dot] at (\x,\ya) {}; \node[dot] at (\x,\yb) {};}
\draw (2.93,1.15)--(3.01,1.15)--(3.01,-0.7)--(2.93,-0.7);
\node[lab,anchor=west] at (3.03,0.22) {$\supp\zeta_\alpha$};
\end{scope}
\begin{scope}[shift={(13.6,0)}]
\node[lab,anchor=west] at (-0.1,1.85) {(d)\ \ $\Ug[2]{2\pi_{e_i}\pi_{\zeta_\alpha}}$};
\draw (0,1.2)--(2.1,1.2); \node[lab,anchor=west] at (2.28,1.2) {$i$};
\foreach \y in {0.6,0.2,-0.2,-0.6}{\draw (0,\y)--(2.1,\y);}
\foreach \x/\y in {0.4/0.6,0.85/0.2,1.3/-0.2,1.75/-0.6}{
  \draw (\x,1.2)--(\x,\y); \node[dot] at (\x,1.2) {}; \node[dot] at (\x,\y) {};}
\draw (2.18,0.7)--(2.26,0.7)--(2.26,-0.7)--(2.18,-0.7);
\node[lab,anchor=west] at (2.28,0.0) {$\supp\zeta_\alpha$};
\end{scope}
\end{tikzpicture}
\caption{Circuits of the level-two generators. (a) The circuit of the logical gate $\overline{\mathrm S}_a$ consists of an $\mathrm S$ gate on every qubit of the support $\supp z_a$ and a $\CZ$ gate on every pair \eqref{eq:s-gate-implementation}. (b) The circuit of the logical gate $\overline{\CZ}_{ab}$ consists of a $\CZ$ gate on every pair with one qubit in each of the two disjoint supports \eqref{eq:cz-gate-disjoint}. (c) The circuit of (a) on the support of a $Z$ check is a logical identity operator \eqref{eq:parity-gate-implementation}. (d) The $Z$ check $Z(\zeta_\alpha)$ controlled by the $i$-th qubit is a logical identity operator \eqref{eq:controlled-check-implementation}.}
\label{fig:replacement}
\end{figure*}

Each elementary diagonal gate corresponds to one weighted monomial.
At level $m$, the elementary gate $\CPh{\calS}$ in \eqref{eq:elementary-gate} acts as
\begin{align}
 \CPh{\calS}\ket{x}&=\exp\!\left(\frac{2\pi\ii\,2^{|\calS|-1}x^\calS}{2^{m}}\right)\ket{x}
 \label{eq:gate-monomial}\\
 &=\exp\!\left(\frac{2\pi\ii\,2^{|\calS|-1}}{2^{m}}\prod_{i\in \calS}\pi_{e_i}(x)\right)\ket{x},
 \label{eq:gate-parity}
\end{align}
with the standard basis vectors $e_i$, and therefore its phase function is the weighted monomial $2^{|\calS|-1}x^\calS=2^{|\calS|-1}\prod_{i\in \calS}\pi_{e_i}$.
A circuit of such gates has the sum of their phases, and the normal form \eqref{eq:normalform} collects the terms on each support. Therefore, a phase function translates into a circuit gate by gate.

To compare a circuit with the normal form, we rewrite the parities in the same monomials.
The parity function $\pi_z$ expands by \eqref{eq:parityexpansion}, and hence
\begin{align}
 \pi_z(x)&=\sum_{\varnothing\ne \calS\subseteq\supp z}(-1)^{|\calS|-1}2^{|\calS|-1}x^\calS\pmod{2^m},
 \label{eq:parity-mod}\\
 \Ug{\pi_z}&=\prod_{\substack{\calS\subseteq\supp z\\1\leq|\calS|\leq m}}\CPhpow{\calS}{(-1)^{|\calS|-1}},
 \label{eq:parity-circuit}
\end{align}
where the terms with $|\calS|>m$ are divisible by $2^m$ and contribute no gate, and every subset of even size carries the inverse gate.
A product $2^{\ell-1}\pi_{z_1}\cdots\pi_{z_\ell}$ expands in the same way.
The circuits below are these expansions with $z$ a dual representative $z_a$ or a check row $\zeta_\alpha$.

We use this expansion to write down the physical circuits of the logical gates, starting with the logical $\mathrm S$ gate.
The logical phase function of the logical gate $\overline{\mathrm S}_a$ is the logical coordinate $\lambda_a$ (\Cref{sec:clifford}).
On a cycle, $\lambda_a([x])=\pi_{z_a}(x)$. Therefore, as in \eqref{eq:clifflift}, we implement the gate by the physical phase function $\pi_{z_a}$.
Modulo four, the expansion \eqref{eq:parity-mod} reads
\begin{equation}
 \pi_{z_a}(x)=\sum_{i\in\supp z_a}x_i - 2\sum_{\substack{i<j\\i,j\in\supp z_a}}x_ix_j\pmod4,
 \label{eq:s-gate-phase}
\end{equation}
and each term is one gate,
\begin{equation}
 \Ug[2]{\pi_{z_a}}=
 \prod_{i\in\supp z_a}\mathrm{S}_i
 \prod_{\substack{i<j\\i,j\in\supp z_a}}\CZ_{ij}.
 \label{eq:s-gate-implementation}
\end{equation}
The circuit places an $\mathrm{S}$ on every qubit of the support $\supp z_a$ and a $\CZ$ on every pair in it (\Cref{fig:replacement}(a)).
It is the circuit of a $Z$ check in \Cref{sec:clifford} with $\supp z_a$ in place of the support of the check, and it coincides with the decomposition of $\overline{\mathrm S}$ in Ref.~\cite[Sec.~3]{HarunaGauge2025} and with the targeted logical $\mathrm S$ gate of Ref.~\cite{OldBecharMullerHeussen2025}.

The logical $\CZ$ gate is obtained in the same way.
The logical phase function of the logical gate $\overline{\CZ}_{ab}$ is $2\lambda_a\lambda_b$, and we implement it by the physical phase function $2\pi_{z_a}\pi_{z_b}$.
Modulo four, only the single bits in the expansion \eqref{eq:parity-mod} of each parity contribute, and
\begin{equation}
 2\pi_{z_a}(x)\pi_{z_b}(x)=2\sum_{i\in\supp z_a}\sum_{j\in\supp z_b}x_ix_j\pmod4.
 \label{eq:cz-gate-phase}
\end{equation}
When the supports of the representatives $z_a$ and $z_b$ are disjoint, for example when the two logical qubits belong to different code blocks, each term is one $\CZ$ gate, and
\begin{equation}
 \Ug[2]{2\pi_{z_a}\pi_{z_b}}=
 \prod_{i\in\supp z_a}\prod_{j\in\supp z_b}\CZ_{ij}.
 \label{eq:cz-gate-disjoint}
\end{equation}
The circuit places a $\CZ$ on every pair with one qubit in each support (\Cref{fig:replacement}(b)), which is the round-robin circuit of Ref.~\cite{YoderTakagiChuang2016}.
When the supports overlap, \eqref{eq:cz-gate-disjoint} holds with $\CZ_{ii}$ replaced by $Z_i$, because the term $2x_i^2$ equals $2x_i$.

The logical identity operators at level two are obtained in the same way from the parities of the $Z$ checks.
Their generators are the phase functions $\pi_{\zeta_\alpha}$ and $2\pi_{e_i}\pi_{\zeta_\alpha}$ (\Cref{thm:cliffkernel}), which vanish on every cycle because the check parity does.
Expanding the check parity as in \eqref{eq:s-gate-phase} gives
\begin{align}
 \Ug[2]{\pi_{\zeta_\alpha}}&=
 \prod_{i\in\supp\zeta_\alpha}\mathrm{S}_i
 \prod_{\substack{i<j\\i,j\in\supp\zeta_\alpha}}\CZ_{ij},
 \label{eq:parity-gate-implementation}\\
 \Ug[2]{2\pi_{e_i}\pi_{\zeta_\alpha}}
 &=\prod_{j\in\supp\zeta_\alpha}\CZ_{ij}.
 \label{eq:controlled-check-implementation}
\end{align}
The first circuit is \eqref{eq:s-gate-implementation} with the check row $\zeta_\alpha$ in place of the representative $z_a$ (\Cref{fig:replacement}(c)).
The second one is the check $Z(\zeta_\alpha)$ controlled by the $i$-th qubit (\Cref{fig:replacement}(d)).
Its phase function $2x_i\pi_{\zeta_\alpha}(x)$ is zero for $x_i=0$ and equals $2\pi_{\zeta_\alpha}(x)$, the phase function of the check $Z(\zeta_\alpha)$, for $x_i=1$.
When the $i$-th qubit belongs to the support of the check, $\CZ_{ii}$ is replaced by $Z_i$, as for the logical $\CZ$ gate.

At level three, the logical phase function of the logical gate $\overline{\mathrm{T}}_a$ is again the logical coordinate $\lambda_a$.
Modulo eight, the expansion \eqref{eq:parity-mod} of the corresponding parity function $\pi_{z_a}$ reads
\begin{equation}
 \pi_{z_a}(x)=\sum_{i}x_i-2\sum_{i<j}x_ix_j+4\sum_{i<j<k}x_ix_jx_k\pmod8,
 \label{eq:t-gate-phase}
\end{equation}
where the indices run over the support $\supp z_a$, and its physical implementation is given by
\begin{equation}
 \Ug[3]{\pi_{z_a}}=
 \prod_{i\in\supp z_a}\mathrm{T}_i
 \prod_{\substack{i<j\\i,j\in\supp z_a}}\CS_{ij}^{-1}
 \prod_{\substack{i<j<k\\i,j,k\in\supp z_a}}\CCZ_{ijk}.
 \label{eq:t-gate-implementation}
\end{equation}
The term $-2x_ix_j$ gives the inverse of the $\CS$ gate, whose phase function is $2x_ix_j$.
The logical gates $\overline{\CS}_{ab}$ and $\overline{\CCZ}_{abc}$ are obtained in the same way from the phase functions $2\pi_{z_a}\pi_{z_b}$ and $4\pi_{z_a}\pi_{z_b}\pi_{z_c}$ in \eqref{eq:level-three-lift}, and \Cref{fig:level-three} shows the three circuits.
These constructions agree with those by the gauge field formalism~\cite{HarunaGauge2025}.

\begin{figure*}[tb]
\centering
\begin{tikzpicture}[x=1cm,y=1cm,line width=0.5pt,
  dot/.style={circle,fill,inner sep=1.2pt},
  gate/.style={draw,fill=white,inner sep=1.6pt,font=\scriptsize},
  lab/.style={font=\small}]
\def\figCS#1#2#3#4{\draw (#1,#2)--(#1,#3); \node[dot] at (#1,#2) {}; \node[gate] at (#1,#3) {#4};}
\def\figCCZ#1#2#3#4{\draw (#1,#2)--(#1,#4); \node[dot] at (#1,#2) {}; \node[dot] at (#1,#3) {}; \node[dot] at (#1,#4) {};}
\begin{scope}
\node[lab,anchor=west] at (-0.1,1.95) {(a)\ \ $\overline{\mathrm T}_a$};
\foreach \y in {0.9,0.45,0.0}{\draw (0,\y)--(3.1,\y); \node[gate] at (0.35,\y) {$\mathrm T$};}
\figCS{1.0}{0.9}{0.45}{$\mathrm S^\dagger$}
\figCS{1.55}{0.9}{0.0}{$\mathrm S^\dagger$}
\figCS{2.1}{0.45}{0.0}{$\mathrm S^\dagger$}
\figCCZ{2.7}{0.9}{0.45}{0.0}
\draw (3.18,1.0)--(3.26,1.0)--(3.26,-0.1)--(3.18,-0.1);
\node[lab,anchor=west] at (3.28,0.45) {$\supp z_a$};
\end{scope}
\begin{scope}[shift={(5.5,0)}]
\node[lab,anchor=west] at (-0.1,1.95) {(b)\ \ $\overline{\CS}_{ab}$};
\foreach \y in {1.2,0.75,0.1,-0.35}{\draw (0,\y)--(4.15,\y);}
\figCS{0.35}{1.2}{0.1}{$\mathrm S$}
\figCS{0.85}{1.2}{-0.35}{$\mathrm S$}
\figCS{1.35}{0.75}{0.1}{$\mathrm S$}
\figCS{1.85}{0.75}{-0.35}{$\mathrm S$}
\figCCZ{2.45}{1.2}{0.75}{0.1}
\figCCZ{2.9}{1.2}{0.75}{-0.35}
\figCCZ{3.4}{1.2}{0.1}{-0.35}
\figCCZ{3.85}{0.75}{0.1}{-0.35}
\draw (4.23,1.3)--(4.31,1.3)--(4.31,0.65)--(4.23,0.65);
\node[lab,anchor=west] at (4.33,0.97) {$\supp z_a$};
\draw (4.23,0.2)--(4.31,0.2)--(4.31,-0.45)--(4.23,-0.45);
\node[lab,anchor=west] at (4.33,-0.13) {$\supp z_b$};
\end{scope}
\begin{scope}[shift={(12.1,0)}]
\node[lab,anchor=west] at (-0.1,1.95) {(c)\ \ $\overline{\CCZ}_{abc}$};
\foreach \y in {1.2,0.75,0.25,-0.2,-0.7}{\draw (0,\y)--(2.1,\y);}
\figCCZ{0.3}{1.2}{0.25}{-0.7}
\figCCZ{0.8}{1.2}{-0.2}{-0.7}
\figCCZ{1.3}{0.75}{0.25}{-0.7}
\figCCZ{1.8}{0.75}{-0.2}{-0.7}
\draw (2.18,1.3)--(2.26,1.3)--(2.26,0.65)--(2.18,0.65);
\node[lab,anchor=west] at (2.28,0.97) {$\supp z_a$};
\draw (2.18,0.35)--(2.26,0.35)--(2.26,-0.3)--(2.18,-0.3);
\node[lab,anchor=west] at (2.28,0.02) {$\supp z_b$};
\draw (2.18,-0.6)--(2.26,-0.6)--(2.26,-0.8)--(2.18,-0.8);
\node[lab,anchor=west] at (2.28,-0.7) {$\supp z_c$};
\end{scope}
\begin{scope}[shift={(0,-3.9)}]
\node[lab,anchor=west] at (-0.1,1.95) {(d)\ \ $\Ug[3]{\pi_{\zeta_\alpha}}$};
\foreach \y in {1.2,0.75,0.3,-0.15}{\draw (0,\y)--(5.6,\y); \node[gate] at (0.35,\y) {$\mathrm T$};}
\figCS{1.0}{1.2}{0.75}{$\mathrm S^\dagger$}
\figCS{1.5}{1.2}{0.3}{$\mathrm S^\dagger$}
\figCS{2.0}{1.2}{-0.15}{$\mathrm S^\dagger$}
\figCS{2.5}{0.75}{0.3}{$\mathrm S^\dagger$}
\figCS{3.0}{0.75}{-0.15}{$\mathrm S^\dagger$}
\figCS{3.5}{0.3}{-0.15}{$\mathrm S^\dagger$}
\figCCZ{4.1}{1.2}{0.75}{0.3}
\figCCZ{4.5}{1.2}{0.75}{-0.15}
\figCCZ{4.9}{1.2}{0.3}{-0.15}
\figCCZ{5.3}{0.75}{0.3}{-0.15}
\draw (5.68,1.3)--(5.76,1.3)--(5.76,-0.25)--(5.68,-0.25);
\node[lab,anchor=west] at (5.78,0.52) {$\supp\zeta_\alpha$};
\end{scope}
\begin{scope}[shift={(7.6,-3.9)}]
\node[lab,anchor=west] at (-0.1,1.95) {(e)\ \ $\Ug[3]{2\pi_{e_i}\pi_{\zeta_\alpha}}$};
\draw (0,1.45)--(4.5,1.45); \node[lab,anchor=west] at (4.6,1.45) {$i$};
\foreach \y in {0.75,0.3,-0.15,-0.6}{\draw (0,\y)--(4.5,\y);}
\figCS{0.35}{1.45}{0.75}{$\mathrm S$}
\figCS{0.8}{1.45}{0.3}{$\mathrm S$}
\figCS{1.25}{1.45}{-0.15}{$\mathrm S$}
\figCS{1.7}{1.45}{-0.6}{$\mathrm S$}
\figCCZ{2.25}{1.45}{0.75}{0.3}
\figCCZ{2.65}{1.45}{0.75}{-0.15}
\figCCZ{3.05}{1.45}{0.75}{-0.6}
\figCCZ{3.45}{1.45}{0.3}{-0.15}
\figCCZ{3.85}{1.45}{0.3}{-0.6}
\figCCZ{4.25}{1.45}{-0.15}{-0.6}
\draw (4.58,0.85)--(4.66,0.85)--(4.66,-0.7)--(4.58,-0.7);
\node[lab,anchor=west] at (4.68,0.07) {$\supp\zeta_\alpha$};
\end{scope}
\begin{scope}[shift={(14.1,-3.9)}]
\node[lab,anchor=west] at (-0.1,1.95) {(f)\ \ $\Ug[3]{4\pi_{e_i}\pi_{e_j}\pi_{\zeta_\alpha}}$};
\draw (0,1.45)--(1.9,1.45); \node[lab,anchor=west] at (2.0,1.45) {$i$};
\draw (0,1.05)--(1.9,1.05); \node[lab,anchor=west] at (2.0,1.05) {$j$};
\foreach \y in {0.4,-0.0,-0.4,-0.8}{\draw (0,\y)--(1.9,\y);}
\figCCZ{0.3}{1.45}{1.05}{0.4}
\figCCZ{0.7}{1.45}{1.05}{-0.0}
\figCCZ{1.1}{1.45}{1.05}{-0.4}
\figCCZ{1.5}{1.45}{1.05}{-0.8}
\draw (1.98,0.5)--(2.06,0.5)--(2.06,-0.9)--(1.98,-0.9);
\node[lab,anchor=west] at (2.08,-0.2) {$\supp\zeta_\alpha$};
\end{scope}
\end{tikzpicture}
\caption{Circuits of the level-three generators. (a) The circuit of the logical gate $\overline{\mathrm T}_a$ consists of a $\mathrm T$ gate on every qubit of $\supp z_a$, a $\CS^{-1}$ gate on every pair, and a $\CCZ$ gate on every triple \eqref{eq:t-gate-implementation}. (b) The circuit of the logical gate $\overline{\CS}_{ab}$ consists of a $\CS$ gate on every pair with one qubit in each support and a $\CCZ$ gate on every triple with two qubits in one support and one in the other. (c) The circuit of the logical gate $\overline{\CCZ}_{abc}$ consists of a $\CCZ$ gate on every triple with one qubit in each support. (d) to (f) The generators of the logical identity operators are the circuit of (a) on the support of a $Z$ check, the level-two circuit of the check parity controlled by the $i$-th qubit, and the $Z$ check $Z(\zeta_\alpha)$ controlled by the $i$-th and $j$-th qubits. The supports of the logical $Z$ representatives are disjoint, and the control qubits lie outside the support of the check.}
\label{fig:level-three}
\end{figure*}

\begin{figure}[tb]
\centering
\begin{tikzpicture}[x=1cm,y=1cm,line width=0.5pt,
  dot/.style={circle,fill,inner sep=1.2pt},
  gate/.style={draw,fill=white,inner sep=3pt,font=\small},
  lab/.style={font=\small}]
\begin{scope}
\node[lab,anchor=west] at (-0.1,2.45) {(a)};
\foreach \y in {1.8,1.45,0.85,0.5,-0.6,-0.95}{\draw (0,\y)--(1.75,\y);}
\node[lab] at (0.875,-0.02) {$\vdots$};
\foreach \x/\ya/\yb/\yc in {0.3/1.8/0.85/-0.6,0.75/1.45/0.85/-0.6}{
  \draw (\x,\ya)--(\x,0.25); \draw[dotted] (\x,0.25)--(\x,-0.35); \draw (\x,-0.35)--(\x,\yc);
  \node[dot] at (\x,\ya) {}; \node[dot] at (\x,\yb) {}; \node[dot] at (\x,\yc) {};}
\node[lab] at (1.35,1.15) {$\cdots$};
\draw (1.83,1.9)--(1.91,1.9)--(1.91,1.35)--(1.83,1.35);
\node[lab,anchor=west] at (1.93,1.62) {$\supp z_{a_1}$};
\draw (1.83,0.95)--(1.91,0.95)--(1.91,0.4)--(1.83,0.4);
\node[lab,anchor=west] at (1.93,0.67) {$\supp z_{a_2}$};
\draw (1.83,-0.5)--(1.91,-0.5)--(1.91,-1.05)--(1.83,-1.05);
\node[lab,anchor=west] at (1.93,-0.78) {$\supp z_{a_m}$};
\end{scope}
\begin{scope}[shift={(4.3,0)}]
\node[lab,anchor=west] at (-0.1,2.45) {(b)};
\foreach \y in {1.8,1.25}{\draw (0,\y)--(2.5,\y); \node[dot] at (1.25,\y) {};}
\node[lab] at (0.45,1.6) {$\vdots$};
\foreach \y in {0.45,0.1,-0.25,-0.6}{\draw (0,\y)--(2.5,\y);}
\draw (1.25,1.8)--(1.25,0.6);
\node[gate,minimum width=1.9cm,minimum height=1.35cm] at (1.25,-0.075) {$\Ug[m-|\calS|]{\pi_{\zeta_\alpha}}$};
\draw (2.58,1.9)--(2.66,1.9)--(2.66,1.15)--(2.58,1.15);
\node[lab,anchor=west] at (2.68,1.52) {$\calS$};
\draw (2.58,0.55)--(2.66,0.55)--(2.66,-0.7)--(2.58,-0.7);
\node[lab,anchor=west] at (2.68,-0.08) {$\supp\zeta_\alpha$};
\end{scope}
\end{tikzpicture}
\caption{Circuits at level $m$. (a) The circuit of the logical $\mathsf C^{m-1}Z$ gate consists of a $\mathsf C^{m-1}Z$ gate on every tuple with one qubit in each of the $m$ disjoint supports, and two of these gates are drawn. (b) A generator \eqref{eq:identity-generator-circuit} of the logical identity operators is the circuit \eqref{eq:parity-circuit} of the check parity at level $m-|\calS|$ controlled by the qubits in $\calS$, which are drawn outside the support of the check.}
\label{fig:general-level}
\end{figure}

At level $m$, the logical gates are generated by the logical elementary gates $\overline{\CPh{J}}$ with the phase functions $2^{|J|-1}\prod_{a\in J}\lambda_a$, where $J\subseteq[k]$ is a set of logical qubits with $1\leq|J|\leq m$ (\Cref{lem:phase}).
We implement each of them by the same substitution of the parity $\pi_{z_a}$ for the logical coordinate $\lambda_a$,
\begin{equation}
 f_J=2^{|J|-1}\prod_{a\in J}\pi_{z_a}.
 \label{eq:representative-circuit}
\end{equation}
The circuit of the phase function $f_J$ is obtained by expanding the parities in \eqref{eq:representative-circuit} by \eqref{eq:parity-mod}.
For $J=\{a\}$ it is \eqref{eq:parity-circuit} with $z=z_a$, with a gate on every subset of the support $\supp z_a$.
For $|J|=m$, only the terms with one bit from each support survive modulo $2^m$, and therefore the circuit has a $\mathsf C^{m-1}Z$ gate on every tuple with one qubit in each $\supp z_a$, with repeated qubits merged (\Cref{fig:general-level}(a)).

The logical identity operators at level $m$ are constructed in the same way.
The generators of the group $\Trv_m(\Hcode)$ in \Cref{thm:generalkernel} are
\begin{equation}
 2^{|\calS|}\pi_{\zeta_\alpha}\prod_{i\in\calS}\pi_{e_i},
 \label{eq:identity-generator-circuit}
\end{equation}
where the qubits indexed by $\calS$ act as controls.
When the bit $x_i$ equals one for every $i\in\calS$, the generator equals $2^{|\calS|}\pi_{\zeta_\alpha}(x)$, which is the phase function $\pi_{\zeta_\alpha}$ at level $m-|\calS|$. Otherwise, it vanishes.

Expanding the check parity in \eqref{eq:identity-generator-circuit} by \eqref{eq:parity-mod} gives one gate for every subset $\calS'\subseteq\supp\zeta_\alpha$ with $1\leq|\calS'|\leq m-|\calS|$.
This gate acts on the qubits in $\calS\cup\calS'$ and applies the phase $(-1)^{|\calS'|-1}\pi/2^{m-|\calS|-|\calS'|}$ when all of them are in $\ket1$, where a qubit in both sets is counted once.
The circuit is therefore \eqref{eq:parity-circuit} for the check at level $m-|\calS|$, with the qubits in $\calS$ added to every gate as controls (\Cref{fig:general-level}(b)).
In general, if $g\in\Trv_m(\Hcode)$, then $2\pi_{e_i}g\in\Trv_{m+1}(\Hcode)$ for every qubit $i$, and its gate is $\Ug{g}$ controlled by the $i$-th qubit.
\Cref{fig:replacement}(c) and (d) show the two generators at level two, and \Cref{fig:level-three}(d) to (f) show the three generators at level three.

\subsection{Transversality and circuit depth of logical gates}
\label{sec:hier-depth}

\Cref{thm:hier-classification,thm:generalkernel} classify the logical gates composed of diagonal gates, and therefore the transversal and constant-depth implementations of these logical gates can be determined from them.
Two phase functions $f$ and $f'$ in $\Lg_m(\Hcode)$ have the same logical phase function exactly when the difference $g=f-f'$ belongs to the group $\Trv_m(\Hcode)$ (\Cref{thm:hier-classification}), and then the circuit
\begin{equation}
 \Ugd{f'}\Ug{f}=\Ug{g}
 \label{eq:implementation-relative}
\end{equation}
is a product of the circuits of the generators \eqref{eq:identity-generator-circuit} by \Cref{thm:generalkernel}.
Every physical implementation of a logical gate is therefore obtained from the one in \Cref{sec:physical-implementation} by multiplying it by these circuits, and a transversal or constant-depth implementation can be searched for among them.

We first consider transversal gates, and as a concrete case we take the logical gate $\overline{\mathrm S}_a$.
We show that the logical gate $\overline{\mathrm S}_a$ has a transversal implementation if and only if there are a representative $p$ of the logical $Z$ operator $Z(z_a)$ and a binary vector $s$ such that
\begin{equation}
 \sum_{\substack{i<j\\i,j\in\supp p}}x_ix_j=\pair{s}{x}\pmod 2
 \label{eq:transversal-s-criterion}
\end{equation}
for every cycle $x\in Z_1$, that is, the $\CZ$ gates of the circuit \eqref{eq:s-gate-implementation} on the support $\supp p$ act on the code space as the Pauli operator $Z(s)$.

Within one code block, a transversal circuit consists of single-qubit gates~\cite{EastinKnill2009,YoderTakagiChuang2016}, and therefore at level two it is
\begin{equation}
 \Ug[2]{f_{\mathrm{tr}}}=\prod_{i=1}^{n}\mathrm{S}_i^{c_i},
 \qquad f_{\mathrm{tr}}(x)=\sum_{i=1}^{n}c_ix_i,
 \label{eq:implementation-transversal}
\end{equation}
with exponents $c_i\in\Z{4}$.
We write $c_i=p_i+2s_i$ with binary vectors $p$ and $s$, so that the circuit is $Z(s)\prod_{i\in\supp p}\mathrm S_i$.

If this circuit implements $\overline{\mathrm S}_a$, then $\pair{p}{x}=\pair{z_a}{x}$ modulo two for every cycle $x\in Z_1$, and hence the vector $p\oplus z_a$ lies in $Z_1^\perp=B^1$ and the vector $p$ is another representative of the same logical $Z$ operator.
The circuit \eqref{eq:s-gate-implementation} on the support $\supp p$ has the phase function $\pi_p$ and also implements $\overline{\mathrm S}_a$, and its gates $\mathrm S_i$ cancel in the difference $g=\pi_p-f_{\mathrm{tr}}$,
\begin{equation}
\begin{aligned}
 g&=2\sum_{i}s_ix_i+2\sum_{\substack{i<j\\i,j\in\supp p}}x_ix_j,\\
 \Ug[2]{g}&=Z(s)\prod_{\substack{i<j\\i,j\in\supp p}}\CZ_{ij}.
\end{aligned}
 \label{eq:implementation-rr-ratio}
\end{equation}
Since the difference $g$ belongs to $\Trv_2(\Hcode)$, it vanishes on every cycle, which is \eqref{eq:transversal-s-criterion}.

Conversely, if \eqref{eq:transversal-s-criterion} holds for a representative $p$, then the difference $g$ vanishes on every cycle, and hence the circuit $Z(s)\prod_{i\in\supp p}\mathrm S_i$ has the same logical phase function as the parity $\pi_p$ and is a transversal implementation of $\overline{\mathrm S}_a$.
The condition \eqref{eq:transversal-s-criterion} is therefore necessary and sufficient for the logical gate $\overline{\mathrm S}_a$ to have a transversal implementation.
In \Cref{sec:steane}, we confirm that the Steane code satisfies this condition and recover its well-known transversal $\mathrm S$ gate.

Beyond transversal gates, circuits of constant depth can be discussed in the same framework.
The physical implementations of a logical gate differ by the circuits of the generators \eqref{eq:identity-generator-circuit}, and therefore we ask which of them has the smallest depth.
We fix a logical phase function $\lpol f\in\Ph_m(H_1)$ and denote by $f_0$ its physical implementation built from the parities $\pi_{z_a}$ in \Cref{sec:physical-implementation}.
By \Cref{thm:hier-classification}, its physical implementations are the phase functions $f_0+g$ with $g\in\Trv_m(\Hcode)$.
The depth of the logical gate is the minimum over these circuits,
\begin{equation}
 \codedep(\lpol f)=\min_{g\in\Trv_m(\Hcode)}\dep(f_0+g),
 \label{eq:hier-log-depth}
\end{equation}
where $\dep(f)$ on the right-hand side is the minimum circuit depth of the physical circuit $\Ug{f}$.

We define the depth of a circuit as the number of its layers, where a layer consists of gates with pairwise disjoint supports.
Every elementary gate $\CPhpow{\calS}{c}$ with $1\leq|\calS|\leq m$ counts as one gate that fits in a single layer, regardless of its power $c$ and the number of qubits in $\calS$.
The depth is therefore at least the number of gates that meet one qubit.

To express this constraint in terms of the coefficients $c_\calS$ of the phase function $f$ in \eqref{eq:normalform}, we define the maximum incidence
\begin{equation}
 \inc(f)\coloneqq \max_{i\in[n]}|\{\calS:c_\calS\ne0,\ i\in \calS\}|,
 \label{eq:hier-degree}
\end{equation}
counting supports of size one as well~\cite{Rowshan2026}.
The following lemma bounds the circuit depth in terms of the maximum incidence.
\begin{lem}[{\cite{Cabello2011,KellerKueng2026,Rowshan2026}}]\label{lem:depth}
For a phase function $f\in\Ph_m(C_1)$ with $\inc(f)>0$,
\begin{equation}
 \inc(f)\leq \dep(f)\leq m(\inc(f)-1)+1.
 \label{eq:hier-depth}
\end{equation}
\end{lem}
The proof is given in \Cref{app:depth-bounds}, which also gives the exact value of the depth $\dep(f)$ as the minimum number of sets of pairwise disjoint supports into which the supports of the phase function $f$ can be divided.
At level two, the supports of size two are the edges of a graph, and Vizing's theorem improves the upper bound to $\inc(f)+1$~\cite{Vizing1964} (\Cref{thm:depth2}).

For a nontrivial logical phase function $\lpol f$, the same bounds hold after minimizing over its physical implementations.
We define its maximum incidence as
\begin{equation}
\inc(\lpol f) \coloneqq \min_{g\in\Trv_m(\Hcode)}\inc(f_0+g).
\end{equation}
Then \Cref{lem:depth} applied to every phase function $f_0+g$ gives
\begin{equation}
 \inc(\lpol f)\leq \codedep(\lpol f)
 \leq m(\inc(\lpol f)-1)+1,
 \label{eq:hier-coset-depth}
\end{equation}
and therefore, for a fixed level $m$, a logical gate on a code family has implementations of constant depth exactly when its physical implementations can be chosen with uniformly bounded incidence.
As an example of constant-depth gates, we discuss the intra-block logical $\CZ$ gate on the toric code in \Cref{sec:toric}.

\subsection{Chain-map interpretation}
\label{sec:chain-map}

We now relate the classification to the chain-map formulation of logical gates, beginning with level two.
At level two, condition (L2) is a chain-map condition.
It says that the columns of $MH_X^{\T}$ lie in $B^1=\im H_Z^{\T}$. Hence $MH_X^{\T}=H_Z^{\T}\psi$ for a binary matrix $\psi$ sending $X$ checks to $Z$ checks, and transposition gives $H_XM=\psi^{\T}H_Z$ because $M$ is symmetric.
Therefore, the diagram
\begin{equation}
 \begin{CD}
  C_2 @>H_X^{\T}>> C_1 @>H_Z>> C_0\\
  @VV\psi V @VVMV @VV\psi^{\T}V\\
  C^0 @>H_Z^{\T}>> C^1 @>H_X>> C^2
 \end{CD}
 \label{eq:chain-map-diagram}
\end{equation}
commutes, where $C^i$ is the dual of $C_i$. That is, the triple $(\psi,M,\psi^{\T})$ is a chain map from the chain complex of the code to its dual complex.

In terms of the circuit, both matrices describe the Pauli $Z$ operators that the conjugation \eqref{eq:clifford-check-conjugation} attaches to Pauli $X$ operators.
The matrix $\psi$ lists the $Z$ checks attached to each $X$ check, $Z(M\xi_\beta)=\prod_\alpha Z(\zeta_\alpha)^{\psi_{\alpha\beta}}$.
The matrix $M$ sends cycles to cocycles and boundaries to coboundaries, and therefore the operator $Z(M\gamma_b)$ attached to the logical $X$ operator $X(\gamma_b)$ is a logical $Z$ operator.
Up to $Z$ checks, it is the product of the operators $Z(z_a)$ with $\pair{\gamma_a}{M\gamma_b}=1$, which is the logical $Z$ operator that a product of logical $\mathrm S$ and $\CZ$ gates attaches to $X(\gamma_b)$ under conjugation.
The induced map from the homology group $H_1$ to the cohomology group $H^1$ is thus the logical polarization \eqref{eq:logicalpolar}, whose off-diagonal entries are the coefficients $\bar{c}_{ab}$ in \eqref{eq:clifford-logical-coefficients}.

For $\CZ$ circuits between two code blocks, this is the description of Ref.~\cite[Sec.~III.2]{SahayLongKhemani2026}, in which the off-diagonal block of $M$ is a chain map from one block to the dual of the other and the logical gate is given by the induced map on homology.
The chain maps that induce the zero map on homology correspond to the $Z$ checks controlled by single qubits in \Cref{thm:cliffkernel}.

The binary matrix determines the powers of logical $\mathrm S$ only modulo two, and therefore it does not retain the full logical phase function.
Reference~\cite[Sec.~III.4]{SahayLongKhemani2026} treats intra-block $\mathrm S$ gates by lifting the check matrices from $\Z{2}$ to $\Z4$ and leaves a general treatment of intra-block Clifford gates to future work.
Our \Cref{thm:cliffclass} provides this treatment for diagonal gates through the $\Z4$-valued phase functions $\Ph_2(H_1)$ on homology, without a lift of the check matrices.

At higher levels, the chain-map hierarchy of Ref.~\cite{SahayLongKhemani2026} iterates the construction with coefficients in $\F$ and describes the $\mathsf C^{\ell-1}Z$ gates between $\ell$ blocks, while \Cref{thm:hier-classification,thm:generalkernel} keep the exponents $\bar{c}_J\in\Z{2^{m-|J|+1}}$ and therefore also include the gates of $\mathrm T$ and controlled-$\mathrm S$ type and the intra-block gates.
In the recipe for transversal gates of Ref.~\cite[Sec.~IV.1]{SahayLongKhemani2026}, deforming a cycle by stabilizers corresponds to multiplying a circuit by the logical identity operators of \Cref{sec:physical-implementation}, and the sparsification step corresponds to the minimization \eqref{eq:hier-log-depth}.

\section{Applications}
\label{sec:examples}
In this section, we apply the classification to concrete codes and to the choice of physical circuits for a prescribed logical gate.
In \Cref{sec:steane}, we compare the transversal implementations of logical $\mathrm S$ and $\CZ$ gates on the Steane code with the circuits of \Cref{sec:physical-implementation}.
In \Cref{sec:toric}, we determine the minimum depth of logical $\CZ$ on the toric code.
Finally, in \Cref{sec:msd}, we outline how the logical identity operators describe the freedom to modify the non-Clifford circuits used in magic state distillation.

\subsection{Steane code: transversal gates and logical identity operators}\label{sec:steane}

We first compare circuits for the same logical gate on the Steane code, using the transversality criterion and the depth bounds of \Cref{sec:hier-depth}. This example shows how physical implementations of the same logical gate can have different circuit depths.

The Steane code is the $\codepar{7,1,3}$ CSS code with parity-check matrices $H_X=H_Z=H$, where
\begin{equation}
H=\begin{pmatrix}
1&0&0&1&1&0&1\\
0&1&0&1&0&1&1\\
0&0&1&0&1&1&1
\end{pmatrix}.
\label{eq:steane-checks}
\end{equation}

\begin{figure*}[tb]
\centering
\begin{tikzpicture}[x=1cm,y=1cm,line width=0.45pt,
  dot/.style={circle,fill,inner sep=1.0pt},
  g/.style={draw,fill=white,inner sep=1.3pt,font=\scriptsize},
  lab/.style={font=\small},num/.style={font=\tiny},
  idline/.style={draw=idcolor,line width=0.45pt},
  iddot/.style={circle,fill=idcolor,inner sep=1.0pt},
  idg/.style={draw=idcolor,text=idcolor,fill=white,inner sep=1.3pt,font=\scriptsize},
  trline/.style={line width=0.9pt},trdot/.style={circle,fill,inner sep=1.3pt}]
\colorlet{idcolor}{orange!85!black}
\definecolor{idc1}{HTML}{0072B2}\definecolor{idc2}{HTML}{D55E00}\definecolor{idc3}{HTML}{009E73}\definecolor{idc4}{HTML}{AA4499}\definecolor{idc5}{HTML}{8C510A}\definecolor{idc6}{HTML}{56B4E9}\definecolor{idc7}{HTML}{999933}\definecolor{idc8}{HTML}{332288}\definecolor{idc9}{HTML}{CC6677}\definecolor{idc10}{HTML}{44AA99}
\def\qy#1{\ifnum#1<8 -0.26*#1\else -0.26*#1-0.22\fi}
\begin{scope}[yscale=1.4]
\node[lab,anchor=west] at (-0.5,0.2) {(a)};
\foreach \q in {1,...,7}{\draw (0,{\qy{\q}})--(1.2,{\qy{\q}}); \node[num,anchor=east] at (-0.03,{\qy{\q}}) {\q}; \node[g] at (0.6,{\qy{\q}}) {$\mathrm S^\dagger$};}
\end{scope}
\begin{scope}[shift={(3.4,0)},yscale=1.4]
\node[lab,anchor=west] at (-0.5,0.2) {(b)};
\foreach \q in {1,...,7}{\draw (0,{\qy{\q}})--(6.9,{\qy{\q}}); \node[num,anchor=east] at (-0.03,{\qy{\q}}) {\q}; \node[g] at (0.38,{\qy{\q}}) {$\mathrm S^\dagger$};}
\foreach \q/\c in {1/idc1,2/idc2,3/idc3,4/idc4,5/idc5,6/idc6,7/idc7}{\node[idg,draw=\c,text=\c] at (0.95,{\qy{\q}}) {$Z$};}
\foreach \i/\j/\col [count=\c] in {1/2/idc1,1/3/idc1,1/4/idc4,1/5/idc5,1/6/idc6,1/7/idc1,2/3/idc2,2/4/idc2,2/5/idc2,2/6/idc6,2/7/idc7,3/4/idc4,3/5/idc3,3/6/idc3,3/7/idc3,4/5/idc5,4/6/idc4,4/7/idc7,5/6/idc6,5/7/idc5,6/7/idc7}{
  \pgfmathsetmacro{\x}{1.2+0.25*\c}
  \draw[draw=\col,line width=0.6pt] (\x,{\qy{\i}})--(\x,{\qy{\j}}); \node[circle,fill=\col,inner sep=1.1pt] at (\x,{\qy{\i}}) {}; \node[circle,fill=\col,inner sep=1.1pt] at (\x,{\qy{\j}}) {};}
\end{scope}
\begin{scope}[shift={(0,-3.7)}]
\node[lab,anchor=west] at (-0.5,0.3) {(c)};
\foreach \q in {1,...,14}{\draw (0,{\qy{\q}})--(2.0,{\qy{\q}}); \node[num,anchor=east] at (-0.03,{\qy{\q}}) {\q};}
\foreach \i in {1,...,7}{
  \pgfmathtruncatemacro{\j}{\i+7}
  \pgfmathsetmacro{\x}{0.25*\i}
  \draw (\x,{\qy{\i}})--(\x,{\qy{\j}}); \node[dot] at (\x,{\qy{\i}}) {}; \node[dot] at (\x,{\qy{\j}}) {};}
\end{scope}
\begin{scope}[shift={(3.4,-3.7)}]
\node[lab,anchor=west] at (-0.5,0.3) {(d)};
\foreach \q in {1,...,14}{\draw (0,{\qy{\q}})--(12.6,{\qy{\q}}); \node[num,anchor=east] at (-0.03,{\qy{\q}}) {\q};}
\foreach \i in {1,...,7}{
  \pgfmathtruncatemacro{\j}{\i+7}
  \pgfmathsetmacro{\x}{0.25*((\i-1)*7+\i)}
  \draw[trline] (\x,{\qy{\i}})--(\x,{\qy{\j}}); \node[trdot] at (\x,{\qy{\i}}) {}; \node[trdot] at (\x,{\qy{\j}}) {};}
\foreach \i/\jj/\c in {1/2/idc1,1/3/idc1,1/4/idc1,1/5/idc1,1/6/idc2,1/7/idc3,2/1/idc4,2/3/idc5,2/4/idc5,2/5/idc5,2/6/idc2,2/7/idc3,3/1/idc6,3/2/idc5,3/4/idc6,3/5/idc6,3/6/idc2,3/7/idc6,4/1/idc4,4/2/idc5,4/3/idc7,4/5/idc7,4/6/idc7,4/7/idc7,5/1/idc8,5/2/idc5,5/3/idc8,5/4/idc8,5/6/idc8,5/7/idc3,6/1/idc4,6/2/idc9,6/3/idc9,6/4/idc9,6/5/idc9,6/7/idc3,7/1/idc4,7/2/idc10,7/3/idc10,7/4/idc10,7/5/idc10,7/6/idc2}{
  \pgfmathtruncatemacro{\j}{\jj+7}
  \pgfmathsetmacro{\x}{0.25*((\i-1)*7+\jj)}
  \draw[draw=\c,line width=0.6pt] (\x,{\qy{\i}})--(\x,{\qy{\j}}); \node[circle,fill=\c,inner sep=1.1pt] at (\x,{\qy{\i}}) {}; \node[circle,fill=\c,inner sep=1.1pt] at (\x,{\qy{\j}}) {};}
\end{scope}
\end{tikzpicture}
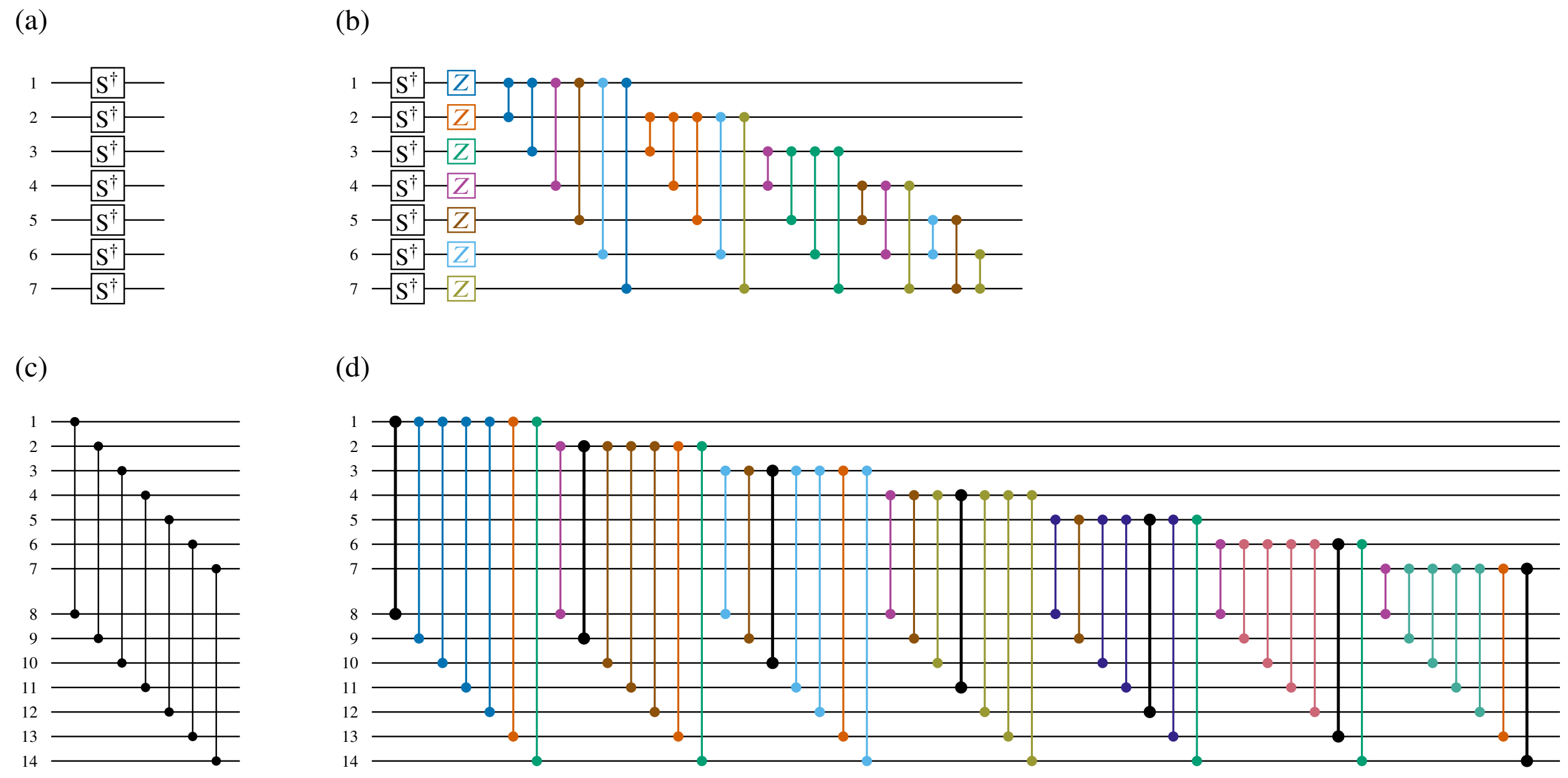
\caption{Circuits on the Steane code. The gates of each color form a logical identity operator, and the remaining gates form the transversal circuit. (a) The transversal gate $(\mathrm S^\dagger)^{\otimes7}$ implements $\overline{\mathrm S}$ on one block with depth one. (b) The circuit \eqref{eq:s-gate-implementation}, $\mathrm S^{\otimes7}\prod_{i<j}\CZ_{ij}$, implements the same gate with depth seven, where each $\mathrm S$ gate is drawn as $\mathrm S^\dagger$ followed by $Z$. (c) The transversal circuit of seven $\CZ$ gates between two blocks implements $\overline{\CZ}_{12}$ with depth one. (d) The round-robin circuit \eqref{eq:cz-gate-disjoint} of all $49$ $\CZ$ gates implements the same gate with depth seven.}
\label{fig:steane}
\end{figure*}

We start with the logical $\mathrm S$ gate on one block.
We choose $\gamma=z=\mathbf{1}_{7}$, the all-one string on the block. The physical implementation $f_0=\pi_z$ of \Cref{sec:physical-implementation} is the circuit \eqref{eq:s-gate-implementation} on all seven qubits (\Cref{fig:steane}(b)),
\begin{equation}
 \Ug[2]{f_0}=\mathrm{S}^{\otimes7} \prod_{1\leq i<j\leq7}\CZ_{ij}.
\end{equation}
Each qubit meets seven gates. Hence $\inc(f_0)=7$, and $\dep(f_0)=7$ by \Cref{lem:depth} and the seven layers given in \Cref{app:examples-steane}.
The transversal gate $(\mathrm{S}^\dagger)^{\otimes7}$ implements the same $\overline{\mathrm{S}}$ with maximum incidence one and depth one (\Cref{fig:steane}(a)).

This is an instance of the criterion \eqref{eq:transversal-s-criterion} with $c_i=3$ and $p=s=\mathbf{1}_{7}$.
The difference \eqref{eq:implementation-rr-ratio} is given by
\begin{equation}
  \label{eq:steane-s-difference-circuit}
\Ug[2]{g}=Z^{\otimes7} \prod_{1\leq i<j\leq7}\CZ_{ij},
\end{equation}
drawn in color in \Cref{fig:steane}(b).
This circuit acts as a logical identity operator.
In the normal form \eqref{eq:cliffkernel-solution} with the rows $\zeta_1,\zeta_2,\zeta_3$ of the matrix $H$, the difference reads
\begin{multline}
 g=2\sum_{j=1}^3 \pi_{\zeta_j}+2\sum_{1\leq j<j'\leq 3}\pi_{\zeta_j}\pi_{\zeta_{j'}}
 +2(\pi_{e_6}+\pi_{e_7})\pi_{\zeta_1} \\
  \quad+2(\pi_{e_5}+\pi_{e_7})\pi_{\zeta_2} + 2(\pi_{e_4}+\pi_{e_7})\pi_{\zeta_3}.
 \label{eq:steane-s-difference}
\end{multline}
The circuit \eqref{eq:steane-s-difference-circuit} is therefore a product of generators of the logical identity operators, and \Cref{fig:steane}(b) shows one such product.
Equivalently, the $\CZ$ part alone acts on the code space as $Z^{\otimes7}$, which is the logical Pauli operator $\overline Z$ (\Cref{app:examples-steane}).

We next consider the logical $\CZ$ gate between two blocks of the Steane code.
We number the qubits of the first block by $1,\ldots,7$ and those of the second block by $8,\ldots,14$, and write a bit string on the two blocks as a pair of its restrictions to the blocks.
As the logical $X$ representatives, we choose $\gamma_1=(\mathbf{1}_{7},\mathbf{0}_{7})$ and $\gamma_2=(\mathbf{0}_{7},\mathbf{1}_{7})$, where $\mathbf{0}_{7}$ is the all-zero string on one block, and as the dual representatives we take $z_a=\gamma_a$.

The logical gate $\overline{\CZ}_{12}$ has the logical phase function $2\lambda_1\lambda_2$.
Its physical implementation in \Cref{sec:physical-implementation} is the round-robin circuit \eqref{eq:cz-gate-disjoint} with the phase function $2\pi_{z_1}\pi_{z_2}$, which places a $\CZ$ gate on all $49$ pairs between the blocks (\Cref{fig:steane}(d)).
Every qubit meets seven gates. Therefore, its maximum incidence is $\inc(2\pi_{z_1}\pi_{z_2})=7$, and $\dep(2\pi_{z_1}\pi_{z_2})=7$. The lower bound is \Cref{lem:depth}, and seven layers that attain it are given in \Cref{app:examples-steane}.

The Steane code is a CSS code with $H_X=H_Z$, and therefore the transversal $\mathrm{CNOT}$ gate between the two blocks and the transversal Hadamard gate on one block are logical gates~\cite{Gottesman1997}.
Conjugating the transversal $\mathrm{CNOT}$ by the transversal Hadamard on the target block gives a transversal logical $\CZ$ gate.
This transversal circuit of seven $\CZ$ gates implements the same $\overline{\CZ}_{12}$ and has the phase function
\begin{equation}
 f(x)=2\sum_{i=1}^7x_ix_{7+i}\pmod4.
\end{equation}
Its gates have disjoint supports, and hence $\inc(f)=1$ and $\dep(f)=1$ (\Cref{fig:steane}(c)).

The difference of the two phase functions is $g=2\pi_{z_1}\pi_{z_2}-f=2\sum_{i\neq j}x_ix_{7+j}$, that is, the $42$ $\CZ$ gates of the round-robin circuit that do not belong to the transversal circuit, drawn in color in \Cref{fig:steane}(d).
We write $\zeta_j$ and $\zeta'_j$ for the $j$-th row of $H$ on the first and on the second block.
In the normal form \eqref{eq:cliffkernel-solution} with these six rows, the difference reads
\begin{multline}
 g=2\sum_{j\neq j'}\pi_{\zeta_j}\pi_{\zeta'_{j'}}
 +2(\pi_{e_{13}}+\pi_{e_{14}})\pi_{\zeta_1}
 +2(\pi_{e_{12}}+\pi_{e_{14}})\pi_{\zeta_2}\\
 +2(\pi_{e_{11}}+\pi_{e_{14}})\pi_{\zeta_3}
 +2(\pi_{e_6}+\pi_{e_7})\pi_{\zeta'_1}
 +2(\pi_{e_5}+\pi_{e_7})\pi_{\zeta'_2}\\
 +2(\pi_{e_4}+\pi_{e_7})\pi_{\zeta'_3}.
 \label{eq:steane-cz-difference}
\end{multline}
The $42$ $\CZ$ gates are therefore a product of generators of the logical identity operators, and \Cref{fig:steane}(d) shows one such product.

In both examples, the transversal circuit is obtained from the circuit supplied by the classification by multiplying it by generators of the logical identity operators.
It would be interesting to search for transversal and constant-depth circuits on other CSS codes in the same way, and we leave this search for future work.

\subsection{Toric code: minimum depth of the intra-block logical \texorpdfstring{$\CZ$}{CZ} gate}
\label{sec:toric}

We next determine the minimum depth of the intra-block logical $\CZ$ gate along the toric code family~\cite{Kitaev2003,Dennis2002}.
Here we consider a single code block of the $L\times L$ toric code, and therefore the transversal $\CZ$ gate between two blocks used in the Steane example is not available for the logical $\CZ$ gate between its two logical qubits.
Several constructions give circuits for this gate~\cite{BreuckmannBurton2024,QuintavalleWebsterVasmer2023,Holmes2026,SahayLongKhemani2026}.
We construct circuits of minimum depth for every $L$, depth one for odd $L$ and depth two for even $L$, and prove their optimality.

The $L\times L$ toric code with $L\geq3$ has $n=2L^2$ qubits on the edges of a square lattice on the torus, with the stars as $X$ checks and the plaquettes as $Z$ checks.
It encodes $k=2$ logical qubits with distance $d=L$, and therefore its parameters are $\codepar{2L^2,2,L}$.
The following proposition states the minimum depth of the intra-block logical $\CZ$ gate in these codes.

\begin{pro}
\label{pro:toric}
In the $L\times L$ toric code with $L\geq3$, the logical gate $\overline{\CZ}_{12}$ between the two logical qubits can be implemented by diagonal circuits of level two as follows.
\begin{enumerate}
\item[(a)] If $L$ is odd, it can be implemented with depth one.
\item[(b)] If $L$ is even, it can be implemented with depth two, whereas it cannot be implemented with depth one, even up to logical $\mathrm{S}$ and $Z$ gates on each logical qubit.
\end{enumerate}
\end{pro}

The proof is given in \Cref{app:toric}, and here we briefly explain its idea.
It uses the general form of a circuit of depth one.
Such a circuit is a layer of gates with disjoint supports, that is, $\CZ$ gates on disjoint pairs of qubits and $\mathrm{S}$, $\mathrm{S}^\dagger$, or $Z$ gates on some of the remaining qubits.
Its phase function $f$ has the form \eqref{eq:quadraticnormal}. By \Cref{thm:cliffcriterion}, it preserves the code space exactly when (L1) and (L2) hold on the stars, and the exponent of the logical $\CZ$ gate is $\bar{c}_{12}=\pair{\gamma_1}{M\gamma_2}$.
For part (a), we therefore construct a pairing of the lattice edges that satisfies these conditions, and for the lower bound in part (b), we show that no pairing satisfies them.

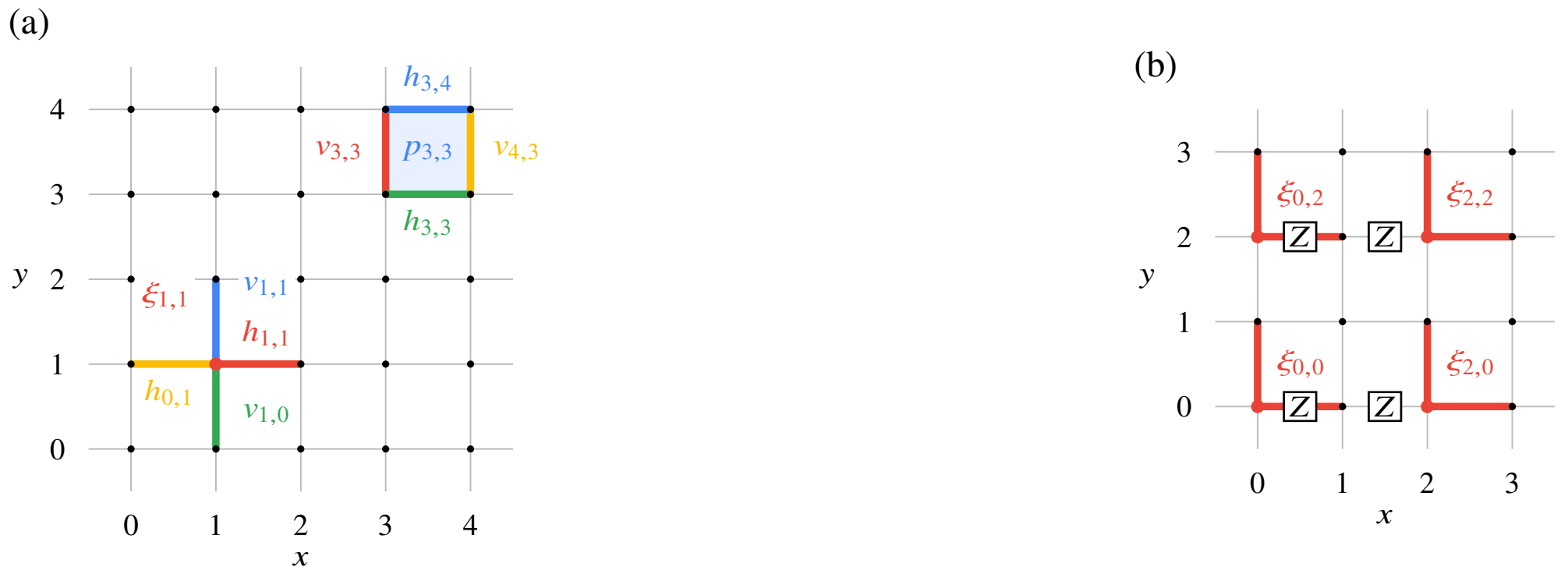
\begin{figure*}[tb]
\centering
\begin{minipage}[c]{0.49\textwidth}
\centering
\begin{tikzpicture}[scale=0.72,every node/.style={font=\scriptsize},
 grid/.style={draw=gray!55,line width=0.35pt},
 ghost/.style={draw=gray!55,line width=0.35pt,dash pattern=on 2pt off 1.5pt},
 hl/.style={line width=1.8pt,line cap=round,line join=round},
 copy/.style={dash pattern=on 3pt off 2pt},
 tag/.style={fill=white,inner sep=1pt}]
\path[use as bounding box] (-1.75,-1.7) rectangle (5.75,5.3);
\begin{scope}
\node[font=\small] at (-1.2,5.0) {(a)};
\fill[Gblue!12] (3,3) rectangle (4,4);

\foreach \x in {0,...,4}{
  \draw[grid] (-0.5,\x) -- (4.5,\x);
  \draw[grid] (\x,-0.5) -- (\x,4.5);}

\draw[hl,Gred] (1,1) -- (2,1);
\draw[hl,Gred] (3,3) -- (3,4);
\draw[hl,Gyellow] (0,1) -- (1,1);
\draw[hl,Gyellow] (4,3) -- (4,4);
\draw[hl,Gblue] (1,1) -- (1,2);
\draw[hl,Gblue] (3,4) -- (4,4);
\draw[hl,Ggreen] (1,0) -- (1,1);
\draw[hl,Ggreen] (3,3) -- (4,3);

\foreach \x in {0,...,4}{\foreach \y in {0,...,4}{
  \fill[black] (\x,\y) circle (1.2pt);}}
\fill[Gred] (1,1) circle (2.2pt);
\node[tag,text=Gred] at (0.4,1.8) {$\xi_{1,1}$};
\node[Gblue] at (3.5,3.5) {$p_{3,3}$};
\node[tag,text=Gred] at (1.6,1.35) {$h_{1,1}$};
\node[tag,text=Gred] at (2.45,3.5) {$v_{3,3}$};
\node[tag,text=Gyellow] at (0.45,0.65) {$h_{0,1}$};
\node[tag,text=Gyellow] at (4.55,3.5) {$v_{4,3}$};
\node[tag,text=Gblue] at (1.6,1.9) {$v_{1,1}$};
\node[tag,text=Gblue] at (3.5,4.35) {$h_{3,4}$};
\node[tag,text=Ggreen] at (1.6,0.40) {$v_{1,0}$};
\node[tag,text=Ggreen] at (3.5,2.65) {$h_{3,3}$};

\foreach \x in {0,...,4}{\node[anchor=north] at (\x,-0.62) {$\x$};}
\foreach \y in {0,...,4}{\node[anchor=east] at (-0.62,\y) {$\y$};}
\node at (2,-1.3) {$x$};
\node at (-1.3,2) {$y$};
\end{scope}
\end{tikzpicture}
\end{minipage}\hfill
\begin{minipage}[c]{0.49\textwidth}
\centering
\begin{tikzpicture}[scale=0.72,every node/.style={font=\scriptsize},
 grid/.style={draw=gray!55,line width=0.35pt},
 ghost/.style={draw=gray!55,line width=0.35pt,dash pattern=on 2pt off 1.5pt},
 hl/.style={line width=1.8pt,line cap=round,line join=round},
 copy/.style={dash pattern=on 3pt off 2pt},
 tag/.style={fill=white,inner sep=1pt}]
\path[use as bounding box] (-1.75,-1.7) rectangle (4.75,4.3);
\begin{scope}
\node[font=\small] at (-1.2,4.0) {(b)};

\foreach \x in {0,...,3}{
  \draw[grid] (-0.5,\x) -- (3.5,\x);
  \draw[grid] (\x,-0.5) -- (\x,3.5);}

\foreach \x in {0,2}{\foreach \y in {0,2}{
  \draw[hl,Gred] (\x,\y) -- ({\x+1},\y);
  \draw[hl,Gred] (\x,\y) -- (\x,{\y+1});}}

\foreach \x in {0,...,3}{\foreach \y in {0,...,3}{
  \fill[black] (\x,\y) circle (1.2pt);}}
\foreach \x in {0,2}{\foreach \y in {0,2}{\fill[Gred] (\x,\y) circle (2.2pt);}}
\foreach \x in {0,1}{\foreach \y in {0,2}{
  \node[draw=black,fill=white,line width=0.5pt,inner sep=1.1pt]
    at ({\x+0.5},\y) {$Z$};}}
\node[tag,text=Gred] at (0.5,0.5) {$\xi_{0,0}$};
\node[tag,text=Gred] at (2.5,0.5) {$\xi_{2,0}$};
\node[tag,text=Gred] at (0.5,2.5) {$\xi_{0,2}$};
\node[tag,text=Gred] at (2.5,2.5) {$\xi_{2,2}$};

\foreach \x in {0,...,3}{\node[anchor=north] at (\x,-0.62) {$\x$};}
\foreach \y in {0,...,3}{\node[anchor=east] at (-0.62,\y) {$\y$};}
\node at (1.5,-1.3) {$x$};
\node at (-1.3,1.5) {$y$};
\end{scope}
\end{tikzpicture}
\end{minipage}
\caption{Pairings of the edges of the $L\times L$ toric code that give the logical $\CZ$ gate.
The edges $h_{x,y}$ and $v_{x,y}$ run from the vertex $(x,y)$ to $(x+1,y)$ and to $(x,y+1)$, respectively.
The lattice has periodic boundary conditions.
(a) The pairing \eqref{eq:toric-odd-pairing-main} for $L=5$ maps the star $\xi_{1,1}$ onto the plaquette $p_{3,3}$, with the two edges of each pair drawn in the same color.
(b) The point reflection \eqref{eq:toric-even-pairing-main} for $L=4$ puts one pair (red) inside each of the four stars with $x,y\in\{0,2\}$, and the $Z$ gates (boxes) connect them.}
\label{fig:toric}
\end{figure*}

We denote the horizontal and vertical edges by $h_{x,y}$ and $v_{x,y}$, respectively, with coordinates modulo $L$ (\Cref{fig:toric}).
In the constructions below, we form pairs of a horizontal edge and a vertical edge so that every star is mapped onto a plaquette, which ensures condition (L2).
We then require every star to contain an even number of pairs, which is condition (L1).

For odd $L$, we consider the pairing $\tau$ of the edge set defined by
\begin{equation}
 \tau(h_{x,y})=v_{-x-1,\,y+\frac{L-1}{2}},\qquad
 \tau(v_{x,y})=h_{-x-1,\,y-\frac{L-1}{2}}.
 \label{eq:toric-odd-pairing-main}
\end{equation}
Applying $\tau$ twice returns every edge to itself, that is, $\tau^2=1$, and hence $\tau$ divides the edges into pairs $\{e,\tau(e)\}$.
No star contains such a pair, and therefore condition (L1) holds (\Cref{fig:toric}(a)).
The corresponding $\CZ$ gates form a circuit of depth one that implements the logical $\CZ$ gate.

For even $L$, we consider the point reflection
\begin{equation}
 \tau(h_{x,y})=v_{-x,-y},\qquad \tau(v_{x,y})=h_{-x,-y},
 \label{eq:toric-even-pairing-main}
\end{equation}
which puts exactly one pair inside each of the four stars at the vertices $(x,y)$ with $x,y\in\{0,L/2\}$, and therefore condition (L1) fails there (\Cref{fig:toric}(b)).
Adding $Z$ gates on $L$ horizontal edges that connect these four vertices restores (L1).
The resulting gate is again a logical $\CZ$. However, it has depth two.

For even $L$, we can show in two steps that no circuit of depth one implements the logical $\CZ$ gate.
First, condition (L2) requires the $\CZ$ gates of a depth-one circuit to pair all the edges so that every star is mapped onto a plaquette, and only two kinds of pairings do so (\Cref{app:toric}).
A pairing of the first kind maps horizontal edges to horizontal edges and vertical edges to vertical edges, and a pairing of the second kind is a point reflection such as \eqref{eq:toric-even-pairing-main}, which exchanges them.
Second, we find that a circuit with a pairing of the first kind has $\bar{c}_{12}=0$ and hence does not implement the logical $\CZ$ gate, and that a circuit with a pairing of the second kind does not preserve the code space.

Previous constructions of this gate use a symmetry of the code or an algorithmic search.
The fold-transversal gates of Breuckmann and Burton use a $ZX$ duality~\cite{BreuckmannBurton2024}, Quintavalle, Webster, and Vasmer pair the two sectors of a symmetric hypergraph product code~\cite{QuintavalleWebsterVasmer2023}, and Sahay, Long, and Khemani fold an inter-block gate onto one block~\cite{SahayLongKhemani2026}.
Holmes found a circuit for every $L\geq3$ by an algorithmic search~\cite{Holmes2026}.
It consists of a layer of $\CZ$ gates and $\mathrm{S}$ gates on paired qubits, which has depth two in the circuit model of \Cref{sec:hier-depth} and attains our bound for even $L$.

Each of these works constructs circuits, whereas a lower bound on the depth requires a statement about all possible circuits.
Our classification provides such a statement because conditions (L1) and (L2) characterize all the code-preserving circuits at level two.
We can therefore exclude every diagonal circuit of depth one for even $L$, and hence our circuits of depth one for odd $L$ and depth two for even $L$ have the minimum depth.

\subsection{Magic state distillation}\label{sec:msd}
Finally, we consider the freedom to choose the physical implementation at level three, where non-Clifford logical gates are used in magic state distillation~\cite{BravyiKitaev2005}.
In magic state distillation based on diagonal gates, a physical non-Clifford circuit must implement the desired logical gate~\cite{BravyiHaah2012}.
For example, the logical gate $\overline{\mathrm{T}}^{\otimes k}$ is implemented by
\begin{equation}
 f_0(x)=\sum_{a=1}^k\pi_{z_a}(x)\pmod8,
\label{eq:msd-target-representative}
\end{equation}
whose circuit is the product of the circuits \eqref{eq:t-gate-implementation}.
By \Cref{thm:hier-classification}, the other physical implementations are $f_0+g$ with $g\in\Trv_3(\Hcode)$, and by \Cref{thm:generalkernel} the phase functions $\pi_{\zeta_\alpha}$, $2\pi_{e_i}\pi_{\zeta_\alpha}$, and $4\pi_{e_i}\pi_{e_j}\pi_{\zeta_\alpha}$ with $i<j$ generate all these changes.
Their circuits (\Cref{fig:level-three}(d) to (f)) change the physical $\mathrm{T}$, $\CS$, and $\CCZ$ gates while keeping the logical gate fixed.
We are investigating magic state distillation protocols that exploit this freedom in the physical implementation, and the details will be reported in a forthcoming paper~\cite{EharaHarunaInPrep}.

\section{Conclusion and outlook}
\label{sec:conclusion}
\label{sec:outlook}

In this work, we have classified the logical diagonal gates of qubit CSS codes at every level $m$ of the Clifford hierarchy.
We described a diagonal circuit by its phase function and showed that the circuit preserves the code space exactly when the phase function satisfies a finite set of equations on the $X$ checks (\Cref{thm:cliffcriterion,thm:hier-finite}).
The logical gate of such a circuit is given by the values of its phase function on representatives of the homology classes.
The code-preserving diagonal circuits of level $m$, taken modulo the logical identity operators, form a group isomorphic to that of the phase functions of level $m$ on the first homology group $H_1$ (\Cref{thm:cliffclass,thm:hier-classification}).
The logical identity operators are generated by the circuits that apply a phase according to the parity of a $Z$ check under the control of at most $m-1$ physical qubits (\Cref{thm:cliffkernel,thm:generalkernel}).

Based on the classification, we also determined all the diagonal circuits without ancillas for each logical gate.
We gave an explicit physical circuit for every logical gate, and every other circuit that implements the same gate is this circuit multiplied by logical identity operators (\Cref{sec:physical-implementation}).
The search for a transversal or constant-depth circuit thus reduces to choosing logical identity operators.
We also derived upper and lower bounds on circuit depth in terms of the maximum incidence, the largest number of gates acting on a single qubit (\Cref{lem:depth}).
On the Steane code, we recovered the transversal implementations of logical $\mathrm S$ and $\CZ$ by choosing suitable logical identity operators (\Cref{sec:steane}).
On the toric code, we determined the minimum depth of the intra-block logical $\CZ$ gate, which is one for odd $L$ and two for even $L$ (\Cref{pro:toric}).

It is an open problem how to find the logical identity operators that give a transversal or constant-depth circuit in a systematic way.
At present, circuits of small maximum incidence and lower bounds on the depth are obtained separately for each code family, as in the qLDPC families with transversal non-Clifford gates~\cite{GolowichLin2025,Zhu2025}.
We also expect our classification to give a better understanding of the known results on these circuits, including the homological obstruction to transversal logical diagonal gates~\cite{Haruna_2026} and the logical gates constructed from cup products and cohomology operations~\cite{BreuckmannCupsGates2024,BreuckmannCupsGatesII2026,Kobayashi_2026,HsinKobayashiZhu2024,ZhuSikanderPortnoyCrossBrown2023}.
Deriving these results within this framework is left for future work.

We expect the classification to extend to wider classes of logical gates and codes.
In \Cref{sec:applicability}, we discussed its extension to CSS codes on prime-dimensional qudits, automorphism gates, and non-CSS stabilizer codes.
For non-CSS codes, the remaining task is to express the logical gate and the depth of its circuits in terms of the original code.
One direction is to derive lower bounds on the depth that also cover non-diagonal gates, in analogy with the restrictions on the logical gates of geometrically local codes~\cite{BravyiKonig2013,Pastawski_2015,Webster_2018}.
Another interesting direction is the extension to non-Abelian codes such as the quantum double models~\cite{Kitaev2003}, for which transversal and constant-depth logical gates have been constructed~\cite{manjunath2026universalquantumcomputationgroup,LiSongMatteiManjunathTiwariEllison2026,WarmanSchaferNameki2025,WarmanSchaferNameki2026,Kobayashi_2026,LiSong2026}.
The first step toward a classification of these gates is to find what replaces the homology group $H_1$ as the label of the encoded basis, for example, the irreducible representations of the Drinfeld double $D(G)$ for the quantum double model on a torus.

For practical use, the circuit model should incorporate the constraints and resources of hardware, such as restricted connectivity, ancillas, and measurements (\Cref{sec:scope}).
Since the diagonal circuits that implement a given logical gate differ from one another by logical identity operators, the gate count and the depth can be optimized over these circuits under connectivity constraints, as with phase polynomials in circuit synthesis~\cite{AmyMosca2019}.
We expect this optimization to be useful for finding transversal and constant-depth circuits on other codes, including qLDPC codes.

\backsection{Acknowledgments}
J.H. thanks Keisuke Fujii, Kohei Yamamoto, and Vieri Mattei for helpful discussions.
J.H. also thanks Nikolas Breuckmann and David Long for fruitful discussions at the 2026 YITP QEC Workshop ``Fault-tolerant logical processing''.

\backsection{Use of AI tools}

Anthropic models (Claude Opus 5.5, Claude Fable 5.1, and Claude Fable 5.2) and OpenAI models (GPT-6 Astra and GPT-6.2 Astra) were used in this work.
They assisted with deriving the proofs, checking calculations, searching the literature, and drafting and editing the manuscript.
The authors directed this work, verified all results, and take responsibility for the final manuscript.

\backsection{Funding}
 K.E. is supported by JST SPRING Grant No. JPMJSP2108 and JST NEXUS Grant No. JPMJNX26C2.
 J.H. is supported by MEXT Quantum Leap Flagship Program (MEXT Q-LEAP) Grant No. JPMXS0120319794, JST COI-NEXT Grant No. JPMJPF2014, and JST CREST Grant No. JPMJCR24I3.

\let\oldaddcontentsline\addcontentsline 
\renewcommand{\addcontentsline}[3]{}
\bibliographystyle{homological-apsrev}

\bibliography{references}

\let\addcontentsline\oldaddcontentsline

\clearpage
\newgeometry{hmargin=1.2in,vmargin=0.8in}

\widetext

\appendix
\crefalias{section}{appsec}
\crefalias{subsection}{appsec}
\renewcommand{\ofsupp}{}
\addtocontents{toc}{\protect\fi}
\begin{center}
{\Large\bfseries Supplemental Materials of}\\[8pt]
{\large\bfseries Homological classification of logical diagonal gates\\[3pt]
in the Clifford hierarchy}\\[10pt]
{\normalsize Koki Ehara$^{1}$ and Junichi Haruna$^{2,3}$}\\[4pt]
{\small\itshape
$^{1}$Department of Basic Science, The University of Tokyo, Tokyo, Japan\\
$^{2}$Graduate School of Informatics, Kyoto University, Sakyo-ku, Kyoto, 606-8501, Japan\\
$^{3}$Center for Quantum Information and Quantum Biology, University of Osaka, Japan}
\end{center}
\tableofcontents

\section{Foundations of the homological classification}
\label{app:supplementary}

In \Cref{app:css-foundations}, we derive the encoded basis and logical coordinates.
In \Cref{app:phase-properties}, we prove the normal form, the substitution rule, and the criterion for code preservation.
In \Cref{app:polarization-details}, we collect the difference identities of phase functions used in the classification at every level and prove that the polarization of the highest order is multilinear.
The theorems of the main text and the depth bounds are proved in \Cref{app:proofs}.
\makeatletter
\renewcommand{\p@subsection}{}
\makeatother
\renewcommand{\thesubsection}{\thesection.\arabic{subsection}}
\subsection{CSS codes, homology, and logical coordinates}
\label{app:css-foundations}

We derive the encoded basis and logical coordinates of \Cref{sec:prelim-css} from the CSS stabilizers~\cite{CalderbankShor1996,Steane1996,BombinHomological2007}.
We use the notation of \Cref{sec:prelim-css}.
The relation $X(u)Z(z)=(-1)^{\pair{u}{z}}Z(z)X(u)$ and the condition $H_XH_Z^{\T}=0$ show that all $X(\xi_\beta)$ and $Z(\zeta_\alpha)$ commute. Their common $+1$ eigenspace is the CSS code $\Hcode$.
The cycles, boundaries, and homology group of the chain complex and the cocycles, coboundaries, and cohomology group are
\begin{align}
 Z_1&=\ker H_Z, & B_1&=\im H_X^{\T}, & H_1&=Z_1/B_1,\label{eq:supp-css-homology}\\
 Z^1&=\ker H_X, & B^1&=\im H_Z^{\T}, & H^1&=Z^1/B^1.\label{eq:supp-css-cohomology}
\end{align}
In particular, $B_1\subseteq Z_1$ and $B^1\subseteq Z^1$.

The $Z$ check conditions restrict the computational support of a code state to $Z_1$.
The $X$ check conditions equate its amplitudes at $x$ and $x\oplus\xi_\beta$, hence at every pair of strings in the same boundary coset.
Thus the code has the orthonormal basis
\begin{equation}
 \ket{[x]}=|B_1|^{-1/2}\sum_{b\in B_1}\ket{x\oplus b},\qquad x\in Z_1.
 \tag{\ref{eq:codebasis}}
\end{equation}
Each vector is normalized because its coset has $|B_1|$ elements, and distinct cosets have disjoint computational supports.
Conversely, every state satisfying the checks is a linear combination of these states.
Consequently,
\begin{equation}
 \dim\Hcode=|Z_1/B_1|=2^k,\qquad
 k=\dim H_1=n-\rank H_Z-\rank H_X.
 \label{eq:supp-css-dimension}
\end{equation}

For a cocycle $z\in Z^1$, the Pauli operator $Z(z)$ acts on the encoded basis as
\begin{equation}
 Z(z)\ket{[x]}=(-1)^{\pair{z}{x}}\ket{[x]},
 \label{eq:supp-logical-z}
\end{equation}
because $\pair{z}{b}=0$ for all $b\in B_1$.

The value $\pair{z}{x}$ is unchanged by adding a boundary to $x$ or a coboundary to $z$, and the binary pairing therefore defines a pairing between the homology group $H_1$ and the cohomology group $H^1$.
This pairing is nondegenerate because $B_1^\perp=Z^1$ and $Z_1^\perp=B^1$.
Indeed, a cycle pairing to zero with every cocycle belongs to $(Z^1)^\perp=B_1$, and a cocycle pairing to zero with every cycle belongs to $Z_1^\perp=B^1$.
We choose a basis $[\gamma_1],\ldots,[\gamma_k]$ of the homology group $H_1$.
Since the pairing is nondegenerate, there are cocycles $z_1,\ldots,z_k$ satisfying $\pair{z_a}{\gamma_b}=\delta_{ab}$, the dual representatives, and every cycle has the form
\begin{equation}
 x=b\oplus\bigoplus_{a=1}^k \lambda_a([x])\gamma_a,\qquad
 b\in B_1,\qquad \lambda_a([x])=\pair{z_a}{x}.
 \label{eq:supp-cycle-coordinates}
\end{equation}
Accordingly, the linear map $x\mapsto(\pair{z_a}{x})_{a=1}^k$ on $C_1$ vanishes on $B_1$ and returns the logical coordinates $\lambda_a([x])$ on $Z_1$.

\subsection{Normal forms and the criterion for code preservation}
\label{app:phase-properties}

The coefficients $c_\calS$ in the normal form \eqref{eq:normalform} of \Cref{lem:phase} are independent, which gives the direct sum of groups
\begin{equation}
\Ph_m(V)\cong\bigoplus_{\ell=1}^{\min(m,\dim V)}
\left(\Z{2^{m-\ell+1}}\right)^{\binom{\dim V}{\ell}}.
\end{equation}
\begin{proof}[Proof of \Cref{lem:phase}]
Every function $f$ of binary coordinates is uniquely written as $f(x)=\sum_\calS a_\calS x^\calS$, where the sum runs over the subsets $\calS$ of coordinates~\cite{Zhegalkin1927}, \cite[Sec.~2.2.4]{Carlet2021}. Its coefficient on a nonempty set $\calS$ is
\begin{equation}
a_\calS=\sum_{R\subseteq \calS}(-1)^{|\calS|-|R|}
 f\left(\bigoplus_{i\in R}e_i\right)
 =\delta_{|\calS|}f((e_i)_{i\in \calS}).
\label{eq:normalform-reconstruction}
\end{equation}
The condition $f(0)=0$ gives $a_\varnothing=0$.

We assume $f\in\Ph_m(V)$. Differences of order greater than $m$ vanish by definition, and hence $a_\calS=0$ for $|\calS|>m$. For $1\leq d=|\calS|\leq m$, we repeat one of the $d$ coordinate directions another $m+1-d$ times. Since $\Delta_v^2=-2\Delta_v$ by \eqref{eq:diffidentities}, this gives
\begin{equation}
(-2)^{m+1-d}a_\calS=0\quad\text{in }\Zm.
\label{eq:supp-repeated-difference-coefficient}
\end{equation}
Thus $a_\calS$ is divisible by $2^{d-1}$ and can be written uniquely as $2^{d-1}c_\calS$ with $c_\calS\in\Z{2^{m-d+1}}$.

Conversely, we consider a weighted monomial $2^{d-1}x^\calS$, $d=|\calS|$. Its $r$-fold difference at $x$ is the coefficient of $t_1\cdots t_r$ after substituting $x\oplus\bigoplus_{j=1}^r t_ju_j$ and reducing $t_j^2=t_j$. Each substituted coordinate is the binary sum of a constant and some of the variables $t_j$. As an integer polynomial, its coefficient of a product of $b\geq1$ distinct variables is divisible by $2^{b-1}$, as follows from
\begin{equation}
t_1\oplus\cdots\oplus t_b
=\frac{1-\prod_{j=1}^b(1-2t_j)}{2}.
\label{eq:supp-affine-parity-expansion}
\end{equation}
Adding the constant $1$ replaces a sum $s$ by $1-s$, which changes the sign of the nonconstant terms and keeps their divisibility. A contribution involving all $r$ variables in a product of $d$ such factors is divisible by $2^{\max(0,r-d)}$. Including $2^{d-1}$ makes it divisible by $2^{r-1}$. In particular, every $(m+1)$-fold difference vanishes modulo $2^m$. Every function in the normal form \eqref{eq:normalform} therefore belongs to $\Ph_m(V)$.

\end{proof}

For integer bit values $x_i\in\{0,1\}$, a parity satisfies
\begin{equation}
 1-2\pi_z(x)=(-1)^{\pair{z}{x}}
 =\prod_{i\in\supp z}(1-2x_i).
 \label{eq:supp-parity-sign}
\end{equation}
Expanding the product and solving for $\pi_z(x)$ gives
\begin{equation}
 \pi_z(x)=\sum_{\varnothing\ne \calS\subseteq\supp z}
 (-2)^{|\calS|-1}x^\calS.
 \tag{\ref{eq:parityexpansion}}
\end{equation}

To verify the substitution rule \eqref{eq:pullback}, we take a linear map $\varphi:V\to W$ and express each coordinate of $\varphi(x)$ as a parity $\pi_z(x)$. In a weighted monomial of degree $d$, we expand these $d$ parity factors using \eqref{eq:parityexpansion}. A product of terms supported on nonempty sets $\calS_1,\ldots,\calS_d$ has a coefficient divisible by
\begin{equation}
2^{d-1}\prod_{j=1}^d2^{|\calS_j|-1}
=2^{\sum_j|\calS_j|-1}.
\end{equation}
After using $x_i^2=x_i$, its support is $\calS_1\cup\cdots\cup \calS_d$. Its coefficient therefore has at least the power of two required by \eqref{eq:normalform}, and terms on more than $m$ coordinates vanish modulo $2^m$. This proves $g\circ\varphi\in\Ph_m(V)$ for $g\in\Ph_m(W)$.

\begin{proof}[Proof of \Cref{thm:logical-clifford}]
For (a), the encoded basis gives
\begin{equation}
\Ug{f}\ket{[x]}=|B_1|^{-1/2}\sum_{b\in B_1}
\exp\!\left(\frac{2\pi\ii f(x\oplus b)}{2^m}\right)\ket{x\oplus b}.
\label{eq:coset-phase-expansion}
\end{equation}
Distinct cosets have disjoint computational supports, and every code state supported on $x\oplus B_1$ is proportional to $\ket{[x]}$. Hence the displayed vector lies in $\Hcode$ exactly when its phases are equal for all $b\in B_1$. The map $a\mapsto\exp(2\pi\ii a/2^m)$ is injective on $\Zm$, and therefore this is precisely constancy of $f$ on that coset, as in Refs.~\cite{Hu2022,WebsterQuintavalleBartlett2023}.
Constancy implies invariance under each $X$ check. Conversely, invariance under all checks implies invariance under their sums, since every intermediate string is still in $Z_1$, and every element of $B_1$ is such a sum.
For (b), the gate acts as the identity on every state of the encoded basis exactly when every phase in \eqref{eq:coset-phase-expansion} equals one. Injectivity of the same exponential map makes this equivalent to $f(x)=0$ for every $x\in Z_1$.
\end{proof}

\subsection{Difference identities and polarization}
\label{app:polarization-details}

The classification at every level uses the polarizations $\delta_d f$ of all orders $d\leq m$. We now collect the required difference identities.
\Cref{lem:pol-sum} is the case $x=0$ of the expansion \eqref{eq:diffexpansion} below.

\begin{lem}[{\cite{Vojtechovsky2004}}]\label{lem:diff}
For $f:\Z{2}^n\to\Zm$, the operators $\Delta_u$ are additive in $f$, commute, and satisfy
\begin{equation}
\Delta_u^2=-2\Delta_u.
\label{eq:diffidentities}
\end{equation}
Their products have the expansion
\begin{equation}
(\Delta_{u_1}\cdots\Delta_{u_d}f)(x)
=\sum_{J\subseteq[d]}(-1)^{d-|J|}
 f\left(x\oplus\bigoplus_{j\in J}u_j\right).
\label{eq:diffexpansion}
\end{equation}
If $f(0)=0$, then
\begin{equation}
f\left(\bigoplus_{a=1}^r v_a\right)
=\sum_{\varnothing\ne J\subseteq[r]}\delta_{|J|}f((v_j)_{j\in J}).
\label{eq:hier-mobius}
\end{equation}
\end{lem}
\begin{proof}
Additivity in $f$ follows directly from the definition. Expanding two differences gives
\begin{equation}
(\Delta_u\Delta_v f)(x)
=f(x\oplus u\oplus v)-f(x\oplus u)-f(x\oplus v)+f(x).
\label{eq:double-difference-expansion}
\end{equation}
This expression is symmetric in $u,v$, proving commutativity. Setting $v=u$ gives $\Delta_u^2=-2\Delta_u$.
For \eqref{eq:diffexpansion}, the case $d=0$ is $f(x)$. Applying one more difference replaces each term by its shifted value minus its original value. The two terms correspond to subsets containing or omitting the new direction, with the stated signs, which proves the formula by induction.
For \eqref{eq:hier-mobius}, we write $T_vf(x)=f(x\oplus v)$, so that $T_v=1+\Delta_v$. Expanding $T_{v_1\oplus\cdots\oplus v_r}=\prod_a(1+\Delta_{v_a})$ and evaluating at zero gives the sum of all polarizations. The empty product contributes $f(0)=0$.
\end{proof}
Applying \eqref{eq:hier-mobius} to the vectors $u_1,\ldots,u_t,\xi_\beta$ and to the vectors $u_1,\ldots,u_t$ and subtracting leaves the terms that contain $\xi_\beta$, which is the expansion \eqref{eq:hier-expansion} of the phase difference used for \Cref{thm:hier-finite}.

For $0\leq r<m$, every element of $2^r\Zm$ is $2^ra$ for a unique $a\in\Z{2^{m-r}}$, and division by $2^r$ returns this $a$.
The following lemma computes the polarization of the highest order from the normal form and proves its multilinearity, which is used in \Cref{sec:hier-logical}.
\begin{lem}\label{lem:top-normalform}
We consider $f\in\Ph_m(\Z{2}^n)$ with coefficients $c_\calS$ in \eqref{eq:normalform}. For $u_1,\ldots,u_m\in\Z{2}^n$,
\begin{equation}
\frac{\delta_m f(u_1,\ldots,u_m)}{2^{m-1}}
=\sum_{\substack{\calS\subseteq[n]\\1\leq|\calS|\leq m}}(c_\calS\bmod2)\times\sum_{h}
\prod_{j=1}^m(u_j)_{h(j)}
\quad\text{in }\Z{2}.
\label{eq:top-normalform}
\end{equation}
Here the inner sum runs over all maps $h$ from $[m]$ onto $\calS$. In particular, $\delta_m f$ takes values in $2^{m-1}\Zm$, and division by $2^{m-1}$ gives a symmetric $m$-linear form.
\end{lem}
\begin{proof}
We introduce binary variables $a_1,\ldots,a_m$. The coefficient of $a_1\cdots a_m$ in
$f(\bigoplus_j a_ju_j)$ is $\delta_m f(u_1,\ldots,u_m)$ by \eqref{eq:hier-pol-sum}. For each physical coordinate $i$, \eqref{eq:parityexpansion} gives
\begin{equation}
\bigoplus_{j=1}^m a_j(u_j)_i
=\sum_{\varnothing\ne A\subseteq[m]}
(-2)^{|A|-1}\prod_{j\in A}a_j(u_j)_i.
\end{equation}
We consider the monomial indexed by $\calS$ in \eqref{eq:normalform}. After reducing $a_j^2=a_j$, a contribution to $a_1\cdots a_m$ selects a nonempty set $A_i$ for every $i\in \calS$, with $\bigcup_{i\in \calS}A_i=[m]$. Its coefficient is divisible by
$2^{\sum_{i\in \calS}|A_i|-1}$. Modulo $2^m$, only selections with $\sum_i|A_i|=m$ can survive. These sets partition $[m]$ and thus correspond to the maps $h$ from $[m]$ onto $\calS$. Dividing by $2^{m-1}$ and reducing modulo two removes their signs and gives \eqref{eq:top-normalform}. Every product in that formula contains exactly one coordinate from each direction $u_j$, and permuting the directions permutes these maps. This proves multilinearity and symmetry.
\end{proof}

The polarization of the highest order is defined on homology.
For $f\in\Lg_m(\Hcode)$ and cycles $u_1,\ldots,u_m$, every binary sum of these cycles is a cycle, and therefore \eqref{eq:hier-pol-sum} gives $\delta_m\lpol f([u_1],\ldots,[u_m])=\delta_m f(u_1,\ldots,u_m)$.
By \Cref{lem:top-normalform} applied to $\lpol f$, division by $2^{m-1}$ gives a symmetric $m$-linear form on $H_1$, which is the multilinear form used in \Cref{sec:hier-logical}.

\section{Proofs of the theorems}
\label{app:proofs}

In this section, we prove the theorems of the main text and the depth bounds of \Cref{lem:depth}.
In \Cref{app:clifford-proof}, we prove \Cref{thm:cliffclass}, the classification at level two.
In \Cref{app:general-proof}, we prove \Cref{thm:hier-classification}, the classification at every level.
In \Cref{app:parity-products}, we prove \Cref{thm:generalkernel}, which gives the generators of the logical identity operators at every level.
In \Cref{app:depth-bounds}, we specify the circuit model and prove \Cref{lem:depth}.
The proof of \Cref{thm:cliffkernel} is given in \Cref{sec:clifford}.

\subsection{Proof of Theorem~\ref{thm:cliffclass}}
\label{app:clifford-proof}

\begin{proof}[Proof of \Cref{thm:cliffclass}]
By \Cref{thm:logical-clifford}(a), restriction to cycles defines the logical phase function $\lpol f([x])=f(x)$. In the logical coordinates of \Cref{app:css-foundations}, it is evaluated as
\begin{equation}
 \lpol f\left(\bigoplus_{a=1}^k \lambda_a[\gamma_a]\right)
 =f\left(\bigoplus_{a=1}^k \lambda_a\gamma_a\right).
 \label{eq:supp-clifford-logical-coordinates}
\end{equation}
The substitution on the right is linear. Therefore, \eqref{eq:pullback} puts this function in $\Ph_2(\Z{2}^k)$ and hence $\lpol f\in\Ph_2(H_1)$. The logical-action map is additive, and \Cref{thm:logical-clifford}(b) identifies its kernel as $\Trv_2(\Hcode)$.

To prove surjectivity, we take any logical phase function $\lpol f\in\Ph_2(H_1)$ with coefficients $\bar{c}_a\in\Z4$ and $\bar{c}_{ab}\in\Z2$ in the logical coordinates. Using the dual representatives $z_a$ from \eqref{eq:logical-coordinates}, we define
\begin{align}
 f_0=\sum_a \delta_1\lpol f([\gamma_a])\,\pi_{z_a} + \sum_{a<b} \delta_2\lpol f([\gamma_a],[\gamma_b])\,\pi_{z_a}\pi_{z_b}.
 \tag{\ref{eq:clifflift}}
\end{align}
Since $\bar{c}_a=\delta_1\lpol f([\gamma_a])$ and $2\bar{c}_{ab}=\delta_2\lpol f([\gamma_a],[\gamma_b])$, this is $f_0=\sum_a \bar{c}_a\pi_{z_a}+2\sum_{a<b}\bar{c}_{ab}\pi_{z_a}\pi_{z_b}$.
It is obtained by substituting the physical parity $\pi_{z_a}$ for each logical coordinate $\lambda_a$, and hence \eqref{eq:pullback} gives $f_0\in\Ph_2(C_1)$.
For $x=b\oplus\bigoplus_a \lambda_a([x])\gamma_a$ as in \eqref{eq:supp-cycle-coordinates}, the relations $\pair{z_a}{\gamma_c}=\delta_{ac}$ and $\pair{z_a}{b}=0$ give $\pi_{z_a}(x)=\lambda_a([x])$, and therefore
\begin{equation}
 f_0(x)=\sum_a \bar{c}_a\lambda_a([x])+2\sum_{a<b}\bar{c}_{ab}\lambda_a([x])\lambda_b([x])=\lpol f([x]).
 \label{eq:supp-clifford-target}
\end{equation}
This value is constant on each boundary coset. \Cref{thm:logical-clifford}(a) then gives $f_0\in\Lg_2(\Hcode)$ and $\lpol{f_0}=\lpol f$.

The map from $\Lg_2(\Hcode)/\Trv_2(\Hcode)$ to $\Ph_2(H_1)$ induced by the logical-action map is therefore an isomorphism.
For fixed representatives $\gamma_a$ and $z_a$, the assignment $\lpol f\mapsto f_0$ is additive.
Every $f\in\Lg_2(\Hcode)$ is therefore uniquely written as $f=g+f_0$ with $g\in\Trv_2(\Hcode)$, which gives the direct-sum decomposition $\Lg_2(\Hcode)\cong\Trv_2(\Hcode)\oplus\Ph_2(H_1)$.
Finally, the uniqueness of the normal form in \Cref{lem:phase} identifies $\Ph_2(H_1)$ with $\Z{4}^{k}\oplus\Z{2}^{\binom{k}{2}}$.
\end{proof}

\subsection{Proof of Theorem~\ref{thm:hier-classification}}
\label{app:general-proof}

\begin{proof}[Proof of \Cref{thm:hier-classification}]
By \Cref{thm:logical-clifford}(a), $\lpol f([x])=f(x)$ is independent of the cycle representative. Its logical-coordinate expression $f(\bigoplus_a \lambda_a\gamma_a)$ belongs to $\Ph_m(\Z{2}^k)$ by \eqref{eq:pullback}, and hence $\lpol f\in\Ph_m(H_1)$.
Since $\lpol f([x])=f(x)$, the map $f\mapsto\lpol f$ is additive, and $\lpol f$ is zero precisely when $f$ vanishes on $Z_1$.
Hence this map is a homomorphism with kernel $\Trv_m(\Hcode)$.

To prove surjectivity, we choose the dual representatives $z_a$ of \Cref{app:css-foundations} and take an arbitrary logical phase function $\lpol f\in\Ph_m(H_1)$.
We write its normal form in the chosen logical coordinates and replace each logical coordinate by its physical parity, which gives
\begin{equation}
\begin{aligned}
 \lpol f\left(\bigoplus_a \lambda_a[\gamma_a]\right)&=\sum_{\substack{J\subseteq[k]\\1\leq|J|\leq m}}
 2^{|J|-1}\bar{c}_J\prod_{a\in J}\lambda_a,\\
 f_0(x)&=\sum_{\substack{J\subseteq[k]\\1\leq|J|\leq m}}
 2^{|J|-1}\bar{c}_J\prod_{a\in J}\pi_{z_a}(x),
\end{aligned}
\label{eq:supp-general-target}
\end{equation}
where $\bar{c}_J\in\Z{2^{m-|J|+1}}$.
The substitution rule \eqref{eq:pullback} gives $f_0\in\Ph_m(C_1)$.
For a cycle $x=b\oplus\bigoplus_a \lambda_a([x])\gamma_a$, we have $[x]=\bigoplus_a \lambda_a([x])[\gamma_a]$.
It follows that $f_0(x)=\lpol f([x])$.
The value is constant on each boundary coset, and hence $f_0$ preserves the code space and implements $\lpol f$.

The map from $\Lg_m(\Hcode)/\Trv_m(\Hcode)$ to $\Ph_m(H_1)$ induced by the logical-action map is therefore an isomorphism.
For fixed representatives $\gamma_a$ and $z_a$, the assignment $\lpol f\mapsto f_0$ is additive.
Every $f\in\Lg_m(\Hcode)$ is therefore uniquely written as $f=g+f_0$ with $g\in\Trv_m(\Hcode)$, which gives the direct-sum decomposition $\Lg_m(\Hcode)\cong\Trv_m(\Hcode)\oplus\Ph_m(H_1)$.
Finally, the uniqueness of the normal form in \Cref{lem:phase} identifies $\Ph_m(H_1)$ with the group in \eqref{eq:hier-classification-structure}, since each subset of $\ell$ logical coordinates carries one coefficient in $\Z{2^{m-\ell+1}}$.
\end{proof}

\subsection{Proof of Theorem~\ref{thm:generalkernel}}
\label{app:parity-products}

\begin{proof}[Proof of \Cref{thm:generalkernel}]
Each generator in \eqref{eq:generalkernel} belongs to $\Ph_m(C_1)$ by the substitution rule \eqref{eq:pullback}.
It vanishes on $Z_1$, hence lies in $\Trv_m(\Hcode)$.
Conversely, we take $f\in\Trv_m(\Hcode)$ and solve the condition $f|_{Z_1}=0$ in the normal form.
As in the proof of \Cref{thm:cliffkernel}, we choose independent rows $\zeta_1,\ldots,\zeta_r$ with $r=\rank H_Z$ and a set $T\subseteq[n]$ of $n-r$ physical qubits such that the parities $\pi_{\zeta_j}$, $j\leq r$, together with the parities $\pi_{e_i}$, $i\in T$, are binary coordinates on $C_1$.
Then $Z_1$ is given by $\pi_{\zeta_1}=\cdots=\pi_{\zeta_r}=0$, and the parities $\pi_{e_i}$, $i\in T$, are binary coordinates on $Z_1$.
By the substitution rule \eqref{eq:pullback} and \Cref{lem:phase}, $f$ has the unique normal form
\begin{equation}
f=\sum_{\substack{A\subseteq[r],\ \calS\subseteq T\\1\leq|A|+|\calS|\leq m}}
2^{|A|+|\calS|-1}a_{A,\calS}\prod_{i\in\calS}\pi_{e_i}\prod_{j\in A}\pi_{\zeta_j}
\label{eq:kernel-normalform}
\end{equation}
with $a_{A,\calS}\in\Z{2^{m-|A|-|\calS|+1}}$.
On $Z_1$, every term with $A\neq\varnothing$ vanishes, and the terms with $A=\varnothing$ form a normal form on $Z_1$,
\begin{equation}
f|_{Z_1}=\sum_{\substack{\calS\subseteq T\\1\leq|\calS|\leq m}}2^{|\calS|-1}a_{\varnothing,\calS}\prod_{i\in\calS}\pi_{e_i}.
\end{equation}
By the uniqueness of the normal form, $f|_{Z_1}=0$ holds exactly when $a_{\varnothing,\calS}=0$ for every $\calS$.
The solution is thus
\begin{equation}
f=\sum_{\substack{A\neq\varnothing\\1\leq|A|+|\calS|\leq m}}
2^{|A|+|\calS|-1}a_{A,\calS}\prod_{i\in\calS}\pi_{e_i}\prod_{j\in A}\pi_{\zeta_j}
\label{eq:kernel-solution}
\end{equation}
with arbitrary coefficients $a_{A,\calS}$, and every term contains the parity of a $Z$ check.

We show that every term of \eqref{eq:kernel-solution} is a combination of the generators \eqref{eq:generalkernel}, and we separate the cases by the number $d=|A|+|\calS|$ of parities in the term.
For $d=1$, we have $A=\{j\}$ and $\calS=\varnothing$, and the term contains only the parity function of a single $Z$ check,
\begin{equation}
a_{\{j\},\varnothing}\,\pi_{\zeta_j}.
\end{equation}
It is a multiple of the generator with $\calS=\varnothing$.
For $d\geq2$, we fix one $j_0\in A$, and the term is the parity function of this $Z$ check multiplied by $d-1$ further parities,
\begin{equation}
2^{d-1}a_{A,\calS}\,\pi_{\zeta_{j_0}}\prod_{i\in\calS}\pi_{e_i}\prod_{j\in A\setminus\{j_0\}}\pi_{\zeta_j}.
\end{equation}
We expand each of the remaining check parities into single bits by \eqref{eq:parityexpansion},
\begin{equation}
2\pi_{\zeta_j}=-\sum_{\varnothing\ne R\subseteq\supp\zeta_j}(-2)^{|R|}\prod_{i\in R}\pi_{e_i}.
\label{eq:hier-anchored}
\end{equation}
Since $\pi_{e_i}^2=\pi_{e_i}$, the term becomes an integer combination of the functions $2^{|R|}\pi_{\zeta_{j_0}}\prod_{i\in R}\pi_{e_i}$ with $R\subseteq[n]$.
They are generators for $|R|\leq m-1$ and vanish modulo $2^m$ for $|R|\geq m$.
Hence the generators \eqref{eq:generalkernel} generate $\Trv_m(\Hcode)$.
The independent checks used here are among the given rows, and therefore including redundant checks leaves the group unchanged.

Finally, the coefficients $a_{A,\calS}$ in \eqref{eq:kernel-solution} are independent.
For $|A|+|\calS|=\ell$, there are $\binom n\ell-\binom{n-r}\ell$ pairs $(A,\calS)$ with $A\neq\varnothing$, each with a coefficient in $\Z{2^{m-\ell+1}}$, which gives \eqref{eq:kernelgroup}.
\end{proof}

\subsection{Proof of Lemma~\ref{lem:depth}}
\label{app:depth-bounds}
\label{sec:scope}

In the circuit model of \Cref{sec:hier-depth}, we allow unrestricted connectivity among the $n$ data qubits and use no ancillas or measurements.
Each layer consists of powers of the diagonal elementary gates \eqref{eq:elementary-gate} on pairwise disjoint supports.
The physical depth $\dep(f)$ is the minimum number of layers that implement $\Ug{f}$ exactly.
We write $\Sigma_f=\{\calS:c_\calS\ne0\}$ for the set of supports of $f$.
We color the supports so that intersecting supports receive different colors and write $\chi(f)$ for the minimum number of colors, with $\chi(f)=0$ for $f=0$.
Circuit scheduling has been described by edge coloring for graph states and $\CZ$ circuits~\cite{MhallaPerdrix2004,Cabello2011,KellerKueng2026} and, in the context of quantum error correction, for logical diagonal gates~\cite{Rowshan2026,SahayLongKhemani2026}.
We give the same coloring argument for phase functions in the normal form \eqref{eq:normalform}, whose elementary gates have supports of sizes up to $m$.

We prove the lemma through the exact value $\dep(f)=\chi(f)$, which holds for every $f\in\Ph_m(C_1)$.
\begin{proof}[Proof of \Cref{lem:depth}]
The unitary $\Ug{f}$ determines its phase function and hence its coefficients uniquely, by \Cref{lem:phase}.
Since elementary gates commute, the coefficient $c_\calS$ of a circuit is the sum of the exponents of its gates with support $\calS$, and therefore every nonzero $c_\calS$ requires a gate with support $\calS$.
We choose one such gate for each support in $\Sigma_f$ and color the support by the layer of this gate.
Intersecting supports occupy different layers, which gives $\chi(f)\leq \dep(f)$.
Conversely, every such coloring schedules the gates $\CPhpow{\calS}{c_\calS}$ in layers of disjoint supports and realizes $\Ug{f}$ exactly.

The supports that contain a given qubit require distinct colors, and therefore their maximum number $\inc(f)$ gives the lower bound.
For the upper bound, we color the supports one at a time.
Each support $\calS$ meets at most $|\calS|(\inc(f)-1)\leq m(\inc(f)-1)$ other supports, and therefore one of $m(\inc(f)-1)+1$ colors is always free.
\end{proof}

At level two, the two-qubit supports form a graph. Applying Vizing's edge-coloring theorem and scheduling the single-qubit gates gives the following bound.
\begin{cor}[{\cite{Vizing1964}}]\label{thm:depth2}
For $f\in\Ph_2(\Z{2}^n)$,
\begin{equation}
\inc(f)\leq \dep(f)\leq\inc(f)+1.
\label{eq:depth2}
\end{equation}
\end{cor}
\begin{proof}

For $m=2$, the supports of size two are the edges of a graph whose maximum degree is at most $\inc(f)$, and Vizing's theorem~\cite{Vizing1964} colors its edges with at most $\inc(f)+1$ colors.
A support of size one at a qubit with $d$ incident edges has $d+1\leq\inc(f)$, and therefore one of the $\inc(f)+1$ colors remains free at that qubit.
This gives $\dep(f)\leq\inc(f)+1$, and the lower bound is the case $m=2$ of \eqref{eq:hier-depth}.
\end{proof}

\section{Details of the examples}
\label{app:examples}

In this section, we give the details of the examples of \Cref{sec:examples}.

\subsection{Transversal and round-robin circuits on the Steane code}
\label{app:examples-steane}

For the examples on the Steane code in \Cref{sec:steane}, we construct the layers that attain depth seven, derive the normal forms \eqref{eq:steane-s-difference} and \eqref{eq:steane-cz-difference}, and describe the sets of gates distinguished by color in \Cref{fig:steane}.

We use the notation of \Cref{sec:steane} and write $|x|$ for the weight of $x$, the number of its nonzero entries.
The seven nonzero vectors of $B_1$, the row space of $H$, have weight four. Therefore, with $\gamma=\mathbf{1}_{7}$, the cycles in the coset $t\gamma\oplus B_1$ have weights $0,4$ for $t=0$ and $3,7$ for $t=1$.

On one block, the circuit \eqref{eq:s-gate-implementation} with $z=\mathbf{1}_{7}$ is $\mathrm S^{\otimes7}\CZ_{[7]}$, where $\CZ_{[7]}=\prod_{1\leq i<j\leq7}\CZ_{ij}$.
Every qubit meets six $\CZ$ gates and one $\mathrm S$ gate, which gives the lower bound seven on the depth of the circuit.
We relabel the qubits by $\Z{7}$. The seven layers
\begin{equation}
 \mathrm S_r\prod_{j=1}^{3}\CZ_{r-j,r+j},\qquad r\in\Z{7},
 \label{eq:steane-one-block-layers}
\end{equation}
consist of gates with disjoint supports and together contain every gate exactly once, because the gate $\CZ_{ij}$ belongs to the unique layer with $2r=i+j$ modulo seven. Hence the depth is seven.
The phase function of $\CZ_{[7]}$ is $2\binom{|x|}{2}$ modulo four. It equals zero for weights $0,4$ and two for weights $3,7$, and hence $\CZ_{[7]}$ acts as $\overline Z$. The Pauli operator $Z^{\otimes7}$ also acts as $\overline Z$, since $|x|=t$ modulo two. Consequently, the circuit \eqref{eq:steane-s-difference-circuit}, $Z^{\otimes7}\CZ_{[7]}$, is a logical identity operator.
Since $\mathrm S^{\otimes7}Z^{\otimes7}=(\mathrm S^\dagger)^{\otimes7}$, the transversal gate $(\mathrm S^\dagger)^{\otimes7}$ implements the same logical gate $\overline{\mathrm S}$ as $\mathrm S^{\otimes7}\CZ_{[7]}$.

We next derive the normal form \eqref{eq:steane-s-difference} of this logical identity operator.
The rows of $H$ restricted to the qubits $1,2,3$ form the identity matrix, and therefore the check parities $\pi_{\zeta_1},\pi_{\zeta_2},\pi_{\zeta_3}$ together with the bits $x_4,x_5,x_6,x_7$ are binary coordinates on $C_1$, with
\begin{equation}
 x_1=\pi_{\zeta_1}\oplus x_4\oplus x_5\oplus x_7,\qquad
 x_2=\pi_{\zeta_2}\oplus x_4\oplus x_6\oplus x_7,\qquad
 x_3=\pi_{\zeta_3}\oplus x_5\oplus x_6\oplus x_7.
 \label{eq:steane-decomposition}
\end{equation}
The phase function of \eqref{eq:steane-s-difference-circuit} is $g=2q$ with $q(x)=\sum_ix_i+\sum_{i<j}x_ix_j$ modulo two.
Substituting \eqref{eq:steane-decomposition} and reducing modulo two gives
\begin{equation}
 q=\sum_{j}\pi_{\zeta_j}+\sum_{j<j'}\pi_{\zeta_j}\pi_{\zeta_{j'}}
 +\pi_{\zeta_1}(x_6+x_7)+\pi_{\zeta_2}(x_5+x_7)+\pi_{\zeta_3}(x_4+x_7),
\end{equation}
and multiplying by two gives \eqref{eq:steane-s-difference}.

On two blocks, the round-robin circuit \eqref{eq:cz-gate-disjoint} places a $\CZ$ gate on all $49$ pairs between the blocks.
Each qubit meets seven gates, which gives the lower bound seven on the depth of the circuit.
We label the qubits of the block $b\in\{1,2\}$ by $(b,i)$ with $i\in\Z{7}$. For $r\in\Z{7}$, the layers
\begin{equation}
 \prod_{i\in\Z{7}}\CZ_{(1,i),(2,i+r)}
 \label{eq:steane-two-block-layers}
\end{equation}
consist of gates with disjoint supports and together contain every inter-block pair exactly once. Hence the depth is seven.

The normal form \eqref{eq:steane-cz-difference} follows in the same way.
It is the normal form of the $42$ $\CZ$ gates of the round-robin circuit that do not belong to the transversal circuit of \Cref{sec:steane}.
We write a string on the two blocks as $(x,y)$ and use \eqref{eq:steane-decomposition} on both blocks, with the check parities $\pi_{\zeta'_j}$ on the second block.
These $42$ gates have the phase function $g=2q$ with $q(x,y)=\sum_{i\neq j}x_iy_j$ modulo two, and the substitution gives
\begin{multline}
 q=\sum_{j\neq j'}\pi_{\zeta_j}\pi_{\zeta'_{j'}}
 +\pi_{\zeta'_1}(x_6+x_7)+\pi_{\zeta'_2}(x_5+x_7)+\pi_{\zeta'_3}(x_4+x_7)\\
 +\pi_{\zeta_1}(y_6+y_7)+\pi_{\zeta_2}(y_5+y_7)+\pi_{\zeta_3}(y_4+y_7).
\end{multline}
Multiplying by two gives \eqref{eq:steane-cz-difference}, in which $y_4,\ldots,y_7$ are written as $\pi_{e_{11}},\ldots,\pi_{e_{14}}$.

The colors in \Cref{fig:steane} group the gates of these two logical identity operators as follows.
In panel (b), each of the seven sets consists of the gate $Z_i$ and the three gates $\CZ_{ij}$ with $j\in\supp\zeta\setminus\{i\}$, where $\zeta$ is a nonzero vector of $B_1$ whose support contains the qubit $i$.
It is the product $Z(\zeta)$ of $Z$ checks controlled by the qubit $i$, with $\CZ_{ii}$ replaced by $Z_i$ as in \Cref{sec:physical-implementation}.
In panel (d), nine sets consist of four gates and are operators $Z(\zeta)$ on one block controlled by a qubit of the other block.
The remaining set consists of the six gates $\CZ_{2,7+j}$ and $\CZ_{j,9}$ with $j\in\{3,4,5\}$, in the numbering of \Cref{sec:steane}.
It is the product of the operator $Z(\zeta'_2\oplus\zeta'_3)$ controlled by the qubit $2$ and the operator $Z(\zeta_2\oplus\zeta_3)$ controlled by the qubit $9$, in which the two gates $\CZ_{2,9}$ cancel.
Each set is therefore a logical identity operator.

\subsection{Minimum depth of logical \texorpdfstring{$\CZ$}{CZ} on the toric code}
\label{app:toric}

We prove \Cref{pro:toric}.
We first fix the coordinates and give a condition for code preservation for circuits that apply $\CZ$ gates to a pairing of the edges, with $Z$ corrections.
We then prove the odd-$L$ construction, and finally we establish the even-$L$ construction and show that depth one is impossible.

\newcommand{\unlistedsubsubsection}[1]{\let\savedaddcontentsline\addcontentsline\renewcommand{\addcontentsline}[3]{}\subsubsection*{#1}\let\addcontentsline\savedaddcontentsline}
\unlistedsubsubsection{Coordinates and circuits on a pairing of the edges}

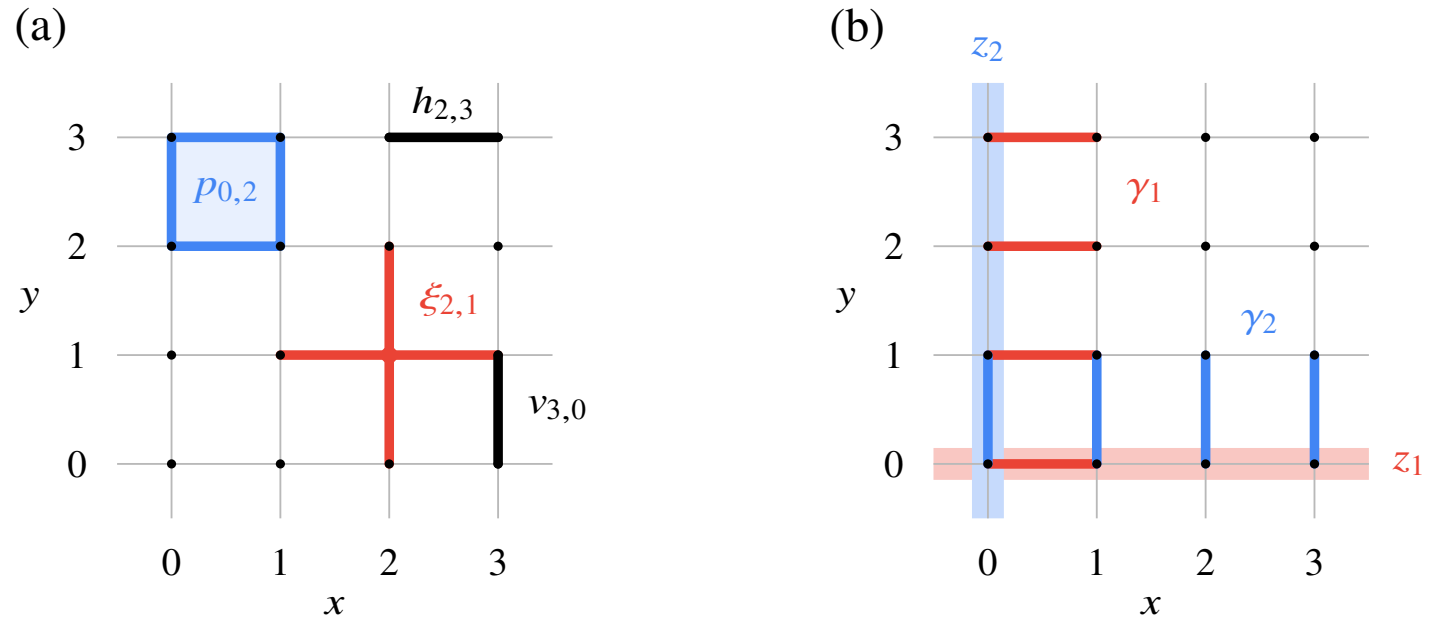
\begin{figure*}[h]
\centering
\begin{tikzpicture}[scale=0.72,every node/.style={font=\scriptsize},
 grid/.style={draw=gray!55,line width=0.35pt},
 hl/.style={line width=1.8pt,line cap=round,line join=round},
 band/.style={line width=6pt,line cap=butt},
 tag/.style={fill=white,inner sep=1pt}]

\path[use as bounding box] (-1.75,-1.7) rectangle (12.25,4.4);

\begin{scope}
\node[font=\small] at (-1.2,4.0) {(a)};
\fill[Gblue!12] (0,2) rectangle (1,3);

\foreach \x in {0,...,3}{
  \draw[grid] (-0.5,\x) -- (3.5,\x);
  \draw[grid] (\x,-0.5) -- (\x,3.5);}
\draw[hl,Gblue] (0,2) -- (1,2) -- (1,3) -- (0,3) -- cycle;
\draw[hl,Gred] (1,1) -- (2,1) -- (3,1);
\draw[hl,Gred] (2,0) -- (2,1) -- (2,2);
\draw[hl,black] (2,3) -- (3,3);
\draw[hl,black] (3,0) -- (3,1);
\foreach \x in {0,...,3}{\foreach \y in {0,...,3}{
  \fill[black] (\x,\y) circle (1.2pt);}}
\fill[Gred] (2,1) circle (2.2pt);
\node[Gblue] at (0.5,2.5) {$p_{0,2}$};
\node[tag,text=Gred] at (2.55,1.5) {$\xi_{2,1}$};
\node[tag] at (2.5,3.3) {$h_{2,3}$};
\node[tag] at (3.55,0.5) {$v_{3,0}$};

\foreach \x in {0,...,3}{\node[anchor=north] at (\x,-0.62) {$\x$};}
\foreach \y in {0,...,3}{\node[anchor=east] at (-0.62,\y) {$\y$};}
\node at (1.5,-1.3) {$x$};
\node at (-1.3,1.5) {$y$};
\end{scope}

\begin{scope}[xshift=7.5cm]
\node[font=\small] at (-1.2,4.0) {(b)};
\draw[band,Gred!30] (-0.5,0) -- (3.5,0);
\draw[band,Gblue!30] (0,-0.5) -- (0,3.5);

\foreach \x in {0,...,3}{
  \draw[grid] (-0.5,\x) -- (3.5,\x);
  \draw[grid] (\x,-0.5) -- (\x,3.5);}
\foreach \y in {0,...,3}{\draw[hl,Gred] (0,\y) -- (1,\y);}
\foreach \x in {0,...,3}{\draw[hl,Gblue] (\x,0) -- (\x,1);}
\foreach \x in {0,...,3}{\foreach \y in {0,...,3}{
  \fill[black] (\x,\y) circle (1.2pt);}}
\node[tag,text=Gred] at (1.45,2.5) {$\gamma_1$};
\node[tag,text=Gblue] at (2.5,1.3) {$\gamma_2$};
\node[text=Gred,anchor=west] at (3.55,0) {$z_1$};
\node[text=Gblue,anchor=south] at (0,3.55) {$z_2$};

\foreach \x in {0,...,3}{\node[anchor=north] at (\x,-0.62) {$\x$};}
\foreach \y in {0,...,3}{\node[anchor=east] at (-0.62,\y) {$\y$};}
\node at (1.5,-1.3) {$x$};
\node at (-1.3,1.5) {$y$};
\end{scope}
\end{tikzpicture}
\caption{Coordinates and logical representatives of \Cref{app:toric}, drawn for $L=4$.
The lattice has periodic boundary conditions.
(a) The edges $h_{x,y}$ and $v_{x,y}$, the star $\xi_{2,1}$ (red), and the plaquette $p_{0,2}$ (blue).
(b) The logical representatives $\gamma_1$ (red lines) and $\gamma_2$ (blue lines) and the dual representatives $z_1$ (red band) and $z_2$ (blue band). Each $\gamma_a$ shares a single edge with $z_a$.}
\label{fig:toric-coordinates}
\end{figure*}

We place the qubits of the $L\times L$ toric code with $L\geq3$ on the horizontal edges $h_{x,y}$ from $(x,y)$ to $(x+1,y)$ and the vertical edges $v_{x,y}$ from $(x,y)$ to $(x,y+1)$, with indices modulo $L$.
We identify each qubit with its edge and write $x_e\in\F$ for the component of a bit string on the edge $e$.
The $X$ checks are the stars $\xi_{x,y}=\{h_{x,y},h_{x-1,y},v_{x,y},v_{x,y-1}\}$, and the $Z$ checks are the plaquettes $p_{X,Y}=\{h_{X,Y},h_{X,Y+1},v_{X,Y},v_{X+1,Y}\}$.
We fix the logical representatives $\gamma_1=\{h_{0,y}\}_{y}$ and $\gamma_2=\{v_{x,0}\}_{x}$ and the dual representatives $z_1=\{h_{x,0}\}_{x}$ and $z_2=\{v_{0,y}\}_{y}$ (\Cref{fig:toric-coordinates}).
Since $\pair{z_a}{\gamma_b}=\delta_{ab}$, the logical gate $\overline{\CZ}_{12}$ has the logical phase function $2\lambda_1\lambda_2$.

We call two edges of a star, or of a plaquette, opposite if they have the same orientation.
Two edges of a star lie in a common plaquette exactly when they have different orientations, and the same holds with stars and plaquettes exchanged.

We fix an involution $\tau$ of the edge set without fixed points, which divides the edges into the pairs $\{e,\tau(e)\}$.
For a set $u$ of edges, we write $\tau(u)=\{\tau(e):e\in u\}$ and $N_\tau(u)$ for the number of pairs $\{e,\tau(e)\}$ contained in $u$.

We denote by $\eta\in\F^n$ the set of edges carrying $Z$ gates and define the phase function $f=f_{\tau,\eta}$ by
\begin{equation}
 f_{\tau,\eta}(x)=2\sum_{\{e,\tau(e)\}}x_ex_{\tau(e)}+2\pair{\eta}{x}\pmod4,
\label{eq:toric-matching-phase}
\end{equation}
where the sum runs over the pairs of $\tau$.
The gate $\Ug[2]{f}$ applies $\CZ$ to every pair of $\tau$ and $Z$ to every edge of $\eta$.
It has depth one if $\eta=\varnothing$ and depth two otherwise, since a $\CZ$ gate and a $Z$ gate on the same qubit occupy different layers.
We write $M$ for the symmetric binary matrix in the quadratic normal form \eqref{eq:quadraticnormal} of $f$.
\begin{lem}\label{lem:toric-matching}
With $\tau$, $\eta$, and $f=f_{\tau,\eta}$ as above, the following hold.
\begin{enumerate}
\item[(a)] $Mx=\tau(x)$ and $f(x)=2\bigl(N_\tau(x)+\pair{\eta}{x}\bigr)$ for every $x\in\F^n$.
\item[(b)] $\Ug[2]{f}$ preserves the code space if and only if $\tau(\xi)\in B^1$ and $N_\tau(\xi)+\pair{\eta}{\xi}$ is even for every star $\xi$.
\end{enumerate}
\end{lem}
\begin{proof}
(a) The $\CZ$ pairs give the off-diagonal entries of $M$, and the single-qubit exponents $c_e=2\eta_e$ are even. Hence $M$ sends each edge $e$ to $\tau(e)$, that is, $Mx=\tau(x)$.
In \eqref{eq:toric-matching-phase}, a pair contributes to the sum exactly when both of its edges lie in $x$, which gives $2N_\tau(x)$.
(b) The stars span $B_1$. Therefore, by \Cref{thm:cliffcriterion} the gate $\Ug[2]{f}$ preserves the code space exactly when $M\xi\in B^1$ and $f(\xi)=0$ in $\Z4$ for every star $\xi$, and (a) turns these into the stated conditions.
\end{proof}

\unlistedsubsubsection{Proof of the odd-$L$ construction}

For odd $L$, we use the pairing $\tau$ in \eqref{eq:toric-odd-pairing-main} and write $c=(L-1)/2$, so that $2c+1=0$ modulo $L$.
\Cref{fig:toric}(a) shows the pairing for $L=5$.

\begin{proof}[Proof of \Cref{pro:toric}(a)]
The pairing $\tau$ is an involution (\Cref{sec:toric}) without fixed points, since it exchanges the orientations.
The image of a star is
\begin{equation}
 \tau(\xi_{x,y})=\{v_{-x-1,y+c},v_{-x,y+c},h_{-x-1,y-c},h_{-x-1,y-c-1}\}=p_{-x-1,y+c},
\end{equation}
because $2c+1=0$ modulo $L$ gives $y-c-1=y+c$ modulo $L$.

A pair inside $\xi_{x,y}$ consists of a horizontal edge $h_{x-i,y}$ and a vertical edge $v_{x,y-j}$ with $i,j\in\{0,1\}$, and $\tau(h_{x-i,y})=v_{-x+i-1,y+c}$ equals $v_{x,y-j}$ only for $c=-j$ modulo $L$.
Since $1\leq c\leq L-2$, we have $N_\tau(\xi)=0$ for every star.
By \Cref{lem:toric-matching}(b) with $\eta=\varnothing$, the circuit preserves the code space.

Each pair contains one edge of each orientation, while $\gamma_1$ consists of horizontal edges and $\gamma_2$ of vertical edges, and hence $N_\tau(\gamma_a)=0$ and $\bar{c}_a=f(\gamma_a)=0$.
The image $\tau(v_{x,0})=h_{-x-1,-c}$ lies in $\gamma_1$ exactly when $x=-1$ modulo $L$. Hence $\bar c_{12}=\pair{\gamma_1}{M\gamma_2}=1$ by \eqref{eq:logicalpolar}, and the circuit implements $\overline{\CZ}_{12}$ with depth one.
\end{proof}

\unlistedsubsubsection{Proof of the even-$L$ case}

In the rest of this subsection, $L\geq4$ is even.

\begin{lem}\label{lem:toric-weight4}
A nonzero element of $B^1$ of weight at most four is a single plaquette.
\end{lem}
\begin{proof}
We write the element as $C=\sum_{x,y}a_{x,y}p_{x,y}$ with $a_{x,y}\in\F$.
Each edge lies in exactly two plaquettes, and therefore an edge belongs to $C$ exactly when the coefficients of these two plaquettes differ.
The horizontal edge $h_{x,y+1}$ lies in $p_{x,y}$ and $p_{x,y+1}$, and the vertical edge $v_{x+1,y}$ lies in $p_{x,y}$ and $p_{x+1,y}$.

For a fixed $x$, the coefficient $a_{x,y}$ changes an even number of times as $y$ runs around the torus, and each change is a horizontal edge of $C$.
Every $x$ for which $a_{x,y}$ depends on $y$ therefore contributes at least two horizontal edges to $C$, and likewise every $y$ for which $a_{x,y}$ depends on $x$ contributes at least two vertical edges.
If $a_{x,y}$ were independent of $y$ for every $x$, it would depend on $x$ for each of the $L$ values of $y$ because $C\neq0$, and $C$ would contain at least $2L>4$ vertical edges.
The same holds with $x$ and $y$ exchanged, and hence $|C|\leq4$ leaves exactly one such $x_1$ and one such $y_1$.

All coefficients other than $a_{x_1,y_1}$ are then equal, and we may take them to be zero because the sum of all plaquettes is zero.
Hence $C=p_{x_1,y_1}$.
\end{proof}

\begin{lem}\label{lem:toric-allornothing}
We consider a code-preserving circuit $\Ug[2]{f}$ of depth one at level two.
Then either $M=0$, or $M$ is the permutation matrix of an involution $\tau$ of the edge set, possibly with fixed points, that maps every star to a plaquette.
\end{lem}
\begin{proof}
A circuit of depth one consists of $\CZ$ gates on disjoint pairs of edges, together with phases $\mathrm{S}^{c_e}$ on the remaining edges.
Hence $M$ sends an edge of a pair to its partner, a remaining edge with odd $c_e$ to itself, and every other edge to zero.
We write $W$ for the set of edges with nonzero image, on which $M$ is injective.

For a star $\xi$, the vector $M\xi$ therefore has weight $|\xi\cap W|\leq4$.
Condition (L2) of \Cref{thm:cliffcriterion} and \Cref{lem:toric-weight4} give $|\xi\cap W|\in\{0,4\}$, and therefore every star is contained in $W$ or disjoint from $W$.
Since adjacent stars share an edge and the lattice is connected, $W$ is empty or the whole edge set.

In the first case, $M=0$.
In the second case, $M$ is the permutation matrix of an involution $\tau$, and $M\xi=\tau(\xi)$ is an element of $B^1$ of weight four, hence a plaquette by \Cref{lem:toric-weight4}.
\end{proof}

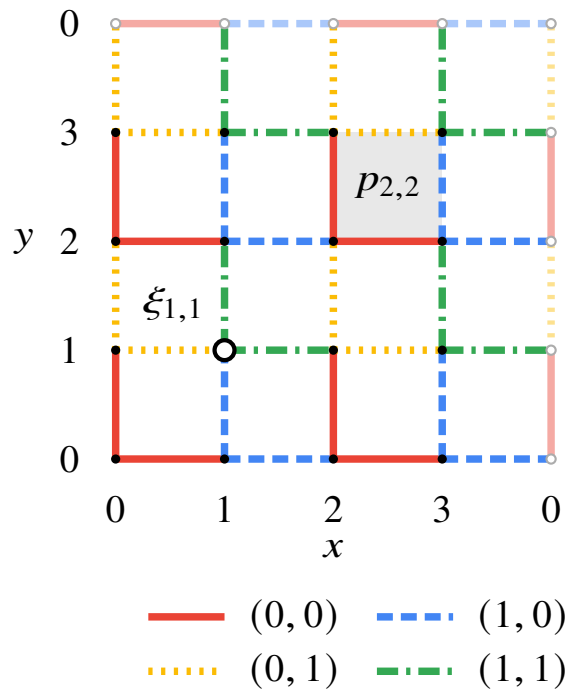
\begin{figure}[h]
\centering
\begin{tikzpicture}[scale=0.72,every node/.style={font=\scriptsize},
 grid/.style={draw=gray!55,line width=0.35pt},
 ghost/.style={draw=gray!55,line width=0.35pt,dash pattern=on 2pt off 1.5pt},
 hl/.style={line width=1.8pt,line cap=round,line join=round},
 copy/.style={dash pattern=on 3pt off 2pt},
 tag/.style={fill=white,inner sep=1pt}]
\path[use as bounding box] (-1.25,-2.3) rectangle (4.6,4.5);
\begin{scope}
\fill[gray!18] (2,2) rectangle (3,3);
\draw[line width=1.4pt,Gred] (0,0) -- (1,0);
\draw[line width=1.4pt,Gred] (0,0) -- (0,1);
\draw[line width=1.4pt,Gred,opacity=0.45] (0,4) -- (1,4);
\draw[line width=1.4pt,Gred,opacity=0.45] (4,0) -- (4,1);
\draw[line width=1.4pt,Gyellow,dash pattern=on 0.8pt off 1.6pt] (0,1) -- (1,1);
\draw[line width=1.4pt,Gyellow,dash pattern=on 0.8pt off 1.6pt] (0,1) -- (0,2);
\draw[line width=1.4pt,Gyellow,dash pattern=on 0.8pt off 1.6pt,opacity=0.45] (4,1) -- (4,2);
\draw[line width=1.4pt,Gred] (0,2) -- (1,2);
\draw[line width=1.4pt,Gred] (0,2) -- (0,3);
\draw[line width=1.4pt,Gred,opacity=0.45] (4,2) -- (4,3);
\draw[line width=1.4pt,Gyellow,dash pattern=on 0.8pt off 1.6pt] (0,3) -- (1,3);
\draw[line width=1.4pt,Gyellow,dash pattern=on 0.8pt off 1.6pt] (0,3) -- (0,4);
\draw[line width=1.4pt,Gyellow,dash pattern=on 0.8pt off 1.6pt,opacity=0.45] (4,3) -- (4,4);
\draw[line width=1.4pt,Gblue,dash pattern=on 3pt off 1.5pt] (1,0) -- (2,0);
\draw[line width=1.4pt,Gblue,dash pattern=on 3pt off 1.5pt] (1,0) -- (1,1);
\draw[line width=1.4pt,Gblue,dash pattern=on 3pt off 1.5pt,opacity=0.45] (1,4) -- (2,4);
\draw[line width=1.4pt,Ggreen,dash pattern=on 4pt off 1.5pt on 0.8pt off 1.5pt] (1,1) -- (2,1);
\draw[line width=1.4pt,Ggreen,dash pattern=on 4pt off 1.5pt on 0.8pt off 1.5pt] (1,1) -- (1,2);
\draw[line width=1.4pt,Gblue,dash pattern=on 3pt off 1.5pt] (1,2) -- (2,2);
\draw[line width=1.4pt,Gblue,dash pattern=on 3pt off 1.5pt] (1,2) -- (1,3);
\draw[line width=1.4pt,Ggreen,dash pattern=on 4pt off 1.5pt on 0.8pt off 1.5pt] (1,3) -- (2,3);
\draw[line width=1.4pt,Ggreen,dash pattern=on 4pt off 1.5pt on 0.8pt off 1.5pt] (1,3) -- (1,4);
\draw[line width=1.4pt,Gred] (2,0) -- (3,0);
\draw[line width=1.4pt,Gred] (2,0) -- (2,1);
\draw[line width=1.4pt,Gred,opacity=0.45] (2,4) -- (3,4);
\draw[line width=1.4pt,Gyellow,dash pattern=on 0.8pt off 1.6pt] (2,1) -- (3,1);
\draw[line width=1.4pt,Gyellow,dash pattern=on 0.8pt off 1.6pt] (2,1) -- (2,2);
\draw[line width=1.4pt,Gred] (2,2) -- (3,2);
\draw[line width=1.4pt,Gred] (2,2) -- (2,3);
\draw[line width=1.4pt,Gyellow,dash pattern=on 0.8pt off 1.6pt] (2,3) -- (3,3);
\draw[line width=1.4pt,Gyellow,dash pattern=on 0.8pt off 1.6pt] (2,3) -- (2,4);
\draw[line width=1.4pt,Gblue,dash pattern=on 3pt off 1.5pt] (3,0) -- (4,0);
\draw[line width=1.4pt,Gblue,dash pattern=on 3pt off 1.5pt] (3,0) -- (3,1);
\draw[line width=1.4pt,Gblue,dash pattern=on 3pt off 1.5pt,opacity=0.45] (3,4) -- (4,4);
\draw[line width=1.4pt,Ggreen,dash pattern=on 4pt off 1.5pt on 0.8pt off 1.5pt] (3,1) -- (4,1);
\draw[line width=1.4pt,Ggreen,dash pattern=on 4pt off 1.5pt on 0.8pt off 1.5pt] (3,1) -- (3,2);
\draw[line width=1.4pt,Gblue,dash pattern=on 3pt off 1.5pt] (3,2) -- (4,2);
\draw[line width=1.4pt,Gblue,dash pattern=on 3pt off 1.5pt] (3,2) -- (3,3);
\draw[line width=1.4pt,Ggreen,dash pattern=on 4pt off 1.5pt on 0.8pt off 1.5pt] (3,3) -- (4,3);
\draw[line width=1.4pt,Ggreen,dash pattern=on 4pt off 1.5pt on 0.8pt off 1.5pt] (3,3) -- (3,4);
\foreach \x in {0,...,3}{\foreach \y in {0,...,3}{
  \fill[black] (\x,\y) circle (1.2pt);}}
\foreach \x in {0,...,4}{\draw[gray!65,fill=white,line width=0.4pt] (\x,4) circle (1.2pt);}
\foreach \y in {0,...,3}{\draw[gray!65,fill=white,line width=0.4pt] (4,\y) circle (1.2pt);}
\draw[black,fill=white,line width=0.8pt] (1,1) circle (2.6pt);
\node at (2.5,2.5) {$p_{2,2}$};
\node[tag] at (0.52,1.42) {$\xi_{1,1}$};
\foreach \x in {0,...,3}{\node[anchor=north] at (\x,-0.18) {$\x$};}
\foreach \y in {0,...,3}{\node[anchor=east] at (-0.18,\y) {$\y$};}
\node[anchor=north] at (4,-0.18) {$0$};
\node[anchor=east] at (-0.18,4) {$0$};
\node at (2.0,-0.82) {$x$};
\node at (-0.84,2.0) {$y$};
\draw[line width=1.4pt,Gred] (0.30,-1.45) -- (1.00,-1.45);
\node[anchor=west] at (1.05,-1.45) {$(0,0)$};
\draw[line width=1.4pt,Gblue,dash pattern=on 3pt off 1.5pt] (2.40,-1.45) -- (3.10,-1.45);
\node[anchor=west] at (3.15,-1.45) {$(1,0)$};
\draw[line width=1.4pt,Gyellow,dash pattern=on 0.8pt off 1.6pt] (0.30,-1.95) -- (1.00,-1.95);
\node[anchor=west] at (1.05,-1.95) {$(0,1)$};
\draw[line width=1.4pt,Ggreen,dash pattern=on 4pt off 1.5pt on 0.8pt off 1.5pt] (2.40,-1.95) -- (3.10,-1.95);
\node[anchor=west] at (3.15,-1.95) {$(1,1)$};
\end{scope}
\end{tikzpicture}
\caption{The parity vector $\omega$ in the proof of \Cref{lem:toric-shape}, drawn for $L=4$, with the colors of its four values listed below the lattice.
The plaquette $p_{2,2}$ (shaded) and the star $\xi_{1,1}$ (circle) each contain exactly two edges with the same value.
Gray vertices and faint edges mark periodic images.}
\label{fig:toric-even}
\end{figure}

\begin{lem}\label{lem:toric-shape}
We consider an involution $\tau$ of the edge set that maps every star to a plaquette.
Then either $\tau$ preserves the orientation of every edge, or there are $a,b\in\Z L$ with
\begin{equation}
 \tau(h_{x,y})=v_{a-x,b-y},\qquad \tau(v_{x,y})=h_{a-x,b-y}
\label{eq:toric-point-reflection}
\end{equation}
for all $x,y$.
\end{lem}
\begin{proof}
Since $\tau$ is a bijection, it maps the $L^2$ stars onto the $L^2$ plaquettes, and $\tau^2=1$ maps every plaquette to a star.
Two edges are opposite in a star exactly when they lie in a common star and in no common plaquette, and the same holds with stars and plaquettes exchanged.
Hence $\tau$ maps opposite edges of a star to opposite edges of a plaquette, and opposite edges of a plaquette to opposite edges of a star.

We assume that some edge changes its orientation. We choose a horizontal edge $h_{x,y}$ whose image is vertical, and write $\tau(h_{x,y})=v_{X,Y}$.
The edges $h_{x,y}$ and $h_{x+1,y}$ are opposite in a star, and $h_{x,y}$ and $h_{x,y+1}$ are opposite in a plaquette, and hence
\begin{equation}
 \tau(h_{x+1,y})\in\{v_{X-1,Y},v_{X+1,Y}\},\qquad
 \tau(h_{x,y+1})\in\{v_{X,Y-1},v_{X,Y+1}\}.
\end{equation}
Repeating this step shows that $\tau$ maps every horizontal edge to a vertical edge. Since $\tau$ is a bijection and both orientations contain $L^2$ edges, $\tau$ exchanges the two orientations.

Since $L$ is even, the parity vector $\omega(h_{x,y})=\omega(v_{x,y})=(x\bmod2,\,y\bmod2)\in\F^2$ is well defined (\Cref{fig:toric-even}).
By the displayed relations, $\omega(\tau(e))+\omega(e)$ takes the same value on $h_{x,y}$, $h_{x+1,y}$, and $h_{x,y+1}$.
It is therefore the same for all horizontal edges $e$, and also for all vertical edges because $\tau$ is an involution.
Hence $\tau$ maps two edges with equal $\omega$ to two edges with equal $\omega$.

We define $\sigma$ on vertices by $\tau(\xi_{x,y})=p_{\sigma(x,y)}$.
The four edges of $p_{X,Y}$ have the values $(X,Y)$, $(X,Y+1)$, $(X,Y)$, and $(X+1,Y)$ modulo two. Therefore, the only two edges with equal $\omega$ in a plaquette are $h_{X,Y}$ and $v_{X,Y}$.
The edges $h_{x,y}$ and $v_{x,y}$ of $\xi_{x,y}$ have equal $\omega$ and exchange their orientations, and hence
\begin{equation}
 \tau(h_{x,y})=v_{\sigma(x,y)},\qquad \tau(v_{x,y})=h_{\sigma(x,y)}.
\label{eq:toric-sigma}
\end{equation}

The two horizontal edges $h_{x,y}$ and $h_{x-1,y}$ of $\xi_{x,y}$ are mapped to the two vertical edges of $p_{\sigma(x,y)}$. Hence $\tau(h_{x-1,y})=v_{\sigma(x,y)+(1,0)}$, and comparing with \eqref{eq:toric-sigma} at $(x-1,y)$ gives $\sigma(x-1,y)=\sigma(x,y)+(1,0)$.
In the same way, the vertical edges $v_{x,y}$ and $v_{x,y-1}$ of $\xi_{x,y}$ give $\sigma(x,y-1)=\sigma(x,y)+(0,1)$.
Hence $\sigma(x,y)=\sigma(0,0)-(x,y)$, and \eqref{eq:toric-point-reflection} holds with $(a,b)=\sigma(0,0)$.
\end{proof}

\begin{lem}\label{lem:toric-defects}
We consider the point reflection $\tau$ in \eqref{eq:toric-point-reflection}.
Then $\tau(\xi_{x,y})=p_{a-x,b-y}$ for every star, exactly four stars contain a pair of $\tau$, and each of them contains exactly one pair.
For $a=b=0$, these are the stars $\xi_{x,y}$ with $x,y\in\{0,L/2\}$, and the pair inside $\xi_{x,y}$ is $\{h_{x,y},v_{x,y}\}$.
\end{lem}
\begin{proof}
Substitution gives $\tau(\xi_{x,y})=\{v_{a-x,b-y},v_{a-x+1,b-y},h_{a-x,b-y},h_{a-x,b-y+1}\}=p_{a-x,b-y}$.

A pair inside $\xi_{x,y}$ consists of $h_{x-i,y}$ and $v_{x,y-j}$ with $i,j\in\{0,1\}$, and $\tau(h_{x-i,y})=v_{a-x+i,b-y}$ equals $v_{x,y-j}$ exactly when $2x= a+i$ and $2y= b+j$ modulo $L$.
Since $L$ is even, exactly one $i$ makes $a+i$ even, and then $2x=a+i$ modulo $L$ has exactly two solutions $x$, which differ by $L/2$.
The same holds for $j$ and $y$, which gives four stars with one pair each.
For $a=b=0$, the solutions are $i=j=0$ and $x,y\in\{0,L/2\}$.
\end{proof}

\begin{proof}[Proof of \Cref{pro:toric}(b)]
We first prove the lower bound.
For a code-preserving circuit $\Ug[2]{f}$ of depth one at level two, the exponent of the logical $\CZ$ gate is $\bar c_{12}=\pair{\gamma_1}{M\gamma_2}$ modulo two by \eqref{eq:logicalpolar}, which vanishes for $M=0$.
For $M\neq0$, \Cref{lem:toric-allornothing,lem:toric-shape} show that $M$ is the permutation matrix of an involution $\tau$ that either preserves every orientation or is a point reflection.

A point reflection has no fixed points, and therefore the $\CZ$ gates occupy every qubit and $f=f_{\tau,\varnothing}$.
\Cref{lem:toric-defects} then gives a star $\xi$ with $N_\tau(\xi)=1$, and hence $f(\xi)=2$ by \Cref{lem:toric-matching}(a), which violates condition (L1).
Therefore, $\tau$ preserves every orientation, $\tau(\gamma_2)$ consists of vertical edges, and $\pair{\gamma_1}{\tau(\gamma_2)}=0$.
Every code-preserving circuit of depth one thus has $\bar c_{12}=0$, and its logical gate is a product of logical $\mathrm{S}$ and $Z$ gates on the individual logical qubits.

For the upper bound, we use the point reflection $\tau$ in \eqref{eq:toric-even-pairing-main}, which is \eqref{eq:toric-point-reflection} with $a=b=0$ (\Cref{fig:toric}(b)). The set $\eta$ consists of the $L$ edges $h_{x,0}$ and $h_{x,L/2}$ with $0\leq x<L/2$, and we set $f=f_{\tau,\eta}$.
By \Cref{lem:toric-defects}, $\tau$ maps every star to a plaquette, and $N_\tau(\xi)=1$ exactly at the four stars named there.
The set $\eta$ is the union of two horizontal paths, from $(0,0)$ to $(L/2,0)$ and from $(0,L/2)$ to $(L/2,L/2)$, and hence $\pair{\eta}{\xi}$ is odd exactly at the stars of these four endpoints.
Hence $N_\tau(\xi)+\pair{\eta}{\xi}$ is even for every star, and $\Ug[2]{f}$ preserves the code space by \Cref{lem:toric-matching}(b).

Since $\tau$ exchanges the orientations, we have $N_\tau(\gamma_1)=N_\tau(\gamma_2)=0$, and the intersections $\gamma_1\cap\eta=\{h_{0,0},h_{0,L/2}\}$ and $\gamma_2\cap\eta=\varnothing$ have even size.
Hence $\bar{c}_a=f(\gamma_a)=0$ by \Cref{lem:toric-matching}(a).
The image $\tau(v_{x,0})=h_{-x,0}$ lies in $\gamma_1$ exactly when $x=0$. Hence $\bar c_{12}=1$.
Since $\eta\neq\varnothing$, the circuit has depth two, and it implements $\overline{\CZ}_{12}$.
\end{proof}

\section{Notation}\label{app:notation}

\begin{table}[!ht]
\caption{Notation used in the manuscript.}\label{tab:notation}
\small
\begin{tabularx}{\linewidth}{@{}l@{\hspace{1.3em}}>{\raggedright\arraybackslash}X@{}}
\toprule
Symbol & Meaning\\
\midrule
\multicolumn{2}{@{}l}{\textbf{CSS codes and logical coordinates}}\\
$\Z2,\ \Zm$ & Binary field and the ring $\mathbb Z/2^m\mathbb Z$ of phase values at level $m$.\\
$n,\ k,\ m$ & Numbers of physical and logical qubits, and level of the Clifford hierarchy.\\
$[n],\ e_i,\ \mathbf{0}_n,\ \mathbf{1}_n$ & Qubit indices $1,\ldots,n$, standard basis vector, all-zero and all-one strings.\\
$\oplus,\ \pair uv,\ W^\perp$ & Binary sum, binary inner product, orthogonal complement.\\
$\supp x,\ |x|$ & Support and weight.\\
$H_X,\ H_Z,\ r_X,\ r_Z$ & Check matrices and their numbers of rows.\\
$\xi_\beta,\ \zeta_\alpha$ & Rows of $H_X$ and $H_Z$, which give the checks $X(\xi_\beta)$ and $Z(\zeta_\alpha)$.\\
$C_2,\ C_1,\ C_0,\ C^i$ & Chain groups $\Z2^{r_X},\Z2^n,\Z2^{r_Z}$, and the dual $C^i$ of $C_i$.\\
$\partial_2,\ \partial_1$ & Boundary maps $H_X^\T:C_2\to C_1$ and $H_Z:C_1\to C_0$.\\
$Z_1,\ B_1,\ H_1$ & Cycles $\ker H_Z$, boundaries $\im H_X^\T$, and homology $Z_1/B_1$ with classes $[x]$.\\
$Z^1,\ B^1,\ H^1$ & Cocycles $\ker H_X$, coboundaries $\im H_Z^\T$, and cohomology $Z^1/B^1$.\\
$\Hphys,\ \Hcode,\ U|_{\Hcode}$ & Physical space, code space, and restriction of $U$ to the code space.\\
$\ket{[x]}$ & Encoded basis state $|B_1|^{-1/2}\sum_{b\in B_1}\ket{x\oplus b}$.\\
$X(u),\ Z(z)$ & Pauli strings with $X(u)\ket x=\ket{x\oplus u}$ and $Z(z)\ket x=(-1)^{\pair zx}\ket x$.\\
$\gamma_a,\ z_a,\ \lambda_a$ & Logical $X$ and $Z$ representatives with $\pair{z_a}{\gamma_b}=\delta_{ab}$, and logical coordinate $\lambda_a([x])=\pair{z_a}{x}$.\\
\addlinespace
\multicolumn{2}{@{}l}{\textbf{Phase functions and logical gates}}\\
$\Ph_m(V),\ \CH^{(m)}$ & Group of phase functions $V\to\Zm$ of level $m$ \eqref{eq:normalform}, and level $m$ of the Clifford hierarchy.\\
$\Delta_u f,\ \delta_d f$ & Difference $f(x\oplus u)-f(x)$ and polarization $(\Delta_{u_1}\cdots\Delta_{u_d}f)(0)$.\\
$\pi_z(x),\ x^\calS$ & Parity $\pair zx$ as an integer in $\{0,1\}$, and monomial $\prod_{i\in\calS}x_i$.\\
$\Ug{f},\ \CPh{\calS}$ & Diagonal gate with phase $\exp(2\pi\ii f(x)/2^m)$ on $\ket x$, and elementary gate on the qubits in $\calS$.\\
$c_\calS,\ \bar c_J$ & Exponents of the physical and logical elementary gates.\\
$\Lg_m(\Hcode),\ \Trv_m(\Hcode)$ & Phase functions of the code-preserving circuits and of the logical identity operators.\\
$\lpol f,\ f_J,\ f_0$ & Logical phase function $\lpol f([x])=f(x)$, and the physical implementations $f_J=2^{|J|-1}\prod_{a\in J}\pi_{z_a}$ and $f_0=\sum_J\bar c_Jf_J$.\\
$(\mathrm L\ell)$ & Condition of order $\ell$ for code preservation \eqref{eq:hier-finite}.\\
$M,\ \psi$ & Symmetric binary matrix of the level-two phase function \eqref{eq:quadraticnormal}, and the matrix with $MH_X^\T=H_Z^\T\psi$.\\
\addlinespace
\multicolumn{2}{@{}l}{\textbf{Circuit depth}}\\
$\inc(f),\ \dep(f)$ & Maximum incidence and minimum depth of $\Ug{f}$ (\Cref{sec:hier-depth}).\\
$\inc(\lpol f),\ \codedep(\lpol f)$ & Their minima over the physical implementations of $\lpol f$.\\
$f_{\mathrm{tr}}$ & Phase function $\sum_ic_ix_i$ of a transversal circuit of $\mathrm S$ gates on one block.\\
\bottomrule
\end{tabularx}
\end{table}

\end{document}